\documentclass[rmp,aps,showpacs]{revtex4}
\usepackage{graphicx}
\begin{document}

\title{A Statistical Description of Parametric Instabilities
with an Incoherent Pump}

\author{D. Pesme,(a) R. L. Berger,(b) E. A. Williams,(b) A.
Bourdier,(c) and A. Bortuzzo-Lesne(d)}

 \affiliation{(a)  Laboratoire Pour L'Utilisation Des Lasers Intenses
(LULI), Unit\'e Mixte de Recherche No. 100, CNRS - Ecole
Polytechnique, 91128 Palaiseau Cedex, France\\
(b)  Lawrence
Livermore National Laboratory, University of California, P.O. Box
808, L-472, Livermore, CA  94551\\
(c)  Commissariat a l'Energie Atomique, Centre d'Etudes de
Bruyeres-le-Chatel, France (also at Laboratoire de Physique des
Milleux Ionises,Ecole Polytechnique, Centre National de la Recherche
Scientifique, 91128 Palaiseau Cedex, France)\\
(d) Laboratoire de Physique Theorique des Liquides, Universite
Pierre et Marie Curie, 75252 Paris Cedex 05, France}

\begin{abstract}
    The effect on parametric instability growth of pump wave incoherence is
treated by deriving a set of equations governing the space-time
evolution of the ensemble-average coupled-mode amplitudes and
intensities.  Particular attention is paid to establishing the
regions of validity of the statistical description.  Thresholds,
growth rates, and amplification rates are given for both spatially
and temporally incoherent pump waves.  Both absolutely and
convectively unstable modes are considered.  The statistical results
are verified where appropriate by numerical integration of the
coupled-mode equations with different models of pump incoherence.
\end{abstract}

\date{June 1995}

\maketitle

\noindent \section{ Introduction}

 The requirements of laser fusion targets for high power lasers with
good laser beam uniformity has driven a quest for new techniques for
smoothing the intensity variations on the target surface.  Early
attempts at beam smoothing[1] were not well characterized but more
systematic techniques[2-6] have demonstrated significant
improvements in beam uniformity.  All techniques involve
introduction of phase nonuniformities which replace the normal beam
pattern, typically containing substantial hotspots, with a
smaller-length-scale speckle pattern.  Further addition of bandwidth
to the laser provides temporal smoothing of the speckle.  The
primary motivation for investing in these smoothing schemes is to
reduce the initial nonuniformities that can seed fluid instabilities
such as Rayleigh-Taylor, yet there is also palpable interest in
reducing the strength of laser plasma instabilities such as
stimulated Raman or Brillouin scattering or two plasmon decay.
Supplying a theoretical framework for understanding these laser
plasma interactions with smoothed laser beams is the task we
undertake in this article.  In a subsequent article[7], the results
obtained here will be applied to particular instabilities in
geometries of interest.

With the usual assumptions about slow variation of parameters with
respect to the frequencies $\omega_j$ and wave vectors k$_j$ of the
wave, the coherent and incoherent problems can be studied within the
context of the coupled mode equations (II.10 or II.7).  The wave
group velocities and damping rates are denoted by V$_j$ and $\nu_j$
respectively; the strength of the coupling between the waves,
$\gamma_0$, proportional to the pump or laser wave amplitude, has
the dimensions of frequency and is the rate at which the waves grow
without damping in an infinite homogeneous plasma. As the reader
will remember, for a coherent driver, both convective and absolute
instability may occur provided the laser intensity, i.e.
$\gamma_0^2$, exceeds certain threshold values set by losses [8]. In
an unbounded plasma, the convective coherent threshold is

$$
\gamma_0^2 = \nu_1 \nu_2 .  \eqno (I.1)
$$

\noindent Absolute instability requires that the decay waves be
oppositely directed $(V_1 V_2 < 0)$, and, the absolute coherent
threshold is

$$
\gamma_0 = \sqrt { \mid V_1 V_2 \mid \over 2}  ( {\nu_1 \over \mid
V_1 \mid} + {\nu_2 \over \mid V_2 \mid} ) \eqno (I.2)
$$

\noindent Temporal and spatial incoherence introduce two additional
parameters, the temporal bandwidth $\Delta \omega_0$ and the spatial
bandwidth (or spread in wavevectors) $\Delta k_0$ which is inversely
related to the correlation length $X_c$.

The effects of both temporal and stationary spatial pump incoherence
on parametric instabilities in homogeneous and inhomogeneous plasma
has been studied extensively both theoretically and experimentally
over the past thirty years [9-68]. The first approach was quite
naturally to consider the purely temporal problem with coupled mode
equations in homogeneous plasma. Zaslavskii and Zakharov [9]
considered the generic undamped decay instability with methods
developed earlier for studying the relaxation of two level molecular
systems driven by an external field [10]. They found that the
convective growth rate was reduced, in the limit $\gamma_0\ll
\Delta\omega$, from $\gamma_0$ to $\gamma_0^2/\Delta\omega$. Both
Valeo and Oberman [11] and Tamor [12] addressed this problem with
different methods and obtained the same result. Tamor included a
damping rate on the Langmuir wave coupled to an acoustic wave and
found a dispersion relation where the damping rate for the Langmuir
wave $nu_p$ is replaced by $\nu_p+2\Delta\omega$. He noted
reasonably that the bandwidth was unimportant unless it exceeded the
damping rate $\nu_p$ but made no comment on the fact that the
bandwidth appeared asymmetrically in his dispersion relation; that
is, the apparent damping of the acoustic wave was unaffected. In a
series of articles [13-15],
 Thomson made a number of important contributions. First, he noted
that for equal damping on both modes, the threshold was
$\gamma_0^2=\nu\Delta\omega$
 if $\Delta\omega\gg\nu$ and incorrectly
speculated that, for unequal damping, the threshold was
$\gamma_0^2=\Delta\omega\nu_1\nu_2/{\rm min}(\nu_1,\nu_2)$. This
guess was based on the assumption that, for growth to occur in
either mode, the average amplitude of both must grow.

The correct solution was presented by Thomson[15] later using an
exactly solvable model [69]. For the purely temporal problem, the
ensemble-averaged mode amplitude equations showed these equations
decouple and have ${\it different}$ thresholds, i.e.

$$
\gamma_0^2 = \Delta \omega_0 \nu_1  , \eqno (I.3)
$$

\noindent and

$$
\gamma_0^2 = \Delta \omega_0 \nu_2 \eqno (I.4)
$$
Thomson noticed and explained the asymmetry as due to averaging over
the rapidly varying phase of the mode amplitude.  He obtained a more
appropriate threshold by solving the equations for the ensemble
averaged intensities. The solution showed that the lower of the
thresholds (I-3,4) was appropriate.  We show, as did Thomson, that
the incoherent growth rate for the average intensity is

$$
\gamma_{inc} = 2 \gamma_0^2 / \Delta \omega_0 \eqno (I.5)
$$
provided $\nu_1 < \gamma_{inc} < \nu_2$.  This answer is expected,
given that the average amplitudes grow with rate $\gamma_0^2 /
\Delta \omega_0$. However, Thomson did not point out the interesting
fact (found also by Laval et al.[16] in the space-time problem when
$|V_1| = |V_2|$) that, if $\gamma_{inc} > \nu_1,\nu_2$, the
incoherent growth rate is
$$
\gamma_{inc} = 4 \gamma_0^2 / \Delta \omega_0 ,  \eqno (I.6)
$$
that is twice the expected rate.  We will comment more on this in
Secs. V-VII. Thomson then considered the space time problem with the
ensemble averaged mode amplitude equations and again found an
asymmetry in the thresholds for the absolute instability when
$\Delta \omega_0 >> \nu_j$. Thomson naturally assumed that the
appropriate intensity threshold, in analogy with the temporal
problem, was the lower one.  Later work showed this to be incorrect.

Laval et al. reconsidered [70] the space -time problem for the
evolution of the intensities both by using the Bourret approximation
and, in the case $V_0 = \infty$, the exactly solvable Kubo-Anderson
process (KAP). They showed that the intensity thresholds are in fact
much lower than the amplitude thresholds when $\Delta \omega_0 >>
\nu_j$ and, in usual cases of interest ($\nu_2 > V_2 \nu_1 / V_1$),
the incoherent absolute threshold equals the product of the
bandwidth and the damping on the slow wave

$$
\gamma_0^2 = \Delta \omega_0 \nu_2 .  \eqno (I.7)
$$
In a later publication [17], Thomson applied the results to
stimulated Raman forward scatter and incorrectly concluded, as we
discuss subsequently, that small amounts of bandwidth, comparable to
its growth rate, can suppress this slowly-growing instability.
During this time period, the nonlinear evolution of instabilities
driven by broadband pumps was investigated by using particle-in-cell
(PIC) simulations [18-20]. These studies demonstrated a reduction in
the stimulated Brillouin reflectivity from 60\% to less than 10\%
with 5\% bandwidth. Because these plasmas were strongly
inhomogeneous, no reduction is expected for modest bandwidth,
$\Delta\omega\simeq\gamma_0$, as shown by Thomson [15]. However,
with $\Delta\omega/\omega_0\simeq .05$, the line separations in
these simulations are larger than an acoustic frequency and each
line acts independently. Kruer et al. [19] also suggest that
bandwidth in an inhomogeneous plasma will be ineffective unless
$\Delta\omega\geq V_0/l$ where $l$ is the interaction length set by
plasma inhomogeneity. This can be rephrased in terms of the
correlation length, $x_c=V_0/\Delta\omega<=l$ where
$l=\gamma_0/\kappa^\prime(|v_1v_2|)^{1/2}$ and
$\kappa^\prime=dk_\Delta(x)/dx$ for $k_\Delta=k_0-k_1-k_2$;
$k_\Delta=0$ is the condition for phase matching. Kruer's condition
is difficult to satisfy for typical laser systems. We return to this
subject in our discussion of spatial incoherence effects. The
correct theory [15] for convective amplification in an inhomogeneous
plasma showed that the growth rate was reduced but the
ensemble-averaged mode amplitude convectively saturates at the same
level as the coherently driven amplitude. (Thomson actually
misstated his result, although his analysis was correct. Later work
[45,49] that solved the coupled mode equations clarified the
answer.)

Estabrook et al. also considered, with PIC simulations, the effect
of laser bandwidth on stimulated Raman scattering-- again in
inhomogeneous plasma. Bandwidth was represented by a series of
equally spaced laser lines. A reduction in the reflectivity was
found when the line separation was greater the growth rate. Each
line acted as an independent pump, and the intensity was low enough
that the instability was not strongly saturated.  Direct comparisons
between homogeneous plasma theory and PIC simulations of SRS in a
plasma slab were made by Forslund et al. [21] with good agreement
for the dependence of the SRS growth rate on the bandwidth of the
frequency modulated laser. These authors pointed out that,
$\Delta\omega/\gamma_0\geq 2$, despite a fourfold reduction in the
growth rate, only a modest reduction in the power reflected was
observed in the final state. A further increase of bandwidth to
$\Delta\omega/\gamma_0\geq 10$ brought the instability below
threshold. Later PIC work [24] with multiple lines also showed good
agreement with theory for the Raman backscatter growth rate when
$\gamma_0/\Delta\omega\ll 1$ at several combinations of density and
laser intensity. These simulations also showed a monotonic decrease
with bandwidth in the absorption into hot electrons due to SRS and
in the SRS reflected power. Less effect on the forward SRS was
observed, consistent with the theory we discuss subsequently.  Other
theoretical work on pump bandwidth effects concerned applications to
specific targets [30], and the application to induced spatial
incoherence (ISI) [44]. The effects of both temporal and spatial
bandwidth in a 1D inhomogeneous plasma [45] for a phase mismatch,
$k_0(x)-k_1(x)-k_2(x)=k_\Delta(x)$, varying linearly,
$k_\Delta(x)=\kappa^\prime x$, or quadratically,
$k_\Delta(x)=\kappa^{\prime\prime} x^2/2$, was treated by analytical
and numerical methods. In reference [47], the Green's function
formalism of Brissaud and Fritsch [69], used by other authors
[15,16,38] was reformulated in the language of effective Hamiltonian
matrices. In a second paper[48], this formalism was used to
investigate the competition between temporal bandwidth and
inhomogeneity for a variety of parametric processes, including two
plasmon decay.  Early experiments [25,26] that attempted to observe
laser bandwidth effects utilized plasmas that were too nonuniform to
expect observable effects. Later experiments used gas jet targets
[27,29] or used microwave plasmas [28]. Clayton et al.  and Giles et
al. divided the laser power into two lines and found a striking
reductions in reflected SBS power compared to a single line with the
same total laser power. The lines were separated by much more than a
growth rate.

It was recognized early [31-33] that random spatial modulation of
the phase mismatch, $\phi=\int dx k_\Delta(x)$, could also reduce
parametric instability growth rates. Using methods similar to the
temporal problem, Kaw et al.  used the steady-state limit of the
coupled-mode equations (II-10) to find a convective gain rate,
$\gamma_0^2/4\Delta|v_1 v_2|$, where $\Delta=<\delta k^2>l_c$. Here
$\delta k$ is the wavevector mismatch induced by the random
variation in the plasma properties and $l_c$ is the correlation
length for the random process. Further work [34-38,40-42] on random
fluctuations of the phase concerned its effect on the growth in
inhomogeneous plasmas where, for example, the stabilization of
absolute modes by linear gradients in the phase could be undone by
these fluctuations. In this article, we concentrate on the effects
in uniform plasmas.

Beam smoothing techniques use not only temporally incoherent but
also spatially incoherent pump waves.  In the focal plane, the laser
wave can be considered a sum of randomly phased plane waves with a
spread in wavevectors, $\Delta k_0$.  In the limit of no temporal
bandwidth, the spatial interference pattern of the pump wave
envelope is stationary and is related to the effect of stationary
plasma turbulence on parametric instability treated by Williams et
al.[39] Using a novel approach (unrelated to the methods used in the
present analysis), they considered the threshold and growth rate for
absolute modes in a finite system of length L.  This analysis finds
that for sufficiently large spatial bandwidth, $\Delta k_0$, the
fastest growth rate is reduced from the coherent absolute rate,
$$
\gamma = \gamma_0 { \sqrt { \mid V_2 V_1 \mid} \gamma_0 \over \mid
V_1 \mid + \mid V_2 \mid} \eqno (I.8)
$$
\noindent when $\nu_j = 0$ to the spatially incoherent absolute rate
$$
\gamma = \gamma_0^2 {\rm ln} ( \mid \Delta k_0 \mid L) / \mid \Delta
k_0 \mid ( \mid V_1 \mid + \mid V_2 \mid ) , \eqno (I.9)
$$

\noindent (To obtain I-9 from Eq. 52 of Williams et al., the reader
is advised that the correct normalization for the growth rate is
$\gamma_0 \sqrt { \mid V_1 V_2 \mid} / ( \mid V_1 \mid + \mid V_2
\mid ).$ Their Eq. 3 is incorrect.)  The incoherent result in Eq.
(I-9) must of course be less than the coherent one in Eq. (I-8). The
unexpected feature of (I-9) is the length dependent logarithmic
factor which is related to the fact that the rate (I-9) is not the
average rate sought in our analysis but the largest rate expected in
a system of length L.  Except for the logarithmic factor, we recover
the scaling of the result of Eq. (I-9). In the present article, we
derive average amplitude equations by using the Bourret
approximation[70] and average intensity equations by using the so-
called random phase approximation (RPA).[71-74]  The major objective
of this article is to provide a unified theory of spatial and
temporal incoherence effects on parametric instabilities.  The
approximations and assumptions necessary to arrive at the RPA
equations (V.32) and the Bourret equations (V.1 4) that form the
basis for theoretical results in Sec. VI are carefully explored and
systematically presented in Sec. III through V.  A comprehensive set
of results is presented for the threshold and growth rate of both
absolute and convective instability in the incoherent limit.  The
domains of validity of the statistical approximations are explicitly
noted and outside these domains the coherent results are shown to
apply.  Moreover the same analysis is done for the spatial
amplification of convectively unstable waves. Finally, we present in
Sec. VII numerical solutions of the fundamental set of equations
(II-10) for particular models of incoherence that illustrate the
meaning of the averaging procedures and verify the main results.

The RPA dispersion relation (VI.1) for an infinite system obtained
from the RPA equations (V.32) form the basis for the analysis of
growth rates and thresholds in Section VI.  There are two parameters
that play a role in the ensemble average equations that measure the
effective bandwidth, $\Delta \omega_j, j = 1,2$ where
$$
\Delta \omega_j \equiv \Delta \omega_0 \mid (1 - V_{jx} /V_0) \mid +
\mid \Delta k_0 \cdot V_j \mid \eqno (I.10)
$$

\noindent In the general case of spatial and temporal incoherence,
there can be a gap between the domain of validity for the RPA
dispersion relation and the coherent one.  The coherent domain is
the whole region where either spectral width $\Delta \omega_j$ is
less than the corresponding damping rate or growth rate.In this
intermediate region, the average amplitude dispersion relations
(VI.2) are valid, and the more unstable one agrees with the RPA
dispersion relation within a factor of two.  It is on this basis
that we argue that the RPA dispersion relation can be used in the
whole domain (denoted the incoherent domain) complementary to the
coherent domain.

In the remainder of Sec. VI, a description of parametric
instabilities is given including the conditions for absolute and
convective instability. Both early time behavior, which is dominated
by convective instability, and long time behavior, which is
dominated by absolute instability (if it exists) or by spatial
amplification are considered.  It is worth noting that, in the
incoherent domain, the average amplitude dispersion relations (VI.2)
allow only convective solutions whereas the RPA dispersion relation
(VI.1) allows the possibility of absolute instability if a threshold
can be exceeded.  However, the simpler average amplitude equations
do provide the convective mode threshold (VI.5), growth rate (VI.7),
and spatial amplification factor (VI.13) and (VI.16) (within better
than a factor of two) in agreement with those obtained in Sec. VI
using the RPA dispersion relation.

Absolute instabilities are of particular practical interest because,
above threshold, the only limit to their growth is the finite laser
energy or other nonlinear effects.  On the other hand, an absolute
instability generally has a larger threshold to overcome losses to
collisional or Landau damping than a convective instability.  For a
coherent laser wave, the threshold laser intensity is determined by
setting the spatial growth rate $\gamma_0 / \mid V_1 V_2 \mid ^{1/2}
$ equal to one-half of the sum of the spatial loss rates $\nu_j /
\mid V_j \mid $ for the decay modes (Eq. I-2). In addition, as given
by Eq. (I-8), the absolute instability growth rate is smaller than
the growth rate $\gamma_0$ by approximately the ratio $(\mid V_2 /
V_1 \mid )^{1/2} $ in the usual case $\mid V_2 / V_1 \mid << 1$. The
exact formula is also given by Eq. (VI.11).  With incoherent laser
beams, the threshold is still determined by setting the effective
spatial growth rate equal to the sum of loss rates but now the
spatial growth rate is ${\rm Max} (\gamma_0^2 /2 \Delta \omega_j
\mid V_j \mid )$, that is, the maximum spatial growth rate for the
average amplitude equations.  The general formula is given by
(VI.9).  Note in the special but interesting case that $\Delta
\omega_1 = \Delta \omega_2 = \Delta \omega_0$, temporal incoherence

increases the absolute threshold if $R_a = (\gamma_0 / \Delta
\omega_0) (\sqrt { \mid V_1 V_2 \mid} / {\rm Min} \mid V_j \mid ) <
1$ which exceeds the naive criterion by the square root of the group
velocity ratio.  We observe that the statement $R_a < 1$ is
equivalent to requiring that the coherent spatial growth length
$\mid V_1 V_2 \mid ^{1/2} / \gamma_0$ exceed the larger coherence
length $\mid V_1 / \Delta \omega_1 \mid$.  A rigorous
application[46] of the Bers and Briggs criteria[75] arrives at this
same condition as a necessary condition (VI.10) for the incoherent
limit to apply.  Above the incoherent threshold, the incoherent
absolute growth rate is roughly $\gamma_0^2 / \Delta \omega_1$, (the
exact formula is given by Eq. (VI.8)) i.e. approximately the same
expression as the incoherent convective growth rate but with a
different domain of validity.

There is a limit where the effective bandwidths $\Delta \omega_j$
are determined only by spatial incoherence.  Then $\mid \Delta
\omega_j / V_j \mid = \Delta k_0$  is independent of the group
velocity and the growth rate,
$$
\gamma ^{abs}_{\langle a^2\rangle} = 4 \gamma_0^2 / \mid \Delta k_0
\mid ( \mid V_1 \mid + \mid V_2 \mid ) , \eqno (I.11)
$$
agrees with the scaling found previously and given in Eq. (I-9).
Thus our general threshold and growth rate formulae indeed recover
the correct limits of purely spatial and purely temporal incoherence
derived previously. The coherent convective threshold, always lower
than that for absolute instability, is given by the requirement that
the growth rate $\gamma_0$ be greater than the mean loss rate
$(\nu_1 \nu_2 ) ^{1/2}$.  The same criterion for an incoherent pump
applies if the laser intensity is reduced by the ratio of line
widths factor $R_c \equiv \nu_1 / \Delta \omega_1 + \nu_2 /\Delta
\omega_2$.  The effective bandwidth $\Delta \omega_j$ must be larger
than the corresponding damping rate $\nu_j$ for each mode to
increase the threshold.  Otherwise the coherent threshold applies
(I-1). Far above the incoherent threshold, the incoherent convective
growth rate is reduced by the factor $\gamma_0 / {\rm Min} (\Delta
\omega_j)$. Note that when $\Delta \omega_j \simeq \Delta \omega_0$
the convective and absolute incoherent growth rates are equal in the
overlapping domain of validity.  More exact expressions are given by
(VI.5) - (VI.7).

For many cases of practical interest $\Delta \omega_j \sim \Delta
\omega_0$ because the spread in laser wavenumbers is sufficiently
small and/or the group velocity is small enough that $\mid \Delta
k_0 \cdot V_j \mid << \Delta \omega_0$ and either $\mid V_j / V_0
\mid << 1$ or $V_j / V_0 < 0$.  Then temporal incoherence is the
dominant stabilizing influence.  However two special cases deserve
mention.  Examination of the expression (VI.8b) for the absolute
instability growth rate above threshold shows that if the maximum
value of the coherence length $V_j / \Delta \omega_j$ occurs for the
minimum group velocity, the growth rate is strongly reduced
$\gamma^{abs} = V_2 \gamma_0^2 / V_1 \Delta \omega_2$ in the usual
case $\mid V_2 \mid << \mid V_1 \mid$.  This case can occur if there
is a large spread in wavenumber accompanied by a weak temporal
incoherence so that $\gamma^{abs} = \gamma_0^2 / \mid \Delta k_0
\cdot V_1 \mid$.  The transient or convective growth rate for the
same case (in the incoherent limit) is larger, $\gamma conv =
\gamma_0^2 / \mid \Delta k_0 \cdot V_2 \mid$. The other unusual case
occurs in the case of temporal incoherence for forward scatter when
$V_1 \sim V_0$.  Then $\Delta \omega_1 = \Delta \omega_0 (1 - V_1 /
V_0) << \Delta \omega_0$ so that the coherent convective growth
still occurs even if $\Delta \omega_0 \sim \gamma_0$.  The case of
practical interest is stimulated Raman forward scatter.[22-24]

The fate of absolutely unstable modes necessarily requires
consideration of nonlinear effects but, in a bounded plasma, the
maximum amplitude of convectively unstable waves may be found by
computing the spatial amplification rate.  In general there are two
roots that are most easily obtained by using the average amplitude
equations (VI.2) but are also accessible from the average intensity
equations (VI.1).  A bandwidth $\Delta \omega_0 > \gamma_0, \nu_j$
is usually sufficient to reduce the spatial growth rate.  For
moderate intensities, i.e., below any absolute threshold, the
spatial gain coefficient $\kappa \sim \gamma_0^2 / \Delta \omega_0
V_1$ (where $\mid V_1 \mid > \mid V_2 \mid$) is reduced by $(
\gamma_0 / \Delta \omega_0 ) (V_2 / V_1)^{1/2}$ in the weak damping
case.  At higher intensities, a second root with gain coefficient
$\kappa \sim \gamma_0^2 / \Delta \omega_0 V_2$ occurs provided
$\Delta \omega_0 > \gamma_0 ( \mid V_1 / V_2 )^{1/2}$ and
$\gamma_0^2 > \Delta \omega_0 \nu_2$, that is, above the absolute
mode threshold when $V_1 V_2 < 0$.  Therefore both a convective and
absolute mode exist; and, interestingly, the convective mode has a
spatial gain rate that exceeds that of the absolute mode.  A more
detailed discussion and exact formulae are given in Eqs. (VI.12) -
(VI.18).

In the general case, it is difficult to summarize the effects of
incoherence on convective and absolute stabilities.  However, it is
useful to consider a common case of practical interest where $\mid
V_1 / V_2 \mid >> 1$ and $\Delta \omega_1 \simeq \Delta \omega_2
\equiv \Delta \omega_0$.  Then diagrams showing stability regions
can be constructed with axes $\Delta \omega_0 / \gamma_0$ and $\nu_2
/ \nu_0$, i.e. moving towards more incoherence in one direction and
towards more damping in the other.  Figure VI.1 is such a diagram
for convectively unstable modes.  There are four regions: coherently
unstable, incoherently unstable with reduced growth rate, coherently
stable, and incoherently stabilized.  In Fig. VI.2, the diagram for
absolutely unstable modes is drawn with four analogous regions. This
figure graphically shows that, in only a small region of parameter
space does incoherence reduce the growth rate but not completely
stabilize. Figure VI.3 shows the different regions for spatial
amplification. Finally, in Fig. VI.4, an overall diagram is shown
for all regions for convective, absolute, and spatial amplification.

To this point, we have presented results for infinite homogeneous
plasma.  For a large enough system, these results are a good guide
to the behavior in finite systems.  Nonetheless, real plasmas are
finite and it is well known that a threshold length is necessary for
absolute instability. Moreover in Sec. VII, numerical solutions of
the coupled mode equations (II.10) are presented in a necessarily
finite system.  Thus partly as a guide to the numerical solutions,
we find the threshold length for absolute instability in the
coherent and incoherent limit as obtained by solving Eq. (II-10) or
Eq. (V.32) as appropriate.  It is an interesting feature that the
normal modes in the slab, sinusoidal in the coherent limit, are
exponential in the incoherent limit.  The threshold length given by
Eq. (VI.30) increases as expected with bandwidth above the coherent
threshold length (VI.29).

Several features of the statistical description that provoked
further analysis were the "unexpected" factor of two that appeared
in the growth rate for the ensemble-average intensity, the question
of the validity of the RPA description in the intermediate domain,
and the applicability of these results to analysis of experiments
using beam smoothing techniques.  These aspects were examined in
Sec. VII by integrating directly the coupled mode equations (II.10)
or (VII.1) with particular choices of random processes to represent
the pump wave incoherence.  For the purely temporal case, an
analytic solution for the distribution function of mode amplitude
(Eq. (VII.10) is obtained which is remarkably broad if the damping
rates are negligible.  In fact, the width of the Gaussian
distribution is equal to the mean.  Thus, higher powers of the
amplitude, e.g., the intensity grow at faster rates than the mean
which gives rise to factor of two mentioned earlier.  We compare
this distribution to a numerically generated one in Fig. 6 for the
same parameters.  On the other hand, with sufficient damping, this
factor of two does not appear; that is, the average intensity and
amplitude grow at corresponding rates.  In this case, the
distribution is strikingly narrowed as also shown in Fig. 6.

In the space-time problem this factor of two also occurs for
absolute instability driven by an incoherent pump when the decay
wave group velocities are equal in magnitude and opposite in sign.
We have verified our supposition by numerical integration that once
again the distribution is broad for $\mid V_1 \mid = \mid V_2 \mid$
but is much narrower if $\mid V_1 \mid >> \mid V_2 \mid$ as shown in
Fig. 7.

The validity of the RPA description in the intermediate domain was
examined numerically in Sec. VII.B, by considering the model of a
spatially incoherent pump driving a pair of decay waves satisfying
the intermediate domain inequalities (VII.13).  Although we did show
that the RPA equations appear valid, we also discovered that the
distribution of mode intensities (Fig. 8) is unusual in that it
consists of a slowly decreasing tail on a distribution with a peak
at nearly zero growth.

We also show in Sec. VII.C that the standard model of an ISI beam,
Eq. (VII.15), which has both phase and intensity variation, can be
treated as an incoherent pump wave provided the temporal and spatial
bandwidth are large enough.  Thus with the appropriate
identification of experimental parameters with $\Delta \omega_0$ and
$\Delta k_0$, the formulae in Sec. VI can be applied to
experiments.[51-65]

A few remarks are in order regarding the derivation of the
statistical equations that form the basics for the results outlined
above.

The analysis begins in Section II with the completely nonlinear
coupled-mode equations (II.2) appropriate for the case when the pump
and decay waves are weakly coupled and weakly damped.  In the linear
analysis of this article, the pump wave is unaffected by the decay
waves and the characteristic growth rate of the parametric
instability is simply related to the pump amplitude at its mean
wavevector (II.5).  Normalization of the decay wave amplitudes to
the average pump wave energy yields the linearized coupled-mode
equation (II.7) in Fourier space.  From these equations, the
envelope equations (II.9) in Fourier space or (II.10) in real space
are obtained by expanding the mode amplitudes about the value at the
mean wavevector.  These envelope equations are used in Sections
III-V to obtain equations for the ensemble-averaged mode amplitudes
and intensities.

If the pump wave has a distribution of wavevectors, then a given
pair of decay waves will be frequency matched to only a portion of
the pump wave spectrum.  Thus one is naturally led to consider the
frequency mismatches (III.1) for a general triplet of wavevectors or
the mismatches (III.2) at the mean value of the decay wavevector.
It is assumed that there exist triplets for which there is exact
matching so that the mismatch near the mean wavevectors is small.
The maximum value of the mismatches at the mean decay wavevector
determines that the interaction is coherent if the mismatches are
both small in the sense defined by III.6.

Two approaches to generating equations governing the ensemble
averaged behavior of the instability are used in this article.  The
first employs the Bourret approximation[70] to obtain the dispersion
relation (IV.5) and (IV.7) for the stability of the
ensemble-averaged mode amplitude $<a_j>$.  Each equation for $<a_j>$
is simple in that it does not involve the other but it does involve
integrals over the spectral density of the pump wave and the
frequency mismatch.  In the incoherent limit or Markov limit defined
by (IV.17), the dispersion relation takes the particularly simple
and well- known form given by Eq. (IV.13) which states that the
coupling between the waves is reduced by the ratio of the
characteristic growth rate to the maximum mismatch $\Delta_{i~max}$
for $i \equiv 1, 2$.  For the one dimensional case with a pump
spectrum that is Lorentzian, e.g. a Kubo-Anderson Process, the
dispersion relations (IV.22) are the familiar ones derived
previously[15-16] and the mismatch $\Delta_{i~max}$ is equivalent to
the spectral widths $\Delta \omega_i$ defined by (IV.20) that are,
within at most a factor of two, equal to the pump wave bandwidth
$\Delta \omega_0$ (except for the special case of forward scatter
where the $\Delta \omega_1 << \Delta \omega_2$).  Note in one
dimension, the problem of spatial and temporal incoherence are not
independent in fact so our restriction to one dimension must yield
the results[16] of Laval et al. However in our numerical models,
this connection is broken for convenience and tractability.

The average amplitude equations (IV.5 and IV.7) are somewhat
unsatisfying because $<a_1>$ and $<a_2>$ evolve independently.  For
the case of temporal incoherence alone, this untidiness was remedied
by obtaining equations for the ensemble averaged intensities which
are coupled and symmetric.  Nevertheless, the result obtained for
convective instabilities with only temporal incoherence is just that
obtained by consideration of the most unstable average amplitude.

In Section V.B, the same procedure can be followed to obtain Bourret
equations for the ensemble-averaged spectral densities (V.25) that
are related to the mode intensities by (V.2).  As with the amplitude
equations (IV.5 and (IV.7), these equations involve integrals over
the pump spectral density and the frequency mismatch; but, in
addition, the integral contains the spectral density of the other
decay wave at the pump shifted wavenumber. Two further
approximations can be made:  first, the Markov assumption valid
provided that the mismatch is larger than the maximum of the growth
rate and damping rates, i.e. ${\rm Min} (\Delta_{i max}) > {\rm Max}
(\gamma_0, \nu_j)$ removes the spatial and temporal growth rate from
the integrand; second, the assumption that the decay wave spectral
densities are slowly varying functions allows these densities to be
evaluated at the mean wavenumber at exact matching.  The first
approximation yields the Random Phase Approximation (RPA) equations
(V.28); and the subsidiary second approximation yields the set
(V.32) with which the instability analysis is done in Section VI. As
they must do, the RPA equations (V.28) are shown to reduce to the
correct intensity equations derived using the Bourret approximation
when the pump wave group velocity is infinite.  For the purely
temporal incoherence modeled by the Kubo-Anderson process[69], this
latter result is exact.

Our conclusions are presented in Sec. VIII.  There, we also give
some examples of current interest and consider the special case when
the temporal bandwidth is larger than one of the mode frequencies.

\section{The Coupled Mode Equations}

\subsection{  The Coupled Mode Equations in Fourier Space.}

In this article we limit our discussion to the coupling of three
wave-packets which are both weakly coupled and weakly damped.  We
assume that each Fourier component $A_k^i$ of the wave packet i
(with i = 0, 1, 2)) behaves in lowest order as a normal mode; i.e.,
it oscillates in time like ${\rm exp}(-i \omega_k^i t)$, where
$\omega_k^i$ is the real part of the frequency $\Omega_k^i \equiv
\omega_k^i - i\nu_k^i$ which characterizes the wave-packet i and
which satisfies the dispersion relation $D_i (\Omega_k^i , k) = 0$.
The assumption of weakly coupled and weakly damped modes can be
written quantitatively as
$$
\mid \omega_k^i \mid >> \nu_k^i , \gamma \eqno (II.1)
$$
where $\nu_k^i$ is the linear damping increment of the Fourier
component $A_k^i$, and where $\gamma$ denotes the inverse of the
characteristic time for nonlinear evolution of the coupled modes.
In this limit the equations describing the coupling of the three
wave packets are of first order in time.  In Fourier space they have
the following general form:
$$
\left[ \partial_t + i \omega_{{\bf k}_0}^0 + \nu_{{\bf k}_0}^0
\right] A_{{\bf k}_0}^0 = - \int d{\bf k}_1 V_{{\bf k}_0,{\bf k}_1}
A_{{\bf k}_1}^1 ( A_{{\bf k}_0 - {\bf k}_1}^2  ) \eqno (II.2a)
$$

$$
\left[ \partial_t + i \omega_{{\bf k}_1}^1 + \nu_{{\bf k}_1}^1
\right] A_{{\bf k}_0}^0 = \int d{\bf k}_0 V_{{\bf k}_0,{\bf k}_1}
A_{{\bf k}_0}^0 ( A_{{\bf k}_0 - {\bf k}_1}^2  )^* \eqno (II.2b)
$$

$$
\left[ \partial_t + i \omega_{{\bf k}_2}^2 + \nu_{{\bf k}_2}^2
\right] A_{{\bf k}_2}^2 = \int d{\bf k}_0 V_{{\bf k}_0,{\bf k}_0}
A_{{\bf k}_0}^0 ( A_{{\bf k}_0 - {\bf k}_2}^1  )^* \eqno (II.2c)
$$

\noindent where $A_{{\bf k}_0}^i$ is proportional to the Fourier
component of the electric field of wave i; wave 0 refers to the pump
wave and waves 1 and 2 to the decay waves.  The coupling constants
$V_{{\bf k}_0,{\bf k}_1}$ are derived from the usual field expansion
of the fluid or the Vlasov equations.

The parametric approximation consists in neglecting the RHS of Eq.
(II.2a) and in taking $A_{{\bf k}_0}^0 = A_{{\bf k}_0}^0 {\rm exp}
(-i \omega_{{\bf k}_0}^0 t$) in the RHS of Eqs. II.2b and II2c.  In
this limit the latter equations describe correctly the parametric
coupling of waves in the usual decay regime (the so-called modified
decay instability and the modulational instability[76] cannot be
described by these equations since they correspond to coupled mode
equations in which the second order partial derivative in time must
be retained).

In this article we derive the conditions for a reduction of the
parametric growth due to the pump wave incoherence.  The incoherence
of the pump wave may be either temporal, or spatial, or both.  In
the case of purely {\it temporal} incoherence, the wave-numbers
$k_0$ of the different components of the pump wave all have the same
direction; if $\Delta \omega_0$ is the spectral width in frequency
space of the pump wave and is small compared to the pump frequency,
$\Delta \omega_0 << \omega_0$, the spread $\Delta K_0$ of the
corresponding wave-number modulus $k_0 \equiv \mid k_0 \mid$ is
given by
$$
\Delta k_0 \simeq \Delta \omega_0 / V_{g0} \eqno (II.3)
$$
Here ${\bf V}_{g0}$ is the characteristic group velocity of the pump
wave and $V_{g0}$ denotes $V_{g0} \equiv \mid {\bf V}_{g0} \mid$.
In the case of purely {\it spatial} incoherence, all the
wave-numbers ${\bf k}_0$ have the same modulus $k_0 \equiv \mid {\bf
k}_0 \mid$ and are spread within a small cone, the {\it half} angle
of aperture of the latter being denoted by $\Delta \Theta_0$.  In
the general case, the mean wave-number and frequency of the pump
wave are denoted by ${\bf K}_0$  and $\omega_0 \equiv \omega _{{\bf
K}_0}^0$ respectively, and the incoherence of the pump wave is
characterized by the spectral width $\Delta k_0 = \omega_0 / V_{g0}$
and by the half-angle spread $\Delta \Theta_0$.  We will assume an
azimuthal symmetry around ${\bf K}_0$ and a cross section of the
total domain of existence of ${\bf k}_0$ is displayed in Fig. 1.

An assumption made implicitly when using Eqs. (II.2) to investigate
the incoherence effects is that the spread in frequency $\Delta
\omega_0$ is much smaller than the frequencies $\omega_{k_i}^i$ of
the various waves. Similarly, the spread in wave number $\mid \Delta
{\bf k}_0 \mid \sim \Delta \omega_0 / V_{g0} + K_0 \Delta \Theta_0$
is assumed to be small compared to wave numbers $\mid {\bf k}_i
\mid$.

We denote by $\alpha_2$ the angle between the propagation directions
of the pump wave and of the wave 1.  For a given direction
$\alpha_2$ of the wave 1, there is a unique couple $({\bf K}_1,
\omega_{{\bf K}_1}^1), ({\bf K}_2, \omega_{{\bf K}_2}^2)$ satisfying
the two resonance conditions,
$$
\omega_{{\bf K}_0}^0 = \omega_{{\bf K}_1}^1 + \omega_{{\bf K}_2}^2 ,
$$
\nobreak
$$
{\bf K}_0 = {\bf K}_1 + {\bf K}_2  . \eqno (II.4)
$$
Throughout this article we investigate the reduction of the
parametric instability corresponding to a given direction $\alpha_2$
and the capital letters ${\bf K}_1$ will be used exclusively for
this wavenumber of wave i which corresponds to the resonance
conditions (II.4), for i = 1, 2; as stated before, ${\bf K}_0$
denotes the mean wavenumber of the pump wave, and the symbol ${\bf
k}_0$  will denote a generic wavenumber of the pump wave; lastly
$\alpha_1$ is the angle between the mean pump wave-number ${\bf
K}_0$ and the wave number ${\bf K}_2$ corresponding to the resonance
conditions II.4 written for an angle $\alpha_2$.

We may now relate the characteristic growth rate of the parametric
instability corresponding to the coherent case, $\gamma_0$, for a
given angle of observation $\alpha_2$, to the mode coupling
constants and the pump amplitude by
$$
\gamma_0^2 = \mid V_{{\bf K}_0,{\bf K}_1} \mid^2 \langle \mid A_0
\mid^2 \rangle \eqno (II.5)
$$
where $\langle \mid A_0 \mid^2 \rangle$ denotes the average energy
of the pump wave in a sense defined in the next section.

Defining the dimensionless coupling constants and wave amplitudes by
$$
v_{{\bf k}_0,{\bf k}_1} \equiv V_{{\bf k}_0, {\bf k}_1} / V_{{\bf
K}_0, {\bf K}_1} \eqno (II.6a)
$$
$$
a_{{\bf k}_{\alpha}}^{\alpha} \equiv A_{{\bf k}_{\alpha}}^{\alpha} /
<
 \mid A_0 \mid^2>^{1/2} \eqno (II.6b)
$$
The dimensionless form of equations is simply
$$
\left[ \partial_t + i \omega_{{\bf k}_1}^1 + \nu_{{\bf k}_1}^1
\right] a_{{\bf k}_1}^1 = \gamma_0 \int d{\bf k}_0 v_{{\bf k}_0,
{\bf k}_1} a_{{\bf k}_0}^0 \left( a_{{\bf k}_0 -{\bf k}_1}^1
\right)^* \eqno (II.7a)
$$
$$
\left[ \partial_t + i \omega_{{\bf k}_2}^2 + \nu_{{\bf k}_2}^2
\right] a_{{\bf k}_2}^2 = \gamma_0 \int d{\bf k}_0 v_{{\bf k}_0,
{\bf k}_0 -{\bf k}_2} a_{{\bf k}_0}^0 \left( a_{{\bf k}_0 -{\bf
k}_2}^1 \right)^*\eqno (II.7b)
$$
where $v_{{\bf k}_0,{\bf k}_{\alpha}}$ is a slowly varying function
of ${\bf k}_0$ and ${\bf k}_{\alpha}$, with $v_{{\bf K}_0, {\bf
K}_{\alpha}} = 1$. Finally, the latter equations are supplemented by
the following equation for the pump wave amplitude
$$
a_{{\bf k}_0}^0 = a_{{\bf k}_0}^0 {\rm exp} (-i \omega_{{\bf k}_0}^0
t) \eqno (II.7c)
$$

\subsection{The Envelope Approximation for the Coupled
Mode Equations in Real Space} At this point we can make a connection
between the coupled mode equations (II.7) written in their general
form in Fourier space and their so-called envelope equation form in
real space.  The envelope approximation for the coupled mode
equations corresponds to the first order expansion of $\omega_{{\bf
k}_i}^i$ in a power series of (${\bf k}_i - {\bf K}_i$).  By writing
$\omega_{{\bf k}_i}^i \simeq \omega_{{\bf K}_i}^i + ({\bf k}_i -
{\bf K}_i) \cdot {\bf V}_{gi}$, and by setting
$$
a_{{\bf k}_i}^i = \hat a_{{\bf k}_i - {\bf K}_i}^i {\rm exp} (-i
\omega_{{\bf K}_i}^i t)  \eqno (II.8a)
$$
or i = 1 and 2, and
$$
a_{{\bf k}_0}^0 = \hat a_{{\bf k}_0 - {\bf K}_0}^0 {\rm exp} \left(
-i [ \omega_{{\bf K}_0}^0 + ({\bf k}_0 - {\bf K}_0) \cdot {\bf
V}_{g0} ] t \right) \eqno (II.8b)
$$
the coupled mode equations can be written as
$$
\left( \partial_t + i {\bf k}_1^\prime \cdot {\bf V}_{g1} + \nu_1
\right) \hat a_{{\bf k}_1^\prime} = \gamma_0 \int d {\bf k}_0^\prime
\hat a_{{\bf k}_0^\prime}^0 \left(\hat a^2 \right)^*_{{\bf
k}_0^\prime-{\bf k}_1^\prime} {\rm exp}-i \left( {\bf k}_0^\prime
\cdot {\bf V}_{g0}\right) t \eqno (II.9)
$$
$$
\left( \partial_t + i {\bf k}_2^\prime \cdot {\bf V}_{g2} + \nu_2
\right) \hat a_{{\bf k}_2 ^\prime} = \gamma_0 \int d {\bf
k}_0^\prime \hat a_{{\bf k}_0^\prime}^0 \left(\hat a^1
\right)^*_{{\bf k}_0^\prime-{\bf k}_2^\prime} {\rm exp}-i \left(
{\bf k}_0^\prime \cdot {\bf V}_{g0}\right) t
$$
where ${\bf k}_i^\prime = {\bf k}_i - {\bf K}_i$ and we neglected
the slow variation of $\nu_{{\bf K}_1 + {\bf k}_i^\prime}$ and of
$v_{{\bf K}_0 + {\bf k}_0^\prime , {\bf K}_1 + {\bf k}_1^\prime}$
with ${\bf k}_1^\prime$ and ${\bf k}_0^\prime$.  By taking the
inverse Fourier transforms of Eqs. II.9, one obtains the
mode-coupling equations in their envelope approximation limit,
namely
$$
\left( \partial_t + {\bf V}_{g1} \cdot \partial_{\bf x} + \nu_1
\right) a_1 ({\bf x},t) = \gamma_0 S \left( {\bf x} - {\bf V}_{g0} t
\right) a_2^*
$$
\nobreak
$$
\left( \partial_t + V_{g2} \cdot \partial_x + \nu_2 \right) a_2
({\bf x},t) = \gamma_0 S \left( {\bf x} - {\bf V}_{g0} t \right)
a_1^* \eqno (II.10)
$$
where the quantity $S({\bf x} - {\bf V}_{g0} t)$ is defined by
$$
S({\bf x} - {\bf V}_{g0} t) = \int d {\bf k}_0^\prime a_{{\bf
k}_0^\prime} {\rm exp} (i {\bf k}_0^\prime \cdot ({\bf x} - {\bf
V}_{g0} t)) \eqno (II.11)
$$

Equations II.10 are the envelope equations which have been
investigated with the statistical methods described in the next
section. These equations are generalizations to three dimensions of
one dimensional equations used in previous treatments of this
subject[16].  For the sake of simplicity the envelope approximation
will henceforth refer to either the coupled mode equations written
in their envelope form II.9, or the first order expansion of
$\omega_{k_i}^i$.

Concerning our notation, ${\bf V}_{g\alpha} ({\bf k}_{\alpha})$ will
denote, in the following the group velocity corresponding to a
generic wavenumber ${\bf k}_{\alpha}$, namely ${\bf V}_{g\alpha}
({\bf k}_{\alpha}) \equiv (\partial \omega^\alpha _{{\bf k}\alpha} /
\partial {\bf k}_{\alpha})$; the notation $V_{g\alpha}$ will be
reserved to the case of exact resonance, i.e., ${\bf V}_{g\alpha}
\equiv {\bf V}_{g\alpha} ({\bf K}_{\alpha})$.

On the other hand, the quantity $a_{\alpha}$ ({\bf x},t) will denote
the Fourier transform of the field  $a_{{\bf k}_{\alpha}}^{(\alpha)}
(t)$ in the case where one considers the mode coupling equations in
their general form (II.7); the notations $\hat a_{\alpha} (x,t)$ and
$\hat a_{k_{\alpha}^\prime}^{(\alpha)}$ will be used by contrast in
the case where the fast time and space variations have been
factorized ab initio as in Eq. (II.8a).  The connection between the
two sets is given by
$$
\hat a_{{\bf k}_{\alpha}^1}^{(\alpha)} \equiv a_{{\bf
K}_{\alpha}+{\bf k}_ {\alpha}^\prime} {\rm exp} (+ i \omega_{{\bf
K}_\alpha}^{(\alpha)} t ) \eqno (II.12a)
$$
and
$$
a_{\alpha} (x,t) = \hat a_{\alpha} (x,t) {\rm exp}~ i ({\bf
K}_{\alpha} \cdot {\bf x} - \omega_{{\bf K}_{\alpha}}^{(\alpha)} t)
+ c. c. \eqno (II.12b)
$$

\section{Statistical Description for Coupled Mode
Equations}
\subsection{The Frequency Mismatches}
In the two
following sections, we derive equations describing the evolution of
average quantities such as the average amplitude $\langle
a_{\alpha}\rangle$, or the average intensity $\langle \mid
a_{\alpha} \mid^2\rangle$ of wave $\alpha$. Here $<\phi>$ denotes
the statistical average of the physical quantity $\phi$, and its
fluctuation is written as $\delta \phi \equiv \phi - <\phi>$.  The
meaning of a statistical average can in some sense be understood as
a time average, and the use of a statistical framework is justified
for the following reasons:  a reduction of the parametric growth can
be expected whenever the spectral width $\mid\mid \Delta {\bf k}_0
\mid\mid$ of the pump wave is large enough (due to a temporal or
spatial incoherence) for its correlation time - as seen by the decay
waves, in a sense to be defined in this section - to be shorter than
the other characteristic times, namely, the damping and the growth
time, $\nu_{\alpha}^{-1}$ and $\gamma_0^{-1}$.  In this case, the
statistical description is justified whenever the correlation length
in K-space of the pump wave $a_{{\bf k}_0}^{(0)}$ is small compared
to its spectral width (e.g., when the number of ISI echelons or
phase elements is large).  In this limit the pump wave electric
field can indeed be regarded as a stochastic process with a short
correlation time and the standard statistical techniques can be
applied. In the case of a purely temporal incoherence, the
statistical description is physically justified only if the pump
wave bandwidth is caused by some stochastic process, e.g., the
spread over several independent lines of a non-monochromatic laser.
Similarly, in the case of a purely spatial incoherence, the natural
spread in angle of the pump wave wavenumbers must follow from the
sum of many independent beamlets.  Such a statistical independence
may result from a random phase shift given to the different beamlets
by means of a transparent phase mask[3]; it may also be due to the
scattering of the incident pump wave upon static random
fluctuations.  Lastly such a spatial incoherence can be achieved in
the so-called ISI technique[2] by a combination of delay increments
$\Delta t$ given to the beamlets by echelon structures and of the
laser temporal incoherence in the case $\Delta \omega_0 \Delta t
\geq 1$.  In these techniques, the laser beam is broken up into a
number of statistically independent "beamlets" that are brought to
focus by an optic of f-number, $f_\#$.  In the focal plane these
beamlets overlap creating a spatially nonuniform pattern with
coherence length across the beam $X_c = \pi/(k_0 {\sin} \theta_f)$
where $\theta_f$ is the half-angle of the optic, i.e. $\theta_f
\simeq (2f_\#)^{-1}$.

We will henceforth restrict ourselves to physical situations where
the statistical independence between the pump wave Fourier
components is satisfied; the pump wave electric field will thus be
regarded as a stochastic variable with zero mean; its statistical
properties will be assumed to be entirely determined by its spectral
density denoted as $n_{{\bf k}_0}^{(0)}$, the latter being itself
characterized by its spectral width $\Delta {\bf k}_0$, that is to
say by $\Delta k_0 = \mid \Delta {\bf k}_0 \mid = \Delta \omega_0 /
{\bf V}_{g0}$ (temporal incoherence) and $\Delta \theta_0$ (spatial
incoherence).

Before deriving the statistical equations for the time evolution of
$<a_i>$ or $<\mid a_i \mid^2>$, let us first consider on a physical
basis the reduction of the parametric growth caused by the pump wave
incoherence. Suppose that we consider the parametric coupling
corresponding to a given angle $\alpha_2$ between the pump wave and
the first decay wave direction; this angle defines the resonant
wavenumbers $K_1$ and $K_2$ of the decay waves corresponding to the
exact resonance condition II.3 with the mean pump wave number ${\bf
K}_0$. If one considers now a generic wavenumber ${\bf k}_0 \not=
{\bf K}_0$ of the pump wave, the resonance conditions II.3 can no
longer be satisfied for the same wave-numbers ${\bf K}_1$ and ${\bf
K}_2$. One is thus led to define the resonance mismatches $\Delta_1$
and $\Delta_2$ as
$$
\Delta_1({\bf k}_0, {\bf k}_2) \equiv \omega_{{\bf k}_0}^{(0)} -
\omega_{{\bf k}_0-{\bf k}_2}^{(1)} - \omega_{{\bf k}_2}^{(2)}  \eqno
(III.1a)
$$
and
$$
\Delta_2 ({\bf k}_0, {\bf k}_1) \equiv \omega_{{\bf k}_0}^{(0)} -
\omega_{{\bf k}_1}^{(1)} - \omega_{{\bf k}_0-{\bf k}_1}^{(2)}  \eqno
(III.1b)
$$

When discussing the effect of the pump incoherence in the case of a
given geometry defining the resonant wave numbers $K_1$ and $K_2$,
it is sufficient to consider the quantities $\Delta_1 ({\bf k}_0,
{\bf K}_2)$ and $\Delta_2 ({\bf k}_0, {\bf K}_1)$. In order to
simplify our notation, we will denote in our paper by $\overline
\theta$ the value of a quantity ${\overline \theta}$ in the case of
the exact resonance ${\bf k}_1 = {\bf K}_1$ and ${\bf k}_2 = {\bf
K}_2$.  Accordingly we define $\overline \Delta_1 ({\bf k}_0)$ and
$\overline \Delta_2 ({\bf k}_0)$ by
$$
\overline \Delta_1 ({\bf k}_0) \equiv \Delta_1 ({\bf k}_0, {\bf
K}_2) \eqno (III.2a)
$$
$$
\overline \Delta_2 ({\bf k}_0) \equiv \Delta_2 ({\bf k}_0, {\bf
K}_1) \eqno (III.2b)
$$

In the case of an incoherent pump wave, the spectral width $\Delta
{\bf k}_0$ gives rise to a characteristic size for $\overline
\Delta_i ({\bf k}_0)$, which will be denoted as $\Delta_{i {\rm
max}}$, i.e.
$$
\Delta_{i ~{\rm max}} = \overline \Delta ({\bf K}_0 + \Delta {\bf
k}_0) \eqno (III.3)
$$

For instance, in the case where the envelope approximation can be
applied, one readily finds $\Delta_{1 ~{\rm max}} = ({\bf V}_{g0} -
{\bf V}_{g1}) \cdot \Delta {\bf k}_0$ which can also be written as
$$
\Delta_{1 ~{\rm max}} = \mid (V_{g0} - V_{g1} {\rm cos}~ \alpha_2)
\Delta {\bf k}_0 - K_0 V_{g1 } {\rm sin}~ \alpha_2 \Delta \Theta_0
\mid \eqno (III.4a)
$$
\noindent or
$$
\Delta_{1 {\rm max}} = \mid (1-V_{g1} {\rm cos}~ \alpha_2/V_{g0})
\Delta \omega_0 - K_0 V_{g1} {\rm sin}~ \alpha_2 \Delta \Theta_0
\mid \eqno (III.4b)
$$

More generally, it has been shown[7] that it is sufficient to expand
$\Delta_{1~{\rm max}}$ as follows:
$$
\Delta_{1 max} = \mid t_1 \Delta k_0 + \sigma_1 \Delta \Theta_0 +
\beta_1\Delta\Theta_0^2 \mid  \eqno (III.5)
$$
with $t_1 = (V_{g0} - V_{g1} {\rm cos} \alpha_2), \sigma_1 = -V_{g1}
K_0 {\rm sin} \alpha_2$.  For backscatter, $\sigma_1$ vanishes and
it is necessary to expand to second order in $\Delta \Theta_0$.  The
coefficient of $\Delta \Theta_0^2, \beta_1,$ can be computed in
terms of the waves parameters $\partial \omega^{\alpha} / \partial
{\bf K}_j$, and $\partial^2 \omega^{\alpha} / \partial {\bf K}_i
\partial {\bf K}_j^1$, and the angles $\alpha_i$ defining the
geometry of the interaction.  The first term in (III.5) corresponds
to temporal incoherence and the last two to spatial incoherence.

In the present article, we will mainly restrict ourselves to some
particular examples convenient for numerical experiments, in which
the 2D character of the pump wave incoherence is modeled by the
introduction of two parameters $\Delta \omega_0$ and $\Delta k_0$,
playing the role of the temporal and spatial incoherence,
respectively, so that $\Delta_{i~{\rm max}}$ is similarly written as
$\Delta_{i~{\rm max}} = \alpha_i \Delta \omega_0 + \beta_i \Delta
k_0$.

Regarding now under which conditions the pump wave incoherence
modifies the parametric coupling, one may first easily admit that a
sufficient condition for neglecting the pump wave incoherence is
that the two following inequalities be both satisfied:
$$
\Delta_{1 max} < \mid D_1 \mid
$$
$$
\Delta_{2 max} < \mid D_2 \mid \eqno (III.6)
$$

In the latter inequalities, $D_i$ denotes the dispersion relation
$$
D_{i, {\bf k}_i} \left( \gamma, k \right) \equiv \gamma + i \left(
\omega_{{\bf k}_i+i\kappa}^{(i) } - \omega_{{\bf k}_i^{(i)}} \right)
+ \nu_i \eqno (III.7a)
$$

In the envelope approximation, $D_i$ reduces to
$$
\hat D_{i,~ {\bf k}_i} \left( \gamma, k \right) \equiv \gamma -
\kappa \cdot {\bf V}_{gi} \left({\bf k}_i \right) + \nu_i \eqno
(III.7b)
$$
where, $\gamma$ and $\kappa$ denote the time and space growth rates.
(It can be noted that our definitions of $D_i$ and $\hat D_i$ are
consistent with each other, in the sense that if one considers {\it
ab initio} the coupled mode equations in their envelope form
(II.10), $D_i$ reduces identically to $\hat D_i$ without any further
approximation: in this case, one has indeed $\omega_{{\bf
\kappa}_i}^{(i)} = {\bf \kappa}_i \cdot {\bf V}_{gi}$, from which
the relation $D_i = \hat D_i$ follows.)

Concerning the inequalities (III.6), we will see more generally that
the coherent or incoherent character of the parametric coupling is
controlled by inequalities between $\Delta_{i max}$ and $\mid D_j
\mid$.  On the other hand, since $\gamma$ and $\kappa$ are related
through the dispersion relation, the size of $D_j$ depends upon the
nature of the instability which is considered, namely convective,
absolute, or spatially amplifying.  It therefore follows that the
coherent or incoherent character of the parametric coupling depends
itself on the nature of the instability.  Defining now as
"incoherent" the domain where the pump wave incoherence induces a
reduction of the parametric growth, one realizes that such a domain
should be specified as "convectively incoherent", or "absolutely
incoherent", or "incoherent for spatial amplification".  For
instance, in the case of a purely temporal incoherence, with
$V_{g0}=\infty$, it has been shown in (Laval, et.al.) that the pump
wave incoherence reduces the convective growth for $\Delta \omega_o
> {\rm Max} \left( \nu_1, \nu_2 \right)$ and $\Delta \omega_0 >
\gamma_0$, whereas in the case of an absolute instability, the
absolute growth can be reduced only if $\Delta \omega_0$ satisfies:
$$
\Delta \omega_ 0 > \gamma_0 \mid {V_1 \over V_2} \mid^{1/2} > \nu_2
\mid {V_1 \over V_2} \mid .
$$
[For simplicity we consider the case where $\mid V_1 \mid >> \mid
V_2 \mid$ and $\nu_2/\mid V_2 \mid >> \nu_1/\mid V_1\mid $.]  One
thus sees with this simple case that the reduction of the absolute
growth is much more difficult to achieve than that of the convective
growth.

We may recall here that the convective growth rate characterizes the
instability in the early stage of its development, by contrast with
its long time evolution which is characterized either by the
absolute growth rate (whenever the conditions for their existence
are satisfied) or by the spatial amplification growth length (in the
opposite case).

In this article we will derive the conditions under which the pump
incoherence gives rise to a reduction of the convective and of the
absolute growth rates and an increase in the spatial amplification
length.  It will be found that generally the conditions for reducing
the absolute instabilities are more severe than those for the
convective ones.  For this reason we will concentrate mainly our
discussions on the "convectively incoherent" domain, which may thus
be regarded as the largest domain in which a reduction of the
parametric growth may be expected in the early stage of the
instability development.  In order to simplify our terminology, the
"incoherent" domain, without any other specification, will
henceforth refer to the previously defined "convectively incoherent"
domain.

Restricting ourselves to the convective instabilities, one may
rewrite inequalities (III.6) as:
$$
\Delta_{1 max} < \nu_1 , \gamma_0
$$
$$
\Delta_{2 max} < \nu_2 , \gamma_0 \eqno (III.8)
$$
in which we used $\gamma \sim \gamma_0$ in the coherent regime. As
said before, these inequalities represent only a {\it sufficient}
condition for the pump wave incoherence being ignorable; the
question is thus whether both of these two inequalities have to be
satisfied, or only one. In order to answer this question, we first
introduce in a brief discussion the validity conditions of the
statistical equations which describe the time evolution of the
quantities $<a_i>$ and $<\mid a_i \mid^2>$.

\subsection{The Incoherent and RPA domains}
A statistical description for the evolution of the decay waves 1 and
2 is provided by the Random Phase Approximation (RPA) equations for
the evolution of the average intensities $<a_i a_i^*>$.  The latter
provide equations for the evolution on a slow time - or slow space -
scale of the spectral densities $n_k^{(i)}$(x,t) of the decay waves
[the latter functions $n_{\bf k}^{(i)}$(x,t)] will be defined in the
general case by Eq. (V.2); at this point it is sufficient to say
that for a purely temporal problem one has the usual relation
$<a_{\bf k}^{(i)} (a_{{\bf k}^\prime}^{(i)})^*
> = \delta ({\bf k}-{\bf k}^\prime)
n_k^{(i)}$].  The RPA equations are nonlinear and they may account
for the pump depletion; the interested reader is referred to
standard textbooks[72] for their derivation in the general case. In
our problem of stability analysis, we face a linear problem; we
could naturally use directly the RPA equations on which we would
make {\it a posteriori} the linear approximation which consists in
neglecting the pump depletion.  In this article we will follow a
different route.  We will first take advantage of the linear
character of the coupled mode equations II.6 which will enable us to
use the so-called Bourret approximation[70] for the evolution of
$<a_i>$; on the other hand, the Bourret approximation for the
evolution of the intensities $<a_ia_i^*>$ is easily tractable within
the envelope approximation only, and then only for the special case
where the pump wave group velocity $V_{g0}$ is infinite.  In the
general case it is necessary to make supplementary approximations to
the Bourret equations, and these approximations are justified in a
domain that will be defined subsequently as the "RPA domain."

Returning now to the condition defining the domain in which the
parametric coupling takes place coherently, we first define the
temporal and spatial growth rates $\gamma_{<a_i>} , {\bf
\kappa}_{<a_i>} ,$ for $\langle\mid a_i \mid^2\rangle$, and
$\gamma_{\langle a^2\rangle}, {\bf \kappa}_{\langle a^2\rangle}$ for
$\langle\mid a_i\mid^2\rangle$, according to:
$$
\langle\hat a_i ({\bf x}, t)\rangle = a_{{\bf K}_i}^{(i)} (0)\, {\rm
exp} \left[\gamma_{\langle a_i\rangle} t - {\bf \kappa}_{\langle
a_i\rangle} \cdot {\bf x} \right] \eqno (III.9)
$$
$$
\langle \mid \hat a_i \mid^2 (x, t) \rangle = \mid a_{{\bf
K}_i}^{(i)} \mid^2(0)\, {\rm exp} ~2 \left[ \gamma_{\langle
a^2\rangle} t - {\bf \kappa}_{\langle a^2\rangle} \cdot {\bf
x}\right]
$$
A factor 2 has been included in the argument of the exponential
defining $\gamma_{<a^2>}$ and $\kappa_{<a^2>}$ so that the equality
$\gamma_{<a>} = \gamma_{<a^2>}$ holds in the coherent case.  More
generally the Schwartz inequality yields the following result
$$
\gamma_{\langle a^2\rangle} \geq \gamma_{\langle a_i\rangle} \eqno
(III.10)
$$
for i = 1 and 2, in the case of a purely temporal growth (and
$\mid\mid {\bf \kappa}_{\langle a^2\rangle} \mid\mid \geq \mid\mid
{\bf \kappa}_{\langle a_i\rangle} \mid\mid$ in the case of a spatial
growth).

Admitting that the pump incoherence may only reduce the parametric
growth (in the case of a convective instability in a homogeneous
plasma), one realizes that, whenever either $\langle a_1\rangle$ or
$\langle a_2 \rangle$ behaves coherently, the two quantities $<\mid
a_1 \mid^2>$ and $<\mid a_2 \mid^2>$ must also behave both
coherently.

In the next section, it will be found that there exists a growth
reduction for $<a_1>$ only if the mismatch $\Delta_{2 ~{\rm max}}$
satisfies the condition $\Delta_{2 ~{\rm max}} > (\nu_2,\gamma_0)$.
From the previous considerations, it follows that the wave energies
$<\mid a_i\mid^2>$ will behave coherently whenever at least one of
the two following inequalities is satisfied:
$$
\Delta_{1 ~{\rm max}} < {\rm Max} (\nu_1, \gamma_0) \eqno (III.11a)
$$
or
$$
\Delta_{2 ~{\rm max}} < {\rm Max} (\nu_2, \gamma_0) \eqno (III.11b)
$$

These inequalities define what is referred to henceforth as the
"convectively coherent" domain.  Accordingly, the "convectively
incoherent" domain - or simply incoherent domain - is defined by the
two conditions
$$
\Delta_{1 ~{\rm max}} > {\rm Max} (\nu_1, \gamma_0) \eqno (III.12a)
$$
and
$$
\Delta_{2 ~{\rm max}} > {\rm Max} (\nu_2, \gamma_0) \eqno (III.12b)
$$

On the other hand, the validity conditions for the RPA are very
stringent.  In Section V they will be found to consist of two sets
of inequalities; the first one corresponds to the inequalities
opposite to (III.6), namely
$$
\Delta_{1 ~{\rm max}} > {\rm Max} \mid D_1 \mid \eqno III.13a)
$$
$$
\Delta_{2 ~{\rm max}} > {\rm Max} \mid D_2 \mid \eqno (III.13b)
$$
and a second set involving the cross inequalities
$$
\Delta_{1 ~{\rm max}} > {\rm Max} \mid D_2 \mid \eqno (III.13c)
$$
$$
\Delta_{2 ~{\rm max}} > {\rm Max} \mid D_1 \mid \eqno (III.13d)
$$

Since the RPA dispersion relation yields $\gamma_{\langle
a^2\rangle} \sim \gamma_0^2 / {\rm min} (\Delta_{i ~{\rm max}})$ for
the convective growth rate, the latter inequalities reduce
essentially to
$$
\Delta_{1 ~{\rm max}} > {\rm Max} (\nu_1, \gamma_0) \eqno (III.14a)
$$
$$
\Delta_{2 ~{\rm max}} > {\rm Max} (\nu_2, \gamma_0) \eqno (III.14b)
$$
$$
\Delta_{1 ~{\rm max}} > {\rm Max} (\nu_2, \gamma_0) \eqno (III.14c)
$$
$$
\Delta_{2 ~{\rm max}} > {\rm Max} (\nu_1, \gamma_0) \eqno (III.14d)
$$
which can also be recast into the more compact form
$$
{\rm Min} (\Delta_{j ~{\rm max}}) > {\rm Max} (\nu_i, \gamma_0)
\eqno (III.15)
$$

The latter inequalities (III.14) or (III.15), define what is called
the (convective) RPA domain.

One can now see that the incoherent domain (III.12) is subdivided
into 1) the RPA domain in which the cross inequalities (III.14b) and
(III.14d) are both satisfied, 2) the intermediate domain
corresponding to the regime where the two inequalities $\Delta_{1
~{\rm max}} > (\nu_1, \gamma_0)$ and $\Delta _{2 max} > (\nu_2,
\gamma_0)$ are both fulfilled whereas at least one of the cross
inequalities (III.14c) and (III. 14d) is not satisfied.

In this intermediate domain there is no theory which is easily
tractable for an explicit computation of the growth rate of the
average energy $<\mid a_\alpha\mid^2>$. On the other hand, the
Bourret approximation makes it possible to calculate the growth rate
of the average amplitudes $<a_1>$ and $<a_2>$:  it will then be
found in Section V that the larger growth rate of $\gamma_{<a_1>}$
and $\gamma_{<a_2>}$ is given (for the convective instabilities), by
the RPA prediction $\gamma^{RPA}$ for $<\mid a_1\mid^2>$ and $<\mid
a_2\mid^2>$ within a numerical factor no larger than two.

For this reason we will argue in Section VI that although the RPA
equations are not in their range of applicability in the
intermediate domain, the convective growth rate $\gamma_{<a^2>}$ for
the intensities remains of the order of $\gamma^{RPA}$.
Accordingly, in the {\it whole} incoherent domain defined by the two
inequalities:
$$
\Delta_{1 ~{\rm max}} > {\rm Max} (\nu_1, \gamma_0) \eqno (III.12a)
$$
and
$$
\Delta_{2 ~{\rm max}} > {\rm Max} (\nu_2, \gamma_0) \eqno (III.12b)
$$
the convective growth rate $\gamma_{<a^2>}$ can be approximated by
the RPA prediction $\gamma^{RPA}$.  The latter approximation is one
of the main results of our paper since it makes it possible to
compute the reduction of the parametric growth induced by the pump
wave incoherence from the simple calculation of the RPA coupling
constants $\gamma_{11}^{RPA}$ and $\gamma_{22}^{RPA}$.

\section{The Bourret Approximation for the Average Amplitudes}
\subsection{Introduction to the Bourret Approximation}
The Bourret approximation[69,70] is a well known equation in
the context of propagation in random media; it deals with stochastic
linear multiplicative equations of the form
$$
\left( \partial_t + L_0 \right) A = \gamma_0 S A \eqno (IV.1) $$

where $L_0$ is a linear deterministic operator, S a linear
stochastic operator with zero mean, $<S> = 0$, and A is the physical
quantity of interest.  The Bourret approximation can be simply
derived as follows:
\begin{itemize}
\item[i)]  by averaging Eq. (IV.1) one first obtains the exact relation
$$
\left( \partial_t + L_0 \right) <A> = \gamma_0 <S\, \delta A> \eqno
(IV.2)
$$
with $\delta A = A - <A>$.
\item [ii)]  by subtracting the latter from Eq. (IV.1), one obtains the
equation for evolution of the fluctuation $\delta A$
$$
\left( \partial_t + L_0 \right) \delta A_0 = \gamma_0 \left[ S<A> +
S\, \delta A - <S\, \delta A> \right]
$$
\item [iii)]  the Bourret approximation consists then in neglecting the
so-called mode coupling term $S\delta A - <S\delta A>$ in the latter
equation; by doing so and neglecting the initial conditions, one
obtains for the fluctuation $\delta A$ the relation $\delta A =
(\partial_t + L_0)^{-1} \gamma_0S<A>$, where ($\partial_t +
L_0)^{-1}$ represents the Green's function of the operator
($\partial_t + L_0$); inserting the latter result into Eq. (IV.2),
one obtains the Bourret equation for $<A>$, namely
$$
\left( \partial_t + L_0 \right) <A> = \gamma_0^2 < S \left(
\partial_t + L_0 \right)^{-1} S > <A> \eqno (IV.3)
$$
\end{itemize}

Defining $\gamma_{\rm eff}$ as $\gamma_{\rm eff} = \gamma_0^2
\mid\mid <S (\partial_t + L_0)^{-1} S> \mid\mid$, it is natural to
introduce the effective bandwidth $\Delta \omega_{\rm eff}$ as
$\Delta \omega _{\rm eff} \equiv \gamma_0^2 /\gamma_{\rm eff}$, and
the validity condition for the Bourret approximation can be
expressed as
$$
\Delta \omega_{\rm eff} > \mid D \mid \equiv \mid \mid \partial_t +
L_0 \mid\mid \eqno (IV.4)
$$

The interested reader is referred to Ref. 69 for a detailed
discussion of these results.

\subsection{The Bourret approximation for the average
amplitude $<a_i>$}

\subsubsection{General three
dimensional result.}
By applying the method outlined just above, one
easily obtains the equation for evolution of the average amplitude
$<a_1>$.  Consistent with (II 12b), we set
$$
a_1 ({\bf x},t) = \hat a_1 \left( {\bf x} \right) {\rm exp }~i \left
( {\rm K}_1 \cdot {\bf x} - \omega_{{\bf K}_1} t \right) + c.c
$$
and the dispersion relation corresponding to the slow time and space
evolution for $<\hat a_1 ({\bf x},t)>$ reads
$$
\left( \gamma + \nu_1 - {\bf \kappa} \cdot {\bf V}_{g1} \right) =
\overline \gamma_1 \eqno (IV.5)
$$
with
$$
\overline \gamma_1 = \gamma_1 \left( {\bf K}_1 \right) \equiv
\gamma_0^2 \int {d{\bf k}_0 n_{k_0} \over -i \overline \Delta_2
\left({\bf k}_0 \right) + \left( \gamma + \nu_2 - {\bf \kappa} \cdot
{\bf V}_{g2} \right) + \epsilon} \eqno (IV.6)
$$
where $\gamma$ and $\kappa$ stand for $\gamma_{<a_1>}$ and
$\kappa_{<a_1>}$, according to (III.9).  In deriving the latter
equation, we neglected for simplicity the slow dependence of
$v_{{\bf k}_0},_{{\bf k}_\alpha}$ and $\nu_{{\bf
k}_\alpha}^{(\alpha)}$ upon ${\bf k}_0$ and ${\bf k}_{\alpha}$. For
the sake of clarity, we also restrict ourselves here to the case of
exact resonance $({\bf k}_1 = {\bf K}_1)$; the general case ${\bf
k}_1 \not= {\bf K}_1$ will be considered in the next Section in the
RPA context; (consistent with our other notation, $\overline
\gamma_1 $ denotes the coupling constant $\gamma_1$ evaluated at
resonance).

In deriving Eq. (IV.6) we made the envelope approximation for the
slow space dependence only, by using $\omega^{(i)}_{{{\bf K}_i} +  i
{\bf \kappa}} \simeq \omega_{{\bf K}_i}^{(i)} + i {\bf \kappa} \cdot
{\bf V}_{g1}$; on the other hand the computation of the resonance
mismatch $\Delta_2$ corresponding to a fast space dependence is not
restricted to the envelope approximation; lastly, the symbol
$\epsilon$ in Eq. (IV.6) represents the usual prescription for the
Laplace transform contour.  The dispersion relation corresponding to
the evolution of the average amplitude $<a_2 (x,t)>$ can be written
in a similar way as
$$
\gamma + \overline \nu_2 - {\bf \kappa} \cdot {\bf V}_{g2} =
\overline \gamma_2 \eqno (IV.7)
$$
with
$$
\overline \gamma_2 = \gamma_0^2 \int {d{\bf k}_0 n_{{\bf k}_0} \over
-i \overline \Delta_1 \left({\bf k}_0 \right) + \left( \gamma +
\nu_1 - {\bf \kappa} \cdot {\bf V}_{g1} \right) + \epsilon} \eqno
(IV.8)
$$

Defining the effective bandwidths $\Delta \omega_{i ~{\rm eff}}$ as
$$
\Delta \omega_{1 ~{\rm eff}} ( \gamma, {\bf \kappa}) \equiv
\gamma_0^2 / \overline \gamma_2 \eqno (IV.9a)
$$
and
$$
\Delta \omega_{2 ~{\rm eff}} ( \gamma, {\bf \kappa}) \equiv
\gamma_0^2 / \overline \gamma_1 \eqno (IV.9b)
$$

The validity condition of the Bourret approximation for the average
amplitude $<a_1>$ reads[16],
$$
\Delta \omega_{2 ~{\rm eff}}>\mid D_2\mid = \mid \gamma + \nu_2 -
{\bf \kappa} \cdot {\bf V}_{g2} \mid \eqno (IV-10)
$$
and
$$
\gamma_0^2 < \Delta \omega_{1 ~{\rm eff}} \Delta \omega_{2 ~{\rm
eff}} \eqno (IV-11)
$$
the same naturally holds for the evolution of $<a_2>$ with
$(1\leftrightarrow2)$.

\subsubsection{Markov limit}
At this
point let us consider the so-called Markov limit of the coupling
constant $\overline \gamma_1$; the latter consists in taking the
limit $\mid \hat D_2 \mid = \mid \gamma + \nu_2 - {\bf \kappa} \cdot
{\bf V}_{g2} \mid << \Delta_{2 max}$, and in using $\left(-i
\Delta_2 ({\bf k}_0) + \epsilon \right)^{\bf -1} - iPP \Delta_2^{-1}
({\bf k}_0) + \pi \delta (\Delta_2 ({\bf k}_0))$.

In this limit the coupling constant $\overline \gamma_1$ is given by
$$
\overline \gamma_1^M \equiv \gamma_0^2 \int d{\bf k}_0 n_{{\bf
k}_0}^{(0)} \pi \delta \left( \overline \delta_2 ({\bf k}_0) \right)
\eqno (IV.12)
$$
where the superscript M stands for "Markov;" the same expression
holds for $\overline \gamma_2^M$ with $(1\leftrightarrow2)$.  The
latter two expressions for $\overline \gamma_1^M$ and $\overline
\gamma^M$ are identical to the RPA coupling constants $\overline
\gamma_{11}^{RPA}$ and $\overline \gamma_{22}^{RPA}$ to be derived
later.  At this point it is sufficient to remark that whenever the
Markov limit can be taken, the orders of magnitude of $\overline
\gamma_1^M$ and $\overline \gamma_2^M$ are
$$
\overline \gamma_1^M \sim \gamma_0^2 / \Delta_{2 ~{\rm max}} \eqno
(IV.13)
$$
$$
\overline \gamma_2^M \sim \gamma_0^2 / \Delta_{1 ~{\rm max}}
$$

These estimates follow simply from the definition (IV-12) for
$\overline \gamma_2^M$ and from the normalization condition $\int
d{\bf k}_0 n_{{\bf k}_0}^{(0)} = 1$. On the other hand it appears to
be convenient, when discussing the continuity between the coherent
and incoherent results, to rewrite the inequalities (III-12)
limiting the incoherent domain in terms of the spectral widths
$\Delta \omega_1$ and $\Delta \omega_2$ defined by the following
relations
$$
\Delta \omega_2 \equiv \gamma_0^2 / \overline \gamma_1^M =
\gamma_0^2 / \overline \gamma_{11}^{RPA} \eqno (IV.14)
$$
$$
\Delta \omega_1 \equiv \gamma_0^2 / \overline \gamma_2^M =
\gamma_0^2 / \overline \gamma_{22}^{RPA}
$$

According to the estimates (IV.13) the order of magnitude of $\Delta
\omega_i$ is given by
$$
\Delta \omega_i \sim \Delta_{i {\rm max}} \eqno (IV.15)
$$

It can also be seen that the quantities $\Delta \omega_i$ can be
expressed as $\Delta \omega_i = \Delta \omega_{i {\rm eff}} (\gamma
= 0, \kappa = 0)$, so that the validity condition (IV-10) for the
Bourret approximation for $<a_1>$ becomes, in the Markov limit,
$\Delta \omega_2 > \mid D_2 \mid$. The spectral widths $\Delta
\omega_i$ are given in Section VII for the examples of the pump wave
correlation function corresponding to numerical solutions.  Their
general expressions are computed in Ref. [7] in terms of the
spectral width parameters $\Delta k_0$ and $\Delta \theta_0$ for the
case of interacting wave- packets; it will be seen that they can be
expanded in a similar way as $\Delta_{i max}$, namely
$$
\Delta \omega_i = \underline {t}_i \Delta k_0 + \underline
{\sigma}_i \Delta \Theta_0 + \underline {\beta}_i \Delta \theta_0^2
\eqno(IV.16)
$$
where the parameters $t_i$, $\sigma_i$ and $\beta_i$ can be
expressed in terms of the waves parameters $\partial \omega^\alpha
/\partial {\bf K}_j$ and $\partial^2 \omega_j^2 /\partial {\bf K}_j
\partial {\bf K}_{j^\prime}$. The latter parameters are all of the
same order of magnitude as the corresponding $t_i$, $\sigma_i$ and
$\beta_i$ of the expression (III-2) for $\Delta_{i ~{\rm max}}$,
although they may differ nonetheless by a numerical constant of
order unity resulting from the averaging procedure over $k_0$
involved in Eq. (III-23), in all cases however the ordering
$$
\Delta \omega_i = O \left( \Delta_{i ~{\rm max}} \right)
$$
holds and the quantity $\Delta_{i ~{\rm max}}$ will henceforth be
replaced by $\Delta \omega_i$ in the inequalities defining the
domain of validity for the incoherent results.  Moreover the
convective growth rate, well above threshold, is easily found to be
$\gamma \sim \gamma_0^2 / \Delta \omega_2$ so that the validity
conditions IV-10 and IV-11 of the Bourret approximations for $<a_1>$
and $<a_2>$ in their Markovian limits become simply
$$
\Delta \omega_2 > \nu_2 , \gamma_0 \eqno (IV.17a)
$$
$$
\Delta \omega_1 > \nu_1 , \gamma_0 \eqno (IV.17b)
$$

The latter inequalities justify the conditions (III-12), written in
the previous Section in terms of $\Delta \omega_{i ~{\rm eff}}$ for
simplicity, defining the incoherent domain, that is to say the
domain where the Bourret approximation is correct for the two
average amplitudes $<a_1>$ and $<a_2>$.

Concerning now the validity condition for the Markovian limit
itself, made when deriving the coupling constant $\gamma_1^M =
\gamma_0^2/\Delta \omega_2$, there is no general criterion, except
in the special case where the envelope approximation is correct - in
this case the validity condition III-27a for the Bourret
approximation for $<a_1>$ in the convective regime justifies {\it a
posteriori} the Markov limit $\Delta \omega_2 >> D_2$.  This result
is discussed in the next subsection; it is shown in particular that
in the case where the envelope approximation is valid, there is no
intermediate regime between the incoherent domain IV-17 and the
coherent domain (III-11); namely, either the Bourret approximation
is correct and the coupling constants take their Markov limits
$\gamma_1^M = \gamma_0^0 /\Delta \omega_2, \gamma_0^2 / \Delta
\omega_1$ in the incoherent domain, or the coherent results apply.
In addition, the two sets of results are continuous from one domain
to the other.

On the other hand in cases where the envelope approximation is not
correct, there is the possibility for an intermediate regime where
neither the Markov limit of the Bourret approximation nor the
coherent result apply; such a situation is discussed in Ref. [7].

In order to illustrate the continuity between the coherent and the
incoherent domains in the case where the envelope approximation is
correct, we consider in the next sub-section the special case of a
Kubo-Anderson process.

\subsection{The Special Case of a Kubo-Anderson Process}
The Kubo-Anderson Process (KAP) is an example of a stochastic
process for which the Bourret approximation is exact, whatever the
spectral width $\Delta \omega_0 = V_{go} \Delta k_o$ is.  In the
case of the coupled mode equations, it makes it possible to compute
the coupling constant $\overline \gamma_1$ (and therefore the growth
rate) as a continuous function of the spectral widths $\Delta
\omega_0$. The interested reader is referred to the
references[14,15,69] for an introduction to the Kubo Anderson
process.  In a one-dimensional geometry for the pump wave, and in
the limit of the envelope approximation, the spectral density of the
pump wave, when it is modeled by a KAP, is given by
$$
n_{l_0}^{(0)} = {\pi^{-1} \Delta k_0 \over \left( {\bf k}_0 - {\bf
K}_0 \right)^2 + \Delta k_0^2} \eqno (IV.18)
$$

Performing the integration over ${\bf k}_0$ in Eq. (IV-6), one
obtains $\overline \gamma_1 = \gamma_0^2 / \Delta \omega_{2 ~{\rm
eff}}$ with
$$
\Delta \omega_{2 ~{\rm eff}} = \gamma + \nu_2 - {\bf \kappa} \cdot
{\bf V}_{g2} + \Delta \omega_2 \eqno (IV.19)
$$
where $\Delta \omega_2$ is given by
$$
\Delta \omega_2 = \mid V_{g0} - V_{g2} {\rm cos} \alpha_1 \mid
\Delta k_0 = \eta_2 \Delta \omega_0 \eqno (IV.20)
$$

The quantity $\eta_2 = \mid 1 - (V_{g2} / V_{g0}) {\rm cos} \alpha_1
\mid$ generalizes the quantity $\eta_2$ defined previously in a
similar framework.[16]  In the latter expressions, $V_{g0} \Delta
k_0$ has simply to be replaced by $\Delta \omega_0$ in the case of a
pump wave with an infinite group velocity $V_{g0} = \infty$; this
limit corresponds to a model in which the pump wave fluctuates in
time only, i.e. where the function $S (x - V_{g0}t)$ in the envelope
equation II-9 is a function $\overline S (t)$ of time characterized
by a correlation time $\tau_c = \Delta \omega_0^{-1}$.  Lastly, in
order to make a connection with the previously introduced
definitions, the expression (IV.20) for $\Delta \omega_2$ can be
seen to correspond to $\underline t_2 = t_2 = \mid V_{g0} - V_{g2}
{\rm cos} \alpha_1 \mid$, following our general expression (IV-16)
for $\Delta \omega_2$ (in the case of a one-dimensional geometry for
the pump, one has naturally $\Delta \theta_0 = 0$). The quantity
$\Delta \omega_{1 eff}$ is defined in a similar way by substituting
($1\leftrightarrow2$).

In order to discuss the continuity between the coherent and the
incoherent domains, it is convenient to introduce as before the
quantities $\hat D_i$ defined by
$$
\hat D_i = \gamma + \nu_i - {\bf \kappa} \cdot {\bf V}_{gi} \eqno
(IV.21)
$$

The dispersion relation for $<a_1>$ is thus
$$
\hat D_1 = {\gamma_0^2 \over \hat D_2 + \Delta \omega_2} \eqno
(IV.22a)
$$
where for $<a_2>$ it is
$$
\hat D_2 = {\gamma_0^2 \over \hat D_1 + \Delta \omega_1} \eqno
(IV.22b)
$$

As stated before, these two dispersion relations are exact for a KAP
and they make it possible to investigate the behavior of the average
amplitude $<a_i>$ as a function of the spectral width $\Delta
\omega_0 $. The first remarkable point is, as first noticed by
Thomson and Karush[14], that the two dispersion relations (IV-22)
are not identical.  Due to phase mixing, $<a_i>$ does not
necessarily behave like $<\mid a_i\mid>$, nor $<a_1>$ like $<a_2>$.

For simplicity we now restrict ourselves to the convective
instability in our discussion concerning the behavior of the average
amplitudes $<a_1>$ and $<a_2>$.  After setting ${\bf \kappa} = 0$,
one can check directly, using the exact dispersion relation (IV-22a)
for $<a_1>$, what has been announced just above. First, the validity
condition $\Delta \omega_{2 ~{\rm eff}} > \mid \hat D_2 \mid$ of the
Bourret approximation necessarily reduces to the condition $\Delta
\omega_2 > \mid \hat D_2 \mid$, and justifies {\it a posteriori} the
Markov limit $\overline \gamma_1 = \overline \gamma_1^M$. Since the
latter condition $\Delta \omega_2 > \hat D_2 \mid$ is itself
equivalent to $\Delta \omega_2 > (\gamma_0, \nu_2)$, it is natural
to define the incoherent domain for $<a_1>$ as the domain $\Delta
\omega_2 > (\gamma_0, \nu_2)$.  (The incoherent domain for $<a_2>$
corresponds naturally to the domain $\Delta \omega_1 > (\gamma_0,
\nu_1)$.)  Second, there is no intermediate regime between the
coherent domain $(\Delta \omega_0 \rightarrow 0)$ and the incoherent
domain for $<a_1>$ in the sense that there is no discontinuity
between the usual dispersion relation of the coherent case $(\hat
D_1 = \gamma_0^2 / \hat D_2)$ and the one corresponding to the
Markov limit of the Bourret approximation $(\hat D_2 = \gamma_0^2 /
\mid \Delta \omega_2 \mid)$; consequently the growth rate and the
threshold are continuous from one domain to the other.  The same
naturally holds for $<a_2>$.

What can we infer regarding the behavior of the intensities $<\mid
a_i \mid^2>$ from the results concerning the average amplitudes?
This problem has already been investigated by several
authors[14-16].  As explained above, from the inequality
$\gamma_{\langle a^2\rangle} \geq \gamma_{\langle
 a\rangle}$, one may assert
that the system behaves necessarily in a coherent way in the
coherent domain $\Delta \omega_1 < (\gamma_0, \nu_1)$ or $\Delta
\omega_2 < (\gamma_0, \nu_2)$. On the other hand, in the
complementary domain defined previously as simply "the incoherent
domain," one may only use the latter inequality: first, it follows
that a lower bound for the threshold for the intensities is given by
the lowest threshold for the average amplitude, namely
$$
\gamma_0^2 \geq {\rm Min} \left[ \Delta \omega_1 \nu_2, \Delta
\omega_2 \nu_1 \right] \eqno (IV.23)
$$
secondly the growth rate $\gamma_{<a^2>}$ satisfies the inequality
$$
\gamma_{\langle a^2\rangle} \geq {\rm Max} \left[ \gamma_{\langle
a_1\rangle} , \gamma_{\langle a_2\rangle} \right] \eqno (IV.24)
$$
which, well above threshold, reduces to
$$
\gamma_{\langle a^2\rangle} \geq \gamma_0^2 / {\rm Min} \left(
\Delta \omega_1 , \Delta \omega_2 \right ) \eqno (IV.25)
$$
in a domain where $\Delta \omega_i > \gamma_0$ for i = 1 and 2. The
question is thus whether these lower bounds are actually attained in
which case it would be sufficient to simply consider the most
unstable average amplitude.  In order to answer the question we
first have to derive statistical equations for the wave intensities,
which we undertake in the next Section.

\section{The RPA Equations
for the Wave Intensities}
The Bourret technique can be applied
exactly in the limit $V_{g0} = \infty$ in order to derive
statistical equations for the wave intensities.  In the case of a
Kubo Anderson Process these equations are exact for any spectral
widths $\Delta \omega_0$, so that they make it possible to
investigate the limit in which they take the form of the RPA
equations.  For the general case $V_{g0} \not= \infty$ additional
approximations have to be made to the Bourret-like equations in
order to obtain the RPA equations.

\subsection{ The Bourret approximation in the limit $V_{g0} =
\infty$.}Let us consider the coupled mode equations in their
envelope form II-9, in which the stochastic function $\hat S \left(
{\bf x} - {\bf V}_{g0} t \right)$ is a Kubo-Anderson process
$\overline S (t)$; the latter is characterized by a correlation
function of the form $\langle \overline S (t) \overline S (t +
\tau)\rangle = exp \left( - \Delta \omega_0 \mid \tau \mid \right)$,
corresponding to a spectral density in frequency space given by
$$
\hat n_{\omega_0}^{(0)} = {\Delta \omega_0 / \pi \over \Delta
\omega_0^2 + \omega_0^2} \eqno (V.1)
$$

In order to derive the Bourret equations for the wave intensities,
following Ref. [16], we consider the two point correlation function
$<\hat a_i (x, t) \hat a_j^* (x^\prime, t)>$ with i and j = 1, 2.
The spectral densitites are defined as
$$
\left< \hat a_i (x, t) \hat a_i^* (x,^\prime t) \right> = \int d{\bf
k}_i^\prime \hat n_{{\bf k}^\prime_i}^{(i)} \left( {{\bf x}_2 + {\bf
x}^\prime \over 2}, t \right) {\rm exp}~i {\bf k}_i^\prime \cdot
\left( {\bf x}-{\bf x}^\prime \right) \eqno (V.2)
$$
where we allow for a slow time and space variation of the spectral
density $\hat n _{{\bf k}^\prime_i}^{(i)}$.  In order to be
consistent with the previous definition (III.9) of growth rates, we
set
$$
\hat n_{{\bf k}^\prime_i}^{(i)} ({\bf x}, t) = \mid a_{{\bf K}_i +
{\bf k}^\prime_i}^{(1)} (0) \mid^2 {\rm exp}~ 2 \left( \gamma t -
{\bf \kappa} \cdot {\bf x} \right) \eqno (V.3)
$$
where $\gamma$ and ${\bf \kappa}$ stand now for $\gamma_{\langle
a^2\rangle}$ and ${\bf \kappa}_{\langle a^2\rangle}$.

The Bourret approximation is easily derived for the set $\langle
\hat a_i ({\bf x}, t) \hat a_j^* ({\bf x}^\prime, t)\rangle$ to give
the following system
$$
\hat D_1 \hat n_{{\bf k}^\prime_1}^{(1)} = \gamma_{11}^B \left( {\bf
k}_1^\prime \right) \left[ \hat n_{{\bf k}^\prime_1}^{(1)} + \hat
n_{-{\bf k}^\prime_1}^{(2)} \right] \eqno (V.4a)
$$
$$
\hat D_2 \hat n_{{\bf k}^\prime_2}^{(2)} = \gamma_{22}^B \left( {\bf
k}_2^\prime \right) \left[ \hat n_{{\bf k}^\prime_2}^{(1)} + \hat
n_{-{\bf k}^\prime_2}^{(2)} \right] \eqno (V.4b)
$$
where $\hat D_i$ is defined as before as $\hat D_i \equiv \gamma +
\nu_i - {\bf \kappa} \cdot {\bf V}_{gi}$; the coupling constants
$\gamma_{ii}^B$ are given by
$$
\gamma_{ii}^B \left( {\bf k}^\prime_i \right) = {\gamma_0^2 \left(
\hat D_1 + \hat D_2 + \Delta \omega_0 \right) \over \left( \hat D_1
+ \hat D_2 + \Delta \omega_0 \right)^2 + \left({\bf k}^\prime_i
\cdot \left( {\bf V}_{g1} - {\bf V}_{g2} \right) \right)^2} \eqno
(V.5)
$$
and the superscript stands for ``Bourret''.
\begin{itemize}
\item [i)]  Let us first consider the Markov limits of the
latter expressions and show that they exactly reproduce the RPA
results. The Markov limit of $\gamma_{ii}^B$ corresponds to the
domain $\mid \hat D_1 + \hat D_2 \mid << \Delta \omega_0$, in which
case the coupling constants $\gamma_{ii}^B$ take the following form
$$
\gamma_{ii}^{B,M} \left( {\bf k}^\prime_i \right) = {\gamma_0^2
\Delta \omega_0 \over \Delta \omega_0^2 + \left( {\bf k}_i^\prime
\cdot \left( {\bf V}_{g1} - {\bf V}_{g2}\right) \right)^2 } . \eqno
(V.6)
$$

On the other hand it will be seen in the next subsection that the
RPA equations take the general following form
$$
\hat D_1 \hat n_{{\bf k}^\prime_1}^{(1)} = \gamma_0^2 \int d{\bf
k}^\prime_0 \hat n_{{\bf k}^\prime_0}^{(0)} \pi \delta \left(
\Delta_2^\prime \left( {\bf k}^\prime_0 , {\bf k}^\prime_1
\right)\right) \left[ \hat n_{{\bf k}^\prime_1}^{(1)} + \hat n_{{\bf
k}^\prime_0 - {\bf k}^ \prime_1}^{(2)}\right] , \eqno (V.7a)
$$
$$
\hat D_2 \hat n_{{\bf k}^\prime_2}^{(2)} = \gamma_0^2 \int d{\bf
k}^\prime_0 \hat n_{{\bf k}^\prime_0}^{(0)} \pi \delta \left(
\Delta_1^\prime \left( {\bf k}^\prime_0 , {\bf k}^\prime_2
\right)\right) \left[ \hat n_{{\bf k}^\prime_2}^{(2)} + \hat n_{{\bf
k}^\prime_0 - {\bf k}^ \prime_2}^{(1)}\right] .\eqno (V.7b)
$$

The quantity $\Delta_2^\prime \left( {\bf k}^\prime_0, {\bf
k}^\prime_1 \right)$, consistent with our notations for the primed
quantities, is given by
$$
\Delta _2^\prime \left( {\bf k}^\prime_0, {\bf k}^\prime_1 \right) =
\Delta_2 \left( {\bf K}_0 + {\bf k}^\prime_0, {\bf K}_1 + {\bf
k}^\prime_1 \right) \eqno (V.8)
$$
where $\Delta_2 \left( {\bf k}_0, {\bf k}_1 \right)$ has previously
been defined as $\Delta_2 \left( {\bf k}_0, {\bf k}_1 \right) =
\omega_{{\bf k}_0}^{(0)} - \omega_{{\bf k}_1}^{(1)} - \omega_{{\bf
k}_0 - {\bf k}_1}^{(2)}$  In the envelope approximation limit
$\Delta^\prime_2 \left( {\bf k}^\prime_0, {\bf k}^\prime_1 \right)$
reduces to
$$
\hat \Delta_2^\prime \left( {\bf k}^\prime_0, {\bf k}^\prime_1
\right) = {\bf k}^\prime_0 \cdot {\bf V}_{g0} - {\bf k}^\prime_1
\cdot {\bf V}_{g1} - \left( {\bf k}^\prime_0 - {\bf k} ^\prime_1
\right) \cdot {\bf V}_{g2} \eqno (V.9)
$$
[It may be noted again that these definitions of $\Delta_i$ and
$\hat \Delta_i$ are consistent with each other in the sense that, if
one considers {\it ab initio} the coupled mode equations in their
envelope form, $\Delta_2 \left( {\bf k}_0, {\bf k}_1 \right)$
reduces identically to $\Delta_2^\prime \left( {\bf k}^\prime_0,
{\bf k}^\prime_1 \right)$ without any further approximation where
${\bf k}^\prime_\alpha = {\bf k}_\alpha - {\bf K}_\alpha$. The same
definition holds for $\Delta_1^\prime \left( {\bf k}^\prime_0, {\bf
k} ^\prime_2 \right)$ with the substitution ($1\leftrightarrow2$).

In the somewhat degenerate limit $V_{g0} = \infty$ considered in
this subsection, ${\bf k}^\prime_0 \cdot {\bf V}_{g0}$ has to be
replaced by $\omega^\prime_0$, corresponding to the limit ${\bf
k}^\prime_0 = 0$ in $\hat \Delta^\prime_i$; accordingly, in the
integral appearing in the RHS of Eq. (V.7), the quantities $n_{{\bf
k}^\prime_0}^{(0)} d{\bf k}^\prime_0$ and $\Delta_i^\prime \left(
{\bf k}^\prime_0, {\bf k}^\prime_j \right)$ have to be replaced by
$\hat n_{\omega_0}^{(0)} d \omega_0^\prime$ and $\hat {\tilde
\Delta}^\prime_i \left( \omega^\prime_0, {\bf k}^\prime_j \right)$
respectively where $\hat  {\tilde \Delta}^\prime_i \left(
\omega^\prime_0, {\bf k}^\prime_j \right) $ denotes $\hat
\Delta^\prime_i \left( {\bf k}^\prime_0, {\bf k}^\prime_j \right)$
in the limit $V_{g0} \rightarrow \infty,  {\bf k}^\prime_0
\rightarrow 0$ with $\omega^\prime_0 = {\bf k}^ \prime_0 \cdot {\bf
V}_{g0}$. One easily finds
$$
\hat {\tilde \Delta}^\prime_2 \left( \omega^\prime_0 , {\bf
k}^\prime_1 \right) = \omega^\prime_0 - {\bf k}^\prime_1 \cdot
\left( {\bf V}_{g1} - {\bf V}_{g2} \right) \eqno (V.10)
$$
and ($1 \leftrightarrow 2$) for $\hat {\tilde \Delta}^\prime_1
\left( \omega ^\prime_0, \vec {\bf k}^\prime_2 \right)$.  It is also
only in the same limit $V_{g0} \rightarrow \infty$ that the
integration over ${\bf k}^\prime_0$ in Eq. (V-7) does not involve
the spectral densities $n_{{\bf k}^\prime_0 - {\bf
k}^\prime_1}^{(2)}$ and $n_{{\bf k}_0 - {\bf k} ^\prime_2}^{(1)}$;
the RPA equations take then a simpler form, namely
$$
\hat D_1 \hat n_{{\bf k}^\prime_1}^{(1)} = \gamma_{11}^{RPA} \left(
{\bf k} ^\prime_1 \right) \left[ n_{{\bf k}^\prime_1}^{(1)} +
n_{-{\bf k}^\prime_1}^{(2)} \right]
$$
$$
\hat D_2 n_{{\bf k}^\prime_2}^{(2)} = \gamma_{22}^{RPA} \left( \vec
{\bf k} ^\prime_2 \right) \left[ n_{-{\bf k}^\prime_2}^{(1)} +
n_{{\bf k}^\prime_2}^{(2)} \right] \eqno (V.11)
$$
where the coupling constants are given by
$$
\gamma_{11}^{RPA} \left( {\bf k}^\prime_1 \right) = \gamma_0^2 \int
d \omega^ \prime_0 n_{\omega^\prime_0}^{(0)} \pi \delta \left( \hat
{\tilde \Delta}^\prime_2 (\omega^\prime_0, {\bf k}^\prime_1) \right)
$$
$$
\gamma_{22}^{RPA} \left( {\bf k}^\prime_2 \right) = \gamma_0^2 \int
d \omega^\prime_0 n_{\omega^\prime_0}^{(0)} \pi \delta \left( \hat
{\tilde \Delta}^\prime_1 (\omega^\prime_0, {\bf k}^\prime_2) \right)
\eqno (V.12)
$$

Performing the integration over $\omega_0$ in the latter expression,
one finds
$$
\gamma_{ii}^{RPA} \left( {\bf k}^\prime_i \right) =
\gamma_{ii}^{B,M} \left( {\bf k}^\prime_i \right) \eqno (V.13)
$$

The RPA equations correspond thus exactly to the Markov limit of the
Bourret approximation for the intensities $\langle\mid a \mid
^2\rangle$.

\item [(ii)]  We now consider the dispersion relation corresponding to the
exact Bourret equations (V-4).  Taking ${\bf k}^\prime_2 = - {\bf k}
^\prime_1$ in Eq. (V-4b), one obtains
$$
\hat D_1 \hat D_2 = \gamma_{11}^B \left( {\bf k}^\prime_1 \right)
\left( \hat D_1 + \hat D_2 \right) \eqno (V.14)
$$
The latter dispersion relation generalizes to three dimensions the
results obtained previously by Laval et al.$^{12}$  Following their
analysis, one may remark that the coupling constants
$\gamma_{11}^{B} \left( {\bf k}^\prime_i \right) = \gamma_{22}^{B}
\left( -{\bf k}^\prime_i \right)$ reach their maximum $\overline
\gamma_{11}^B = \overline \gamma_{22}^B$ for ${\bf k}^\prime_1 = 0$,
corresponding to the exact resonance conditions ${\bf k}_1 = {\bf
K}_1$ and ${\bf k}_2 = {\bf K}_2$ for the two wave packets
considered here. In this case the dispersion relation (V-14) reads
simply
$$
\hat D_1 \hat D_2 = \gamma_0^2 {\hat D_1 + \hat D_2 \over \hat D_1 +
\hat D_2 + \Delta \omega_0} \eqno (V.15)
$$
from which it follows that the dispersion relation $\hat D_1 \hat
D_2 = \gamma_0^2$ of the coherent case is recovered for $\Delta
\omega_0 < \mid \hat D_1 + \hat D_2 \mid$.  The opposite limit
$\Delta \omega_0 > \mid \hat D_1 + \hat D_2 \mid$ corresponds to the
validity domain for the Markov approximation, and therefore for the
RPA equations.  There is thus no intermediate domain between the
coherent and the RPA ones in the case $V_{g0} = \infty$.  This
result can be tracked back to the fact that for $V_{g0} = \infty$,
one has $\Delta \omega_1 = \Delta \omega_2 = \Delta \omega_0$, as
can be seen from the general definition (IV-20) for $\Delta
\omega_i$, so that the cross inequities (III-14c) and III-14d) are
automatically satisfied in the incoherent domain (III-12), and
consequently the RPA equations are exactly applicable in the whole
incoherent domain.  To summarize, the parametric coupling takes
place coherently in the coherent domain
$$
\Delta \omega_0 < {\rm Max} \left(
 \nu_i , \gamma_o \right) \eqno (V.16)
$$
whereas in the complementary - or incoherent - domain
$$
\Delta \omega_0 > \nu_1, \nu_2, \gamma_0 \eqno (V.17)
$$
the coupling constants $\overline \gamma_{11}^B, \overline
\gamma_{22}^B$ may be reduced to their RPA values
$$
\overline \gamma_{ii}^{B,M} = \overline \gamma_{ii}^{RPA} =
\overline \gamma_i^M = {\gamma_0^2 \over \Delta \omega_0} \eqno
(V.18)
$$
for i = 1 and 2.  The corresponding dispersion relation is then
simply
$$
\hat D_1 \hat D_2 = {\gamma_0^2 \over \Delta \omega_0} \left( \hat
D_1 + \hat D_2 \right) \eqno (V.19)
$$

\item [iii)]  Regarding now the comparison between the average amplitudes and
the intensities, one may easily obtain the threshold and the growth
rate for the intensities from the exact dispersion relation III-14.

The threshold reads $\gamma_0^2 > \nu_1 \nu_2 (\nu_1 + \nu_2 +
\Delta \omega_0) / (\nu_1 + \nu_2)$; it reduces to the coherent
result $\gamma_0^2 > \nu_1\nu_2$ for $\Delta \omega_0 < \nu_1 +
\nu_2 \simeq Max (\nu_1, \nu_2)$ and to the RPA result, namely
$\gamma_0^2 / \Delta \omega_0 > \nu_1 \nu_2 / (\nu_1 + \nu_2)
\approx {\rm Min} \left( \nu_1, \nu_2 \right)$ in the opposite case.
These expressions can be easily checked to be identical to the ones
obtained from consideration of the average amplitudes, i.e. by
looking at the most unstable average amplitude, as explained in
Section III-2-2.

Concerning now the growth rates, the growth rate for the intensities
is given, well above threshold, by the usual coherent expression
$\gamma_{<a^2>} = \gamma_0$ for $\gamma_0 > \Delta \omega_0$ and by
the RPA one $\gamma_{<a^2>} = 2\gamma_0^2 / \Delta \omega_0$ in the
opposite case.  It is interesting to notice that the correct value
for the growth rate ($2 \gamma_{<a^2>}$) of the intensities is
exactly {\it twice} the lower bound (IV-25) obtained from
consideration of the average amplitudes.  The square amplitudes
behave indeed as $\langle a_1\rangle^2 \sim \langle a_2\rangle^2
\sim {\rm exp}~ {(2 \ gam ma_0^2 / \Delta \omega_0)} t$, whereas the
RPA equations predict in the same incoherent regime $\langle \mid
a_1 \mid^2 \rangle \sim \langle \mid a_2 \mid^2 \rangle \sim {\rm
exp} {(4 \gamma_0^2 / \Delta \omega_0)} t$. This result is most
easily understood by remembering that, in a stochastic process, such
as this one, no two realizations are exactly the same, and there is
probability associated with a given mode's history in amplitude
space.  Given that probability distribution function, one can
compute the average mode amplitude or the average mode intensity. If
the distribution is narrow, then the intensity growth rate will be
close to twice the amplitude growth rate.  However, in this case,
the fluctuations about the mean are comparable to the mean in the
incoherent limit so that higher order moments of the mode amplitude
will have a more than proportionately larger growth rate.  This
facet of the problem is discussed more completely in Sec. VII.  This
result is not unexpected and is actually a general feature of the
fully incoherent regimes:  in an incoherent regime, the growth rate
of an average amplitude is indeed given quite generally by
$\gamma_{<a>} = \gamma_0^2 \tau_c$, where $\gamma_0$ is the coherent
growth rate and $\tau_c$ the correlation time of the stochastic
process.  Since the coherent growth rate of the quantity $\mid a
\mid^2$ is $2 \gamma_0$, we deduce that the growth rate of the
average intensity is given by $\gamma_{\langle a^2\rangle} = \left(
2 \gamma_0 \right)^2 \tau_c = 2 \left( 2 \gamma_{\langle
a\rangle}\right)$, corresponding thus to twice the lower bound $2
\gamma_{\langle a\rangle}$ obtained from Eq. III.10.  More generally
it will be argued in Section VI that the thresholds and convective
growth rate regarding the intensities can be obtained quite
accurately from the results corresponding to the average amplitudes,
whenever one has $\gamma_{11} >> \gamma_{22}$ or $\gamma_{22} >>
\gamma_{11}$, and at worst within a factor of two in the case
$\gamma_{11} \approx
 \gamma_{22}$.
\end{itemize}

\subsection{ The RPA equations} In this subsection
we derive the RPA equations in the general case from the Bourret
equations for the intensities.  Following the definition (V-2) for
the spectral densities, we set
$$
\langle a_1 \left( {\bf x}, t \right) a_1^* \left( {\bf x}^\prime, t
\right) \rangle = \int d{\bf k}_1 n_{{\bf k}_1}^{(1)} \left( {{\bf
x} + {\bf x}^\prime \over 2} , t\right) {\rm exp}~ i {\bf k}_1 \cdot
\left( {\bf x} - {\bf x}^\prime \right)
$$
and we Laplace transform in space the slowly varying spectral
density $n_{{\bf k}_1}^{(1)}$ ({\bf x}, t) by setting $n_{{\bf
k}_1}^{(1)} ({\bf x}, t) = \int d {\bf \kappa}~ {\rm exp}-2 {\bf
\kappa} \cdot x ~ n_{{\bf k}_1}^{(1)} ({\bf \kappa}, t)$. The
quantity $n_{{\bf k}_1}^{(1)} ({\bf \kappa}, t)$ is easily checked
to be given by
$$
n_{{\bf k}_1}^{(1)} ({\bf \kappa}, t) = \langle a_{{\bf k}_1 +
i\kappa}^{(1)} \left( a_{k_1 - i\kappa}^{(1)} \right)^* \rangle
\eqno (V.20)
$$

The equation for the evolution of $a_{{\bf k}_1+i \vec{\bf
\kappa}}^{(1)} \left( a_{{\bf k}_1 - i {\bf \kappa}}^{(1)}
\right)^*$ can be obtained from the coupled mode equation (II.7a)
and reads
$$
\left[ \partial_t + i \left( \omega_{{\bf k}_1 + i{\bf
\kappa}}^{(1)} - \omega_{{\bf k}_1-i{\bf \kappa}}^{(1)} \right) + 2
\nu_1 \right] a_{{\bf k} _1+i{\bf \kappa}}^{(1)} \left( a_{{\bf
k}_1-i {\bf \kappa}}^{(1)} \right)^*
$$
$$
= \gamma_0 \int d{\bf k}_0 a_{{\bf k}_0}^{(0)} \left( a_{{\bf
k}_0-({\bf k} _1 + i{\bf \kappa})} \right)^* \left( a_{{\bf
k}_1-i{\bf \kappa}}^{(1)} \right)^*
$$
$$
+ \left[ c \cdot c, {\bf \kappa} \rightarrow - {\bf \kappa} \right]
\eqno (V.21)
$$
where we neglected as before the slow variations of $\nu_{{\bf
K}_1}^{(1)}$ and $v_{{\bf k}_0, {\bf k}_1, {\bf k}_0 - {\bf
k}_1}^{(1)}$ upon ${\bf k}_0$ and ${\bf k}$.  Similar equations can
be written for the quantities
$$
a_{{\bf k}_0-{\bf k}_1+i{\bf \kappa}}^{(2)} \left( a_{{\bf
k}_1-i{\bf \kappa}} ^{(1)} \right)^* ~{\rm and} a_{{\bf k}_2+i {\bf
\kappa}}^{(2)} \left( a_{{\bf k}_2-i{\bf \kappa}}^{(2)} \right)^*
$$
so that the complete set for the quantities $a_i a_j*$ has the form
of a stochastic linear multiplicative system (where the random
variable is the set $a_{{\bf k}_0}^{(0)}$), upon which the Bourret
approximation can be performed. It is convenient at this stage to
Laplace transform in time the resulting equation by setting
$$
n_{{\bf k}_1}^{(1)} \left( {\bf \kappa}, t \right) = \int d \gamma
n_ {{\bf k}_1}^{(1)} \left( {\bf \kappa}, \gamma \right) {\rm exp}~
2 \gamma t
$$
and similarly for the other quantities.  The Bourret equations take
the form
$$
D_1^+ n_{{\bf k}_1}^{(1)} \left( {\bf \kappa}, \gamma \right) =
\gamma_0^2 \left[ \int d{\bf k}_0 n_{{\bf k}_0}^{(0)} R_2 \left(
{\bf k} _0,{\bf k}_1 \right) \right] \left[ n_{{\bf k}_1}^{(1)}
\left( {\bf \kappa}, \gamma \right) + n_{{\bf k}_0-{\bf k}_1}^{(2)}
\left( {\bf k}, \gamma \right) \right] \eqno (V.22)
$$
where the resonance function $R_2 \left( {\bf k}_0, {\bf k}_1
\right)$ is given by
$$
R_2 \left( {\bf k}_0, {\bf k}_1 \right) = {D_1 + D_2 \over \left(
D_1 + D_2 \right)^2 + \left( \Delta_2 \left( {\bf k}_0, {\bf k}_1
\right) \right)^2} \eqno (V.23)
$$
and where the operator $D_1^+ (\gamma, {\bf \kappa})$ denotes the
Hermitian part of $D_1 \left( \gamma, {\bf \kappa} \right)$, i.e.
$$D_1^+ \left(\gamma, {\bf \kappa}\right) = {1\over2} \left[ D_1
\left( \gamma, {\bf \kappa} \right) + \left( D_1 \left( \gamma^*,
{\bf \kappa}^*\right) \right)^*\right].$$ In the general case
$D_1^+$ is given by
$$
D_{1{\bf k}_1}^+ \left( \gamma, {\bf \kappa} \right) = \gamma + i
\left( \omega_{{\bf k}_1+i\kappa}^{(1)} - \omega_{{\bf k}_1-i{\bf
\kappa}} ^{(1)} \right) + \nu_1 \eqno (V.24)
$$
which reduces to $\hat D_{1,{\bf k}_1}^+= \hat D_{1,k_1} = \gamma +
\nu_1 - {\bf \kappa} \cdot {\bf V}_{g1} \left({\bf k}_1\right)$ in
the envelope approximation.  As a matter of fact, the envelope
approximation can be made for the {\it slow} variation associated
with the variables $\gamma$ and ${\bf \kappa}$; thus the $D_1$ and
$D_2$ appearing in the definition of $R_2 \left({\bf k}_0,{\bf k}_1
\right)$ can be reduced to $\hat D_{1,{\bf k} _1}$ and and $\hat
D_{2,{\bf k}_0 - {\bf k}_1}$ respectively, and similarly the
operator $D_1^+$  can be replaced by $\hat D_{1,{\bf k}_1}$ (by
contrast the quantity $\Delta_2 \left({\bf k} _0,{\bf k}_1 \right)$
contains the fast dependence on the integration variable ${\bf k}_0$
and cannot be reduced to its envelope approximation in the general
case.]

Making the envelope approximation for the slow dependence, one
obtains the following set of equations.
$$
\left( \hat D_1 - \gamma_{11}^B ({\bf k}_1) \right) n_{{\bf
k}_1}^{(1)} = \gamma_0^2 \int d{\bf k}_0 \hat R_2 \left( {\bf
k}_0,{\bf k}_1 \right) n_{{\bf k}_0} ^{(0)} n_{{\bf k}_0-{\bf
k}_1}^{(2)}
$$
$$
\left( \hat D_2 - \gamma_{22}^B ({\bf k}_2) \right) n_{{\bf
k}_2}^{(1)} = \gamma_0^2 \int d{\bf k}_0 \hat R_1 \left({\bf
k}_0,{\bf k}_1 \right) n_{{\bf k}_0} ^{(1)} n_{{\bf k}_0-{\bf
k}_1}^{(1)} \eqno (V.25)
$$
where the "diagonal" coupling constants $\gamma_{ii}^B$ are given by
$$
\gamma_{11}^B \left( k_1 \right) = \gamma_0^2 \int d{\bf k}_0
n_{{\bf k}_0} ^{(0)} \hat R_2 \left( {\bf k}_0,{\bf k}_1
\right)\eqno (V.26)
$$
with
$$
\hat R_2 \left( {\bf k}_0,{\bf k}_1 \right) = {\hat D_1 + \hat D_2
\over \left( \hat D_1 + \hat D_2 \right)^2 + \left( \Delta_2 ({\bf
k}_0,{\bf k}_1) \right)^2} \eqno (V.27)
$$
and ($1\rightarrow2$) for $\gamma_{22}^B ({\bf k}_2)$.

The set (V.25) constitutes the Bourret equations for the intensities
$\langle \mid a_1 \mid^2 \rangle$ and $\langle \mid a_2
\mid^2\rangle$. This set is still very difficult to solve since, in
addition to the convolution integral on the RHS, it contains various
quantities which are non-Markovian. The coupling constants
$\gamma_{ii}^B$ can be further simplified in the Markov
approximation, denoted by $\gamma_{ii}^{B,M}$. In this
approximation, i.e. $\mid \Delta_2 \mid >> \mid \hat D_1 + \hat D_2
\mid, R_2 \left( k_0, k_1
 \right)$ can be replaced by
$\pi \delta \left( \Delta_2 ({\bf k}_0,{\bf k}_1) \right)$ so that
the system (V.25) becomes
$$
\hat D_1 n_{{\bf k}_1}^{(1)} = \gamma_0^2 \int d{\bf k}_0 n_{{\bf
k}_0}^{(0)} \pi \delta \left( \Delta_2({\bf k}_0,{\bf k}_1) \right)
\left[ n_{{\bf k}_1}^{(1)} + n_{{\bf k}_0-{\bf k}_1}^{(2)}\right]
\eqno (V.28a)
$$
$$
\hat D_2 n_{{\bf k}_2}^{(2)} = \gamma_0^2 \int d{\bf k}_0 n_{{\bf
k}_0}^ {(0)} \pi \delta \left( \Delta_1({\bf k}_0,{\bf k}_2) \right)
\left[ n_{{\bf k}_2}^{(2)} + n_{{\bf k}_0-{\bf k}_2}^{(1)}\right]
\eqno (V.28b)
$$

The latter set constitutes the RPA equations for the coupled mode
equations neglecting pump depletion and written in the Laplace
variables $\gamma$ and ${\bf \kappa}$. In the real space the
operators $\hat D_\alpha$ have simply to be replaced by
$$
{1 \over 2} \left( \partial_t + {\bf V}_{g\alpha} \cdot
\partial_{\bf x} + 2 \nu_\alpha \right) .
$$

In order to make a closer connection with the special case
considered in the previous subsection (V.7), one can easily check
that the latter set reduced identically to the system (V.7), when
dealing with primed quantities (i.e. after factorization of the fast
variation).  One may also remark that the mismatch $\Delta_2 \left(
{\bf k}_0,{\bf k}_1 \right)$ reduces to $\hat \Delta_2 \left({\bf
k}_0,{\bf k}_1 \right) = {\bf k}_0 \cdot V_{g0} - k_1 \cdot V_{g1}-
\left( k_0 - k_1 \right) \cdot V_{g2}$ in the envelope
approximation, in which limit the set (V.28) can be recognized to be
the 3-D generalization of the Laval et al. results.$^{12}$

The validity condition for taking the Markov limit of the Bourret
equations (V.25) is {\rm Min} $\Delta \omega_i > \mid D_1 + D_2 \mid
\simeq {\rm Max} \mid D_i \mid$.  On the other hand, by using the
fact that well above threshold the growth rate is of the order of
$\gamma \simeq \gamma_0^2 / {\rm Min} (\Delta \omega_i)$, one easily
finds that the latter inequality reduces to
$$
{\rm Min} (\Delta \omega_j) > {\rm Max} (\nu_i, \gamma_0) \eqno (V.2
9)
$$
that is to say the condition (III.15) previously written in terms of
$\Delta_{j{\rm Max}}$. This condition entails in turn the inequality
$\gamma_0^2 < \Delta \omega_1 \Delta \omega_2$ which ensures the
validity of the Bourret approximation.  Therefore, the condition
(V.29) defines entirely the RPA domain.

The RPA set (V.28) remains a system difficult to solve due to the
integration over $k_0$ on the RHS.  One can however make some
additional approximations which are consistent with the validity
condition (V.29).  By assumption the spectral density
$n_{k_0}^{(0)}$ of the pump wave peaks when ${\bf k}_0$ is equal to
the mean wave number of the pump wave ${\bf K}_0$; consequently, for
a for a given angle of observation defining the direction of the
scattered wave, the RPA coupling constant
$$
\gamma_{11}^{RPA} ({\bf k}_1) \equiv \gamma_{11}^{B,M} ({\bf k}_1) =
\gamma_0^2 \int d{\bf k}_0 n_{{\bf k}_0}^{(0)} \pi \delta \left(
\Delta_2 ({\bf k} _0,{\bf k}_1) \right) \eqno (V.30)
$$
is maximum for ${\bf k}_1 = {\bf K}_1$, corresponding to the exact
resonance. In this case $\gamma_{11}^{RPA}$is given by
$$
\overline \gamma_{11}^{RPA} \equiv \gamma_{11}^{RPA} ({\bf K}_1) =
\gamma_0^2 \int d{\bf k}_0 n_{{\bf k}_0}^{(0)} \pi \delta \left(
\overline \Delta_2( {\bf k}_0) \right) \eqno (V.31)
$$
corresponding to the result announced in Section IV, namely
$\overline\gamma ^{RPM} = \overline \gamma_i^M$. If one now takes
${\bf k}_1 = {\bf K}_1$ in Eq. (V.28a) the relation $\Delta_2
\left({\bf k}_0,{\bf K}_1 \right) = 0$ defines a surface in ${\bf
k}_0$ space that contains the point ${\bf k}_0 = {\bf K}_0 ({\rm in}
1-D, \Delta_2 \left( {\bf k}_0,{\bf K}_1 \right) =0$ would require
${\bf k}_0 = {\bf K}_0$ uniquely; as a matter of fact, the function
$n_{{\bf k}_0}^{(0)}$ attains its maximum at this point.  Due to the
large spectral width imposed by the RPA validity conditions (V.29),
the function $n_{{\bf k}_0-{\bf K}_1}^{(2)}$ is a slowly varying
function of ${\bf k}_0$, so that one does not make a significant
error in the integral appearing in the RHS of Eq. (V.28a) by
replacing $n_{{\bf k}_0-{\bf K}_1}^{(2)}$ with $n_{{\bf K}_2}^{(2)}$
(an approximation which is actually exact in 1D), and similarily
$n_{{\bf k}_0 - {\bf K}_2}^{(1)}$ by $n_{{\bf K}_1}^{(1)}$ in Eqs.
(V28b), and obtaining
$$
\hat D_1 \overline n_1 = \overline \gamma_{11}^{RPA} \left(
\overline n_1 + \overline n_2 \right)
$$
$$
\hat D_2 \overline n_2 = \overline \gamma_{22}^{RPA} \left(
\overline n_1 + \overline n_2 \right) \eqno (V.32)
$$
where the RPA coupling constants $\overline \gamma_{ii}^{RPA}$ are
given by
$$
\gamma_{11}^{RPA} = \gamma_0^2 / \Delta \omega_1
$$
and
$$
\gamma_{22}^{RPA} = \gamma_0^2 / \Delta \omega_2
$$
according to our definition (IV.14).  The quantities $\overline
n_\alpha$ denote, for the for the sake of simplicity in the
notations, the spectral density $n_{{\bf k}_\alpha}^{(\alpha)}$ at
the point of exact resonance ${\bf k}_{\alpha} = {\bf K}_{\alpha}$.
Equations (V.32) provide the desired system with which the stability
of the intensities $<\mid  a_\alpha \mid^2>$ is analyzed in the
remaining of the paper.

\section{Thresholds and Growth Rates for the Convective and
the Absolute Instabilities and Spatial Amplification}  In this
section we derive the expressions of the convective and absolute
growth rates for the parametric instabilities as a function of the
two quantities $\Delta \omega_1$ and $\Delta \omega_2$
characterizing the pump incoherence; we also compute the rate of
spatial amplification. Let us first recall the physical meaning of
these various growth rates by considering the Green function, that
is to say the response of the parametric coupling to an
infinitesimal impulse given to the system at t = 0 and x = 0. For
simplicity we restrict ourselves to a one dimensional problem, and
we denote as $x_m(t)$ the point where the Green function G(x,t) is
maximum.

For an infinite system, the {\it convective} growth rate, denoted as
$\gamma^{{\rm conv}}$, is defined to be the time growth rate of
$G_\infty (x_m (t),t)$, i.e.
$$ G_\infty (x_m(t),t)~_{t\rightarrow \infty}^\sim {\rm exp} (\gamma^{
{\rm conv}}t)
$$
Quite generally the point $x_m(t)$ moves in time according to
$x_m(t) = V_m t$, where $V_m (= V_1+V_2)/2)$ is the group velocity
of the maximum. For a finite system, the convective growth rate
characterizes therefore the transient behavior only; although the
Green function $G_L$ for a plasma slab of length L does not reduce
to the Green function $G_B$ of the infinite case, it is usually
admitted that the infinite system convective growth rate
$\gamma^{{\rm conv}}$ does properly describe the early time behavior
for t$ \leq L/V_m$.

Regarding now the long time behavior of the parametric coupling, one
has first to look for the existence of absolute instabilities.  For
an infinite system, when such instabilities exist, the {\it
absolute} growth rate is the time growth rate of $G_\infty (x,t)$,
i.e.
$$
\mid G_\infty (x,t) \mid~ _{t\rightarrow \infty}^\sim {\rm exp}
\left[ \gamma^{{\rm abs}} t - \kappa^{{\rm abs}} x \right]
$$
for finite values of $x$; in the latter expression $\kappa^{abs}$
denotes the space growth rate associated with the absolute
instability.  For a finite system, there exists in general a
critical length Labs above which there are unstable normal modes, so
that the system will behave asymptotically in time as the most
unstable normal mode.  Similarly it is usually admitted that
whenever the plasma length L satisfies the condition $L >> L_{abs}$,
the growth rate of the most unstable normal mode is well
approximated by the absolute growth rate $\gamma^{abs}$ of the
infinite case.

When absolute instabilities do not exist but the system is
convectively unstable, the long term behavior is determined by
spatial amplification. The latter corresponds to setting a source
$s_o {\rm exp} - i\omega_st$ at x = 0 and looking for the $\omega_s$
which maximizes the {\it spatial amplification} rate
$\kappa(\omega_s)$; Bers and B riggs have indeed shown that in such
a case the time asymptotic behavior of an infinite system in the
presence of the source goes like
$$
\left[ G_\infty s_o {\rm exp} \left(- i \omega_s t
\right)\right]_{\mid x \mid \rightarrow \infty} \sim {\rm exp}~
\left(\epsilon \kappa (\omega_s) x\right)
$$
where $\epsilon = \pm 1$, depending upon the direction for spatial
amplification.  Such behavior also indicates that a significant
spatial amplification may be expected in a finite plasma slab
whenever the condition $\mid \kappa^{sa}L \mid >> 1$ is fulfilled,
where $\kappa^{sa}$ denotes the maximum $\kappa (\omega_s)$.

Lastly there is also the possibility for a mixed situation where
there is the coexistence of an absolute instability and spatial
amplification. Indeed such a case will be seen to be provided by the
R.P.A. dispersion relation.  In this case the short and long time
behaviors are still given by the convective and absolute growth
rates, for the reasons explained just above.  In the case however
where the spatial amplification growth rate $\mid \kappa^{sa} \mid$
is significantly larger than the space growth rate $\mid
\kappa_{abs} \mid$ associated with the absolute instability, one may
expect the existence of an intermediate regime in the time evolution
during which the spatial amplification feature dominates that of the
absolute instability.

\subsection{Convective instabilities. }
\subsubsection{Domain of applicability of the RPA equations.}
To begin with, let us first justify our conjecture concerning the
applicability of the RPA results in the intermediate domain where
neither the coherent nor the RPA equations could {\it a priori} be
used.  To do so we compare the RPA predictions with those for the
average amplitudes.

The dispersion relation corresponding to the RPA equations V.32. is
$$
\left( \hat D_1 - \gamma_1 \right) \left( \hat D_2 - \gamma_2
\right) = \gamma_1 \gamma_2 \eqno (VI.1)
$$
and the dispersion relation for the average amplitude $\langle
a_\alpha \rangle$ Eq. (IV.5) is
$$
\hat D_\alpha - \gamma_\alpha = 0 \eqno (VI.2)
$$
where $\gamma_\alpha$ denotes the RPA coupling constant $\overline
\gamma_ {\alpha\alpha}^{RPA}$ at exact resonance, namely $\gamma_1
\equiv \gamma_0^2 / \Delta \omega_2$ and $\gamma_2 = \gamma_0^2 /
\Delta \omega_1$.  The spectral widths $\Delta \omega_j$ are defined
by Eq. (I-10) and (IV.20).

As explained in Section III, the intermediate domain is defined as
the domain where equations (VI.2) are correct for the two waves
$\alpha$ = 1 and 2 (i.e. the diagonal inequalities $\Delta
\omega_\alpha > {\rm Max} (\nu_\alpha , \gamma_0)$ are both
satisfied) whereas the RPA equations are a priori not applicable
(because one of the cross inequalities $\Delta \omega_\alpha >
\nu_\beta$, with $\alpha \not= \beta$, is not satisfied).  In this
domain one can only use the inequality (III.10), namely {\rm
Max}$(\gamma_{\langle a_i \rangle}\rangle \leq \gamma_{\langle a^2
\rangle}$ which states that the actual growth rate $\gamma_{\langle
a^2\rangle}$ is no less than the most unstable average amplitude
(the latter being reduced as compared with the coherent growth rate
since Eq. (VI.2) predicts a reduction of the parametric growth in
its domain of applicability).  The question is thus whether this
lower bound is attained or whether $\gamma_{\langle a^2\rangle}$
remains of the order of the coherent growth rate.  To answer this
question, it is natural to first compare the RPA predictions with
those for the average amplitudes in their common domain of
applicability, namely the RPA domain.  It will be seen in the next
subsection that they are identical with regard to the convective
threshold and convective growth rate to within a numerical factor no
larger than two.

The reason for this identity can be tracked back as follows.  The
RPA dispersion relation VI.1 can be approximated by $\hat D_1 -
\gamma_1 \simeq 0$ if $\gamma_2 << \gamma_1$ or by $\hat D_2 -
\gamma_2 \simeq 0$ if $\gamma_2 >> \gamma_1$, so that the two
quantities  $\langle \mid a_1 \mid^2 \rangle \sim \langle \mid a_2
\mid^2 \rangle$ have necessarily a growth rate $\gamma _{\langle a_2
\rangle}$ which is of the order of Max $\gamma_{\langle a_i\rangle}
$. More precisely, denoting by $\gamma_{RPA}$ the growth rate
obtained from the dispersion relation VI.1, and by $\gamma_{{\rm
max}}$ the largest convective growth rate given by the Bourret
dispersion relation VI.2, it is the matter of a simple calculation
to show that the following inequalities
$$
\gamma_{{\rm max}} \leq \gamma_{RPA} \leq 2 \gamma_{{\rm max}} \eqno
(VI.3)
$$
hold formally, i.e. independently of the applicability of the RPA
and Bourret equations (the upper inequality occurs in the case
$\gamma_1 = \gamma_2$).  Since in the RPA domain one has $\gamma_
{\langle a^2 \rangle} = \gamma_{RPA}$ and Max $\gamma_{\langle a_i
\rangle} = \gamma_{{\rm max}}$, one deduces that the latter
inequality can be written
$$
{\rm Max} \left( \gamma_{\langle a_i \rangle} \right) \leq \gamma_
{\langle a^2 \rangle} \leq 2 {\rm Max} \left( \gamma_{\langle a_i
\rangle} \right) \eqno (VI.4)
$$
in the RPA domain.  This result means that the phase mixing effects
on $\langle a_\alpha \rangle$ are negligible, at least for the most
unstable average amplitude.  Since this property is satisfied in the
RPA domain, which can be regarded as the most incoherent domain,
i.e. the domain in which the phase mixing effects are expected to be
the more important, we make the conjecture that the inequality
(VI.4) is satisfied in the intermediate domain as well.  On the
other hand, in this domain one has Max $\left( \gamma_{\langle a_i
\rangle} \right) = \gamma_{{\rm max}}$, so that this conjecture
means $\gamma_{{\rm max}} \leq \gamma_{\langle a^2 \rangle} \leq 2
\gamma_{{\rm max}}$; on comparing this inequality with (IV.3), one
realizes that one can approximate in the intermediate domain, the
actual growth rate $\gamma_{\langle a^2 \rangle}$ by the RPA
prediction $\gamma_{RPA}$, with an error no larger than a factor of
two.  As anticipated in Section III, these arguments allow us to
apply the RPA results in the entire incoherent domain $\Delta
\omega_1 > \nu_1, \gamma_0$ and $\Delta \omega_2 > \nu_2, \gamma_0$
including the intermediate domain.

\subsubsection{ Convective instabilities expressions}
The threshold $\gamma_{0 {\rm conv}}$ for the convective
instabilities is easily obtained from the RPA dispersion relation
(VI.1); it reads
$$
\gamma_{0 {\rm conv}} = \left[ \nu_1 \nu_2 \ \left( \nu_1 / \Delta
\omega_1 + \nu_2 / \Delta \omega_2 \right) \right] ^{1/2} \eqno
(VI.5)
$$
and can be approximated by
$$
\gamma_{0 {\rm conv}} = \left[ {\rm Min} \left(\Delta \omega_1 \nu_2
, \Delta \omega_2 \nu_1 \right) \right] ^{1/2} \eqno (VI.6)
$$

It is easily checked that the latter approximation corresponds to
what can be deduced from the behavior of the average amplitudes,
[namely from the condition $\gamma_{{\rm max}} \geq 0$].  We may
also notice that the expression VI.6 for the threshold in the
incoherent domain goes continuously into the usual coherent
threshold expression $\gamma_0^2 = \nu_1 \nu_2$ at the boundary
between the two domains, namely $\Delta \omega_1 = \nu_1$ and
$\Delta \omega_2 > \nu_2$, or $\Delta \omega_2 = \nu_2$ and $\Delta
\omega_1 > \nu_1$ (again within a numerical factor no larger than
two).

The convective growth rate, well above threshold, is given by
$$
\gamma_{\langle a^2 \rangle}^{{\rm conv}} = \gamma_0^2 \left( {1
\over \Delta \omega_1} + {1 \over \Delta \omega_2}\right) \eqno
(VI.7)
$$
which is of the order of $\gamma_{{\rm max}} = \gamma_0^2 /
\left({\rm Min} \Delta \omega_i\right)$, again within a factor no
larger than two [one has $\gamma_{\langle a^2 \rangle} = 2
\gamma_{{\rm max}}$ for $\Delta \omega_1 = \Delta \omega_2$ only,
which occurs for instance in the case $V_{g0}= \infty$ encountered
in Section III.2].  One can again show that the expression for the
growth rate in the incoherent domain goes continuously into the
coherent one $ \gamma_{\langle a^2 \rangle} = \gamma_0$ at the
boundary $\Delta \omega_1 = \gamma_0$, and $\Delta \omega_2 >
\gamma_0$ and $\Delta \omega_1 > \gamma_0$ (once more within a
numerical factor no larger than two).

\subsection{ Absolute instabilities}
The absolute growth rate is obtained from the RPA dispersion
relation VI.1 by looking for a double root for $\kappa$ and by
imposing the well-known Bers and Briggs criterion, i.e. tracing the
roots in the K-plane to insure that the double root arises from
roots coming from opposite sides of the real K axis as one varies
the growth rate from $\gamma_{\langle a^2 \rangle} ^{{\rm abs}}$ to
$\gamma_{\langle a^2 \rangle}^{{\rm abs}} + \infty$. One finds that
this occurs for $V_1 V_2 < 0$ and that the absolute growth rate is
given by
$$
\gamma_{\langle a^2 \rangle}^{{\rm abs}} = \left( \mid V_1 \mid +
\mid V_2 \mid \right) ^{-1} \left[ \gamma_0^2 - \left( \gamma_{0
{\rm abs}} \right) ^2 \right] \left[ \left( {V_1 \over \Delta
\omega_1} \right)^{1/2} + \left( {V_2 \over \Delta \omega_2}
\right)^{1/2} \right]^2 \eqno (VI.8a)
$$
where $\gamma_{0 {\rm abs}}$ denotes the absolute threshold
$$
\gamma_{0 {\rm abs}}^2 = \mid V_1 V_2 \mid {\left( {\nu_1 \over \mid
V_2 \mid} + {\nu_2 \over \mid V_2 \mid} \right) \over \left[ \left(
{V_1 \over \Delta \omega_1} \right) ^{1/2} + \left( {V_1 \over
\Delta \omega_1} \right) ^{1/2} \right]^2} . \eqno (VI.9a)
$$
Well above threshold the absolute growth rate can be approximated by
$$
\gamma_{\langle a^2 \rangle} ^{{\rm abs}} \simeq {\gamma_0^2 \over
\left( {\rm Max} \mid V_i \mid \right) \left( {\rm Min} \left(\Delta
\omega_j / \mid V_j  \mid \right)\right)} \eqno (VI.8b)
$$
and the absolute threshold is of the order of
$$
\gamma_{0 {\rm abs}}^2 \simeq \mid V_1 V_2 \mid \left( {\rm Max}
{\nu_i \over \mid V_i \mid} \right) \left( {\rm Min} {\Delta
\omega_j \over \mid V_j \mid} \right) . \eqno (VI.9b)
$$

Concerning now the domain of validity of the RPA expressions for the
absolute instabilities, let us first consider the case $\Delta
\omega_1 = \Delta \omega_2 \equiv \Delta \omega_0$; in this case the
cross inequalities (III.13c-d) reduce to $\Delta \omega_\alpha \geq
\mid \hat D_\alpha \mid$, so that it would appear that the RPA
validity domain is given by the inequalities $\Delta \omega_0 > \mid
\hat D_\alpha \mid$ with a = 1,2, where $\hat D_\alpha =
\gamma_{\langle a^2 \rangle}^{{\rm abs}} - \kappa^{{\rm abs}}
V_{g\alpha} + \nu_\alpha$ is computed for $\gamma = \gamma_{\langle
a^2 \rangle} ^{{\rm abs}}$ (with $\kappa$ evaluated at its double
root $\kappa_{{\rm abs}}$). Actually a detailed calculation[46]
performed with the exact dispersion relation (V.15) for the case
$V_{g0} = \infty$, shows that the limiting RPA expressions are
applicable in a smaller region, obtained by imposing the RPA
validity condition not at the single point ($\gamma_{\rm abs},
\kappa_{\rm abs}$) but on the entire path needed to apply the Bers
and Briggs procedure:  in fact one requires that $\Delta \omega_0 >
\mid \hat D_ \alpha \mid = \mid \gamma - \kappa_\pm(\gamma)
V_{g\alpha} + \nu_\alpha \mid$ for $\gamma$ varying from the
absolute growth rate $\gamma_{\langle a^2 \rangle} ^{{\rm abs}}$ to
the value at which one of the two roots $\kappa_\pm (\gamma)$
crosses the real axis.  We assume such a procedure is also required
in the the general case $\Delta \omega_1 \not= \Delta \omega_2$. One
then finds $\mid \hat D_1 \mid \simeq \gamma_0^2 / \mid V_2 \mid
{\rm Min} (\Delta \omega_1 / V_1)$ and $\mid \hat D_2 \mid \simeq
\gamma_0^2 / \mid V_1 \mid {\rm Min} \left( \Delta \omega_i / \mid
V_i \mid \right)$, so that the condition $\mid \Delta \omega_\alpha
\mid > \mid \hat D_\alpha \mid$ reduces to the simple inequality
$$
{\rm Min} \left( {\Delta \omega_i \over \mid V_i \mid}\right) >
{\gamma_0 \over \sqrt {\mid V_1 V_2 \mid}} . \eqno (VI.10)
$$

On the other hand, by looking at the behavior of the average
amplitudes $\langle a_i \rangle$, one easily finds that in the
complementary domain ${\rm Min} \left(\Delta \omega_i / \mid V_i
\mid \right)< \gamma_0 /\sqrt {V_1 V_2}$, the parametric system
behaves coherently for the absolute instabilities and the growth
rate reduces to the coherent limit given by
$$
\gamma_{coh}^{{\rm abs}} = {2 \sqrt {\mid V_1 V_2 \mid} \over \mid
V_1 \mid + \mid V_2 \mid } \left( \gamma_0 - \gamma_{0 {\rm
abs}}^{coh} \right) , \eqno (VI.11a)
$$
$$
\gamma_{0, {\rm abs}}^{coh} = {1 \over 2} \left( \nu_1 \sqrt{\mid
{V_2 \over V_1} \mid} + \nu_2  \sqrt{\mid {V_1 \over V_2}
\mid}\right) .  \eqno (VI.11b)
$$
For this reason we can define the incoherent domain for the absolute
instabilities by the inequality (VI-10).

With regard to the two cross inequalities $\mid \Delta \omega_1 \mid
> \mid \hat D_2 \mid $ and $\Delta \omega_2 > \mid \hat D_1 \mid$,
it can be seen that, because they are not necessarily fulfilled when
the condition VI.10 is satisfied, there is again the possibility for
an intermediate domain in which neither the RPA nor the coherent
results apply for the absolute instabilities.  One can verify
however that the single validity condition VI-10 ensures the
continuity (within a factor 2) of the RPA expression (VI-8) for the
absolute growth rate with the usual coherent expression
$\gamma^{{\rm abs}} \simeq 2 \gamma_0 \left({\rm Min} \mid V_i \mid
/ {\rm Max} \mid V_i \mid \right) ^{1/2}$.  For this reason we
conjecture that the RPA expression (VI.8) is actually applicable in
the whole incoherent domain.  As already noticed[16], it is
interesting to remark that the behavior of $\langle a \rangle$ and
$\langle \mid a \mid^2 \rangle$ are significantly different as far
as absolute instabilities are concerned: in the incoherent domain
(VI.10) the average amplitudes are essentially stable, whereas the
intensities are absolutely unstable as long as $\gamma_0$ exceeds
the threshold (VI.9).

In conclusion, therefore, the absolute growth rate is given by the
minimum of $\gamma_{\langle a^2 \rangle}^{{\rm abs}}$ given by Eq.
(VI.8a) and the coherent growth rate $\gamma_{coh}^{{\rm abs}}$
given by Eq. (VI.11).

\subsection{Spatial Amplification}
The spatial amplification growth rate $\kappa^{sa}$ is obtained from
the RPA dispersion by taking $\gamma = i\omega$, with $\omega$ real,
and by looking for a maximum of $\mid {\rm Re} ~\kappa \mid$ which
satisfies the Bers and Briggs criterion[75] for spatial
amplification namely, the root $\kappa$ in the $\kappa$-plane has to
cross the imaginary axis as one varies $\gamma$ from $i \omega$ to
$i \omega + \infty$.  Although the procedure is straightforward, the
general expressions of the different unstable roots are tedious due
to the large number of the independent parameters $V_\alpha,
\nu_\alpha, \Delta \omega_\alpha$. For the sake of simplicity we
will restrict ourselves in this subsection to the cases where the
following ordering is satisfied:
$$
\mid V_1 \mid >> \mid V_2 \mid \eqno (VI.12a)
$$
$$
{\nu_2 \over \mid V_2 \mid} >> {\nu_1 \over \mid V_1 \mid} \eqno
(VI.12b)
$$
$$
{\Delta \omega_2 \over \mid V_2 \mid} >> {\Delta \omega_1 \over \mid
V_1 \mid} \eqno (VI.12c)
$$

Such an ordering is always satisfied in the case of SRS and SBS
instabilities and quite often for two plasmon decay.  It is easy to
find that in the above limit there are two roots corresponding to
spatial amplification.  The spatial amplification growth rate for
the first root is given by
$$
\kappa_1^{sa} = - \gamma_0^2 / \Delta \omega_2 V_1 \eqno (VI.13)
$$
and its domain of existence corresponds to the inequalities
$$
\Delta \omega_1 \Delta \omega_2 > \gamma_0^2 \eqno (VI.14a)
$$
$$
\Delta \omega_1 > \nu_1 \eqno (VI.14b)
$$
$$
\Delta \omega_2 > \nu_2 \eqno (VI.14c)
$$
$$
\Delta \omega_2 \nu_1 < \gamma_0^2 < \Delta \omega_1 \nu_2 \eqno
(VI.15)
$$

The inequalities (VI.14) are the RPA validity conditions, whereas
the inequalities (VI.15) represent the condition for satisfying the
Bers and Briggs criterion, i.e., the condition for spatial
amplification (this condition implies in particular the inequality
$\Delta \omega_2 \nu_1 < \Delta \omega_1 \nu_2$).

The spatial amplification growth rate for the second root is given
by
$$
\kappa_2^{sa} = - \gamma_0^2 / \Delta \omega_1 V_2 \eqno (VI.16)
$$
and its domain of existence corresponds to the following
inequalities:
$$
\Delta \omega_1 > \nu_1 \eqno (VI.17a)
$$
$$
\Delta \omega_2 > \nu_2 \eqno (VI.17b)
$$
$$
\Delta \omega_1 > \gamma_0 \mid {V_1 \over V_2}\mid^{1/2} \eqno
(VI.17c)
$$
$$
\gamma_0^2 > \Delta \omega_1 \nu_2 \eqno (VI.18)
$$
again where inequalities (VI.17) are the RPA validity conditions for
this root, and as the inequality (VI.18) is the Bers and Briggs
condition for spatial amplification.  It is interesting to remark
that the inequalities (VI.17c) and (VI.18), reduce exactly to the
conditions (VI.10) and (VI.9b) for the existence of absolute
instabilities in the RPA regime, in the case $V_1 V_2 < 0$ and where
the ordering (VI.12) holds.  In this case one is thus in a mixed
situation where an absolute instability and spatial amplification
coexist together.  It can also be easily shown that the ordering
(VI.12) implies the inequality $\mid \kappa^{sa} \mid > \mid
\kappa^{{\rm abs}} \mid$, so that spatial amplification may dominate
the absolute instability for an intermediate stage in time or for an
unbounded plasma.

Lastly, it is worth mentioning that the two roots $\kappa_1^{sa}$
and $\kappa_2^{sa}$ can be interpreted {\it a posteriori} as
corresponding to spatial amplification for the average amplitudes
$\langle a_1 \rangle$ and $\langle a_2 \rangle$ respectively. For
instance the expression for $\kappa_2^{sa}$ follows simply from $D_2
= \gamma_2$, the inequalities (VI.17) are simply the Markov
condition $\mid D_\alpha \mid < \Delta \omega_\alpha$, and the
condition (VI.18) is the condition for instability for the average
amplitude $\langle a_2 \rangle$. We may therefore conclude that the
main features corresponding to spatial amplification - and to the
convective growth rate - for the intensities $\langle \mid a
 \mid^2 \rangle$ may be obtained from
consideration of the average amplitudes $\langle a_\alpha \rangle$,
in contrast with the absolute instabilities the features of which
can be found from the dispersion relation for $\langle \mid a \mid^2
\rangle$ only.

\subsection{Comparison of Stability Domains in the Case
$V_{g0} = \infty$}
In this subsection we repeat the various
stability domains obtained in the previous subsections in the
special case $V_{g0} = \infty$, for which one has $\Delta \omega_1 =
\Delta \omega_2 = \Delta \omega_0$, in order to exemplify the
different kinds of stabilization that can be expected from the pump
wave incoherence.  To do so, it is convenient to introduce the
following dimensionless quantities
$$
\overline {\Delta \omega_0} \equiv \Delta \omega_0 / \gamma_0
$$
$$
\overline {\nu_\alpha} \equiv \nu_\alpha / \gamma_0
$$
$$
v_2 \equiv V_2 / V_1
$$

We assume the ordering (VI.12) to hold, and in addition we suppose
$\nu_1 < \nu_2$ so that we restrict ourselves to the situations
where the two inequalities
$$
\mid v_2 \mid << 1 \eqno (VI.19a)
$$

$$
\overline \nu_1 < \overline \nu_2 \eqno (VI.19b)
$$
are satisfied.

\subsubsection{Convective Instability}
In Fig. 2 we reproduce the stability diagram corresponding to the
reduction of the convective instability growth rate induced by the
pump wave incoherence.  This diagram has to be understood as
representing the effect of pump bandwidth $\Delta \omega_0$ as a
function of $\nu_2$, for $\gamma_0$ and $\nu_1 < \nu_2$ given.  As
explained in the introduction to this Section, the convective
instability represents the early or transient time behavior of the
parametric coupling.

The domain of applicability of the statistical theory corresponds to
the inequalities
$$
\Delta \omega_0 > \gamma_0 ,
$$
$$
\Delta \omega_0 > {\rm Max} \left( \nu_1, \nu_2 \right) = \nu_2 ,
$$
that is to say,
$$
\Delta \overline {\omega_0} > 1 , \eqno (VI.20a)
$$

$$
\Delta \overline {\omega_0} > \overline \nu_2  . \eqno (VI.20b)
$$
The domain corresponding to these inequalitites is referred to as
"VR" in Fig. 2, and stands for "convective growth rate reduction;"
the solid line presents the boundary between the coherent and
incoherent domains.  The threshold VI.5 for convective instabilities
is approximately $\gamma_{0 {\rm conv}} = \left( \Delta \omega_0
\nu_1 \right) ^{1/2}$, so that the pump wave incoherence induces a
complete stabilization of the convective growth rate in the domain
referred to as "VS" in Fig. 2,
$$
\Delta \overline \omega_0 > \overline \nu_1 ^{-1} \eqno (VI.21)
$$
The boundary of the convectively stable region is the dotted line as
obtained from Eq. VI.5.  Lastly the parametric system is coherently
stable in the domain $\overline \nu_2 > \overline \nu_1^{-1}$
(corresponding to the usual coherent condition for stability
$\gamma_0^2 < \nu_1 \nu_2$), the boundary of which is represented by
the dashed-dot-dot line.

\subsubsection{Absolute Instability}
In Fig. 3, we reproduce the stability diagram corresponding to the
reduction of the absolute instability growth rate due to the pump
incoherence, i.e. the growth reduction of the long time behavior of
the parametric coupling in the case of existence of absolute
instabilities.  The domain of applicability of the statistical
theory of absolute instabilities (VI.10) reads here $\Delta \omega_0
> \gamma_2 \mid V_1 / V_2 \mid ^{1/2}$; on the other hand the
threshold (VI.9b) is $\gamma_{0 {\rm abs}} = \left( \Delta \omega_0
\nu_2 \right)^{1/2}$.  The intersection between the two lines
$\Delta \omega_0 = \gamma_0 \left( V_1 / V_2 \right) ^{1/2}$ and
$\Delta \omega_0 = \gamma_0^2 / \nu_2$ corresponds to $\nu_2 =
\gamma_0 \left( V_2 / V_1 \right)^{1/2}$, which is smaller by a
factor two than the usual coherent threshold for absolute
instabilities, namely $\nu_2 > 2 \gamma_0 \left( V_2 / V_1
\right)^{1/2}$.  Since the applicability condition for statistical
theory has been obtained from strong inequalities (whereas the
threshold condition $\gamma_{0{\rm abs}} = \left( \Delta \omega_0
\nu_2 \right)^{1/2}$ results from solving the RPA dispersion
relation) the two limits could be reasonably made continuous between
each other by writing the RPA validity condition as $\Delta \omega_0
> \gamma_0 \left( V_1 / V_2 \right)^{1/2} / 2$ for $\nu_2 < 2
\gamma_0 \left( V_2 / V_1 \right)^{1/2}$.  By doing so, the domain
corresponding to a reduction of the absolute growth rate (henceforth
referred to as the "AR" domain) would be approximated by the
inequality
$$
\Delta \overline {\omega_0} > \left( 2 \mid v_2 \mid ^{1/2}
\right)^{-1} \eqno (VI.22a)
$$
and the domain for complete stabilization of the absolute growth
(referred to as the "AS" domain) would be given by:
$$
\Delta \overline {\omega_0} > \mid \overline \nu_2 \mid^{-1} \eqno
(VI.23a)
$$

Actually a more detailed calculation[46] based upon the exact
dispersion relation (VI.15) shows that the AR domain corresponds to
the inequality
$$
\overline {\Delta \omega_0} > \mid v_2 \mid^{-1/2} \eqno (VI.22b)
$$
for $\overline \nu_2 < \mid v_2 \mid^{1/2}$, and that the complete
stabilization occurs when inequality (VI.23a) is satisfied in the
range $\overline \nu_2 < \mid v_2 \mid^{1/2}$, and when the
inequality
$$
\Delta \overline \omega_0 > 2 \mid v_2 \mid^{-1/2} - \overline \nu_2
\mid v_2 \mid^{-1}
$$
is fulfilled in the range $\mid v_2 \mid ^{-1} < \overline \nu_2 < 2
\mid v_2 \mid^{1/2}$.

In Fig. 3 the boundary of the AR domain is represented by a
short-dashed line, and the boundary of the AS domain, computed from
the incoherent threshold VI.9a, by a long-dashed line.  One may see
that the two boundaries almost coalesce at $\overline \nu_2 = \mid
v_2 \mid^{1/2}$, $\Delta \omega_0 = \mid V_2 \mid^{-1/2}$; this
simply means that the AR domain shrinks to zero in the limit $\mid
v_2 \mid \rightarrow 0$ for $\overline \nu_2 > \mid v_2 \mid ^{1/2}$
(for this reason the short-dashed line, the long-dashed line, and
the verticle dashed- dot-dot line bound a very small region for
$\overline \nu_2 > \mid v_2 \mid^{1/2}$).  The verticle line is the
absolute coherent threshold Eq. VI.11b.

It can be easily checked that the domains corresponding to the
approximate expression (VI.22a) and (VI.23a) do not significantly
differ from the exact domains (VI.22b) and (VI.23b), in the range
$\overline \nu_2 << \mid v_2 \mid^{1/2}$, i.e. far from the absolute
threshold.  This result justifies in particular all the approximate
expressions obtained in the remainder of our paper.  On the other
hand, in the range close to the absolute threshold, namely for $\mid
v_2 \mid^{1/2} < \overline \nu_2 < 2 \mid v_2 \mid^{1/2}$, this
exact solution exhibits the additional AS domain $2 \gamma_0 \mid
V_1 / V_2 \mid^{1/2} - \nu_2 \mid V_1 / V_2 \mid < \Delta \omega_0 <
\gamma_0^2 / \nu_2$, corresponding to a regime where a small $\Delta
\omega_0$ may reduce the absolute growth rate.  Such an effect is
not really surprising since it means that a small frequency
bandwidth may have a large effect when the parametric coupling is
close to the coherent threshold, namely it converts an absolute
instability into a convective one (which in turn gives rise to a
spatial amplification, as it will be seen in VI.D3), by decreasing
the usual coherent absolute threshold $\overline \nu_2 = 2 \mid v_2
\mid^{1/2}$ into the absolute threshold (VI.23b) modified by
incoherency effects, namely $\overline \nu_2 = 2 \mid v_2 \mid^{1/2}
- \mid v_2 \mid \Delta \omega_0$.

\subsubsection{Spatial Amplification}
In Fig. 4, we reproduce the stability diagram corresponding to the
reduction of the spatial amplification growth rate $\kappa^{sa}$.
The first root (VI.13) for the spatial amplification growth rate is
given in terms of dimensionless quantities as
$$
\overline \kappa_1 = - \left( \Delta \overline \omega_0
\right)^{-1}\eqno (VI.24)
$$
where the dimensionless growth rates $\overline \kappa$ are defined
as $\overline \kappa
 = \kappa^{sa} V_1/\gamma_0$.  The
domain of existence of this root, referred to in Fig. 4 as "AR1"
(for "spatial amplification reduction"), is defined by inequalities
(VI.14) and (VI.15), namely
$$
\overline {\Delta \omega}_0 > 1 \eqno (VI.25a)
$$
$$
\overline {\Delta \omega}_0 > \overline \nu_2 \eqno (VI.25b)
$$
$$
\overline {\Delta \omega}_0 > \overline \nu_2^{-1} \eqno (VI.25c)
$$
The boundary of this domain is represented by the dotted-dashed line
in Fig. 4.

The threshold condition (VI.15) can be written as
$$
\overline {\Delta \omega}_0 < \overline \nu_1^{~-1} \eqno (VI.26)
$$
The boundary of this domain "AS" is given more exactly by the dotted
line as in Fig. 2.

The spatial amplification growth rate for the second root is
$$
\overline \kappa_2 = - \left( \overline {\Delta \omega}_0 v_2
\right)^{-1} \eqno (VI.27)
$$
and its domain of existence, referred to in Fig. 4 as "AR2,"
corresponds to the inequalities (VI.17) and (VI.18), i.e.
$$
\Delta \omega_0 < \overline \nu_2^{~-1} \eqno (VI.28a)
$$
$$
\Delta \omega_0 > \mid v_2 \mid^{-1/2} \eqno (VI.28b)
$$
in addition to inequalities in (VI.25a) and (VI.25b).  The boundary
of this domain is represented by the decreasing dotted-dashed line
and the dashed line in Fig. 4.  It is interesting to remark that in
the domain $\overline \nu_2 > 2 \mid v_2 \mid^{1/2}$, the coherent
spatial amplification growth rate is $\overline \kappa_{coh} = - 1 /
\overline \nu_2$ so that $\overline \kappa$ varies continuously from
its coherent value to $\kappa_1 = -1 / \Delta \overline \omega_0$,
along the boundary $\Delta \overline \omega_0 = \overline \nu_2$
(the dotted-dashed line for $\overline \nu > 1$), whereas $\overline
\kappa$ is strongly reduced from $\overline \kappa_{\rm coh}$ to
$\overline \kappa_1$ along the boundary $\overline {\Delta \omega}_0
= 1 / \overline \nu_2$ (the dotted-dashed line for $\overline \nu_2
< 1$).

Similarly, in the domain $\overline \nu_2 << 2 \mid v_2 \mid^{1/2}$,
and $V_1V_2 > 0, \overline \kappa$ varies continuously from its
coherent expression $\overline \kappa = -1 / v_2^{1/2}$ to
$\overline \kappa_2 = -1 / \Delta \overline \omega_0v_2$ along the
boundary $\Delta \overline \omega_0 = (V_2)^{-1/2}$ between the
coherent domain and the AR2 domain, whereas $\overline \kappa$ is
strongly reduced from $\overline \kappa_2$ to $\overline \kappa_1$
along the boundary $\Delta \overline \omega_0 = 1 / \overline \nu_2$
between the AR2 and AR1 domains.

Let us consider now the last case, corresponding to the domain
$\overline \nu << 2 \mid v_2 \mid^{1/2}$ for $V_1V_2 < 0$. In this
case there is no spatial amplification in the coherent regime, but
an absolute instability characterized by a spatial growth rate
$\overline \kappa^{abs} \sim - 1 / \mid v_2 \mid^{1/2}$. We thus
find the same continuity as the one observed in the previous case
$V_1V_2 > 0$ along the boundary $\Delta \overline \omega_0 = \mid
v_2 \mid^{1/2}$ between the spatial growth rate $\overline
\kappa^{abs}$, corresponding to the coherent {\it absolute}
instability, and the incoherent {\it spatial amplification} growth
rate $\overline \kappa_2$; on the other hand $\overline \kappa$ is
strongly reduced from $\overline \kappa_2$ to $\overline \kappa_1$
along the boundary between the AR2 and AR1 domains, as in the case
$V_1V_2 > 0$. We may thus conclude that in all cases the boundary
$\Delta \overline \omega_0 = 1 / \overline \nu_2$, i.e. $\Delta
\omega_0 = \gamma_0^2 / \nu_2$ in physical units, corresponds to a
large reduction of the spatial amplification growth rate.

Lastly, we may give for the sake of completeness the expression of
the space growth rate $\overline \kappa^{abs}$ associated with the
incoherent absolute instability in the domain AR2.  As stated
before, for $V_1V_2 < 0$, the domain AR2 corresponds to a mixed
situation where there is the coexistence of a spatial amplification
characterized by $\overline \kappa_2 = 1 / \Delta \overline \omega_0
\mid v_2 \mid$, (for $V_1 > 0$ and $V_2 < 0$) and of an absolute
instability the space amplification
 growth rate of which is
$\overline \kappa^{abs} = 1 / \Delta \overline \omega_0 \mid v_2
\mid^{1/2}$; one has thus $\mid \overline \kappa^{abs} \mid << \mid
\overline \kappa_2 \mid$.

\subsubsection{Comparison of the Short Time and Long Time
behaviors}
In Fig. 5 we reproduce the three boundaries found in the
previous stability diagrams at which there is a reduction of the
parametric growth due to the pump wave incoherence; the solid line
corresponds to the boundary of the domain VR where the pump wave
incoherence reduces the convective growth rate, i.e. where it
modifies the short time behavior; the short- dashed line represents
the boundary of the domain AR where the pump wave incoherence
reduces the absolute growth rate, and the dotted-dashed line (which
coincides with the solid line for $\overline \nu_2 > 1$) where it
reduces the spatial amplification growth rate, i.e. the two domains
where the long time behavior is modified.  One immediately realizes
that all the domains for which there is a growth reduction in the
long time behavior are contained in the set for which there is a
reduction of the initial growth, whereas the converse is not true.
Thus the domain B in Fig. 5 is a domain where the spatial
amplification is not affected by the pump wave incoherence although
the convective growth is reduced in the initial stage; the same
remark applies to the domain A for $V_1V_2 > 0$; for $V_1 V_2 <0$
this latter domain corresponds to a regime where the absolute
instability growth rate, although much smaller than the convective
growth rate, is not modified by the pump wave incoherence whereas
the initial convective growth rate is reduced.  We may have
therefore the surprising conclusion that the reduction of the
parametric growth is more difficult to achieve in the long time
limit than in the initial stage.

\subsubsection{Finite Length Thresholds for Absolute
Modes:} The thresholds set by the physical damping of the modes have
been given for convective modes by (VI.5) and for absolute modes by
(VI.9) in the incoherent limit or by (VI11b) in the coherent limit.
In addition to satisfying these thresholds, the solutions must
satisfy a threshold length criterion.

In the case of absolute instability, the length criterion in the
coherent limit is known to be
$$
{\rm tan} \left( {\gamma_0^2 - \gamma_{abs}^2 \over \mid V_1 V_2
\mid} \right)^{1/2}  L = - \left( {\gamma_0^2 \over \gamma_{abs}^2}
-1 \right)^{1/2} \eqno (VI.29a)
$$
which in the absence of damping, $\gamma_{abs}$ sets
$$
L = {\pi \over 2} \sqrt {\mid V_1 V_2 \mid } \ \gamma_0 . \eqno
(VI.29b)
$$
A similar condition is obtained by solving the set (V.32) with
$\nu_j = 0$ and with the boundary condition $n_1(1) = n_2(L) = 0$
for $V_1V_2 , 0$ and $V_1 < 0$.  This RPA threshold condition is
given by the unusual expression,
$$
L_{th}^{RPA} = {\mid V_1V_2 \mid \over 2 \gamma_0^2} \left( {\mid
V_1 \mid \over \Delta \omega_2} - {\mid V_2 \mid \over
\Delta\omega_1} \right)^{-1} {\rm ln} \left( \mid {V_1 \over V_2}
\mid {\Delta \omega_1 \over \Delta \omega_2} \right) , \eqno (VI.30)
$$
which, not unexpectedly, does not reduce to the coherent expression
as $\Delta \omega_j \rightarrow 0$.  In the limit of purely temporal
incoherence, (VI.30) reduces to
$$
L_{th}^{RPA} = {1 \over 2} L_g^2 {\Delta \omega \over V_1 - V_2}
{\rm ln} \mid {V_1 \over V_2} \mid , \eqno (VI.31)
$$
whereas, for purely spatial incoherence, (VI.30) reduces to
$$
L_{th}^{RPA} = {\Delta K L_g^2 {\rm ln} \mid {V_1 \over V_2} \mid
\over \left( \mid {V_1 \over V_2} \mid - \mid {V_2 \over V_1} \mid
\right)} . \eqno (VI.32)
$$
The main point is that the ensemble averaged absolute growth rate
for the infinite system given by VI.8a) in the RPA theory can be
obtained only if the plasma size L sufficiently exceeds a length
given by the maximum of (VI.29) and (VI.30).

\section{Distribution of Mode Amplitudes and Intensities}
The statistical analysis has obtained dispersion relations from
which growth rates, thresholds, and amplification rates in different
domains of incoherence were obtained for the ensemble averaged
behavior of parametric instabilities with incoherent pump waves.  In
this analysis, some questions arose such as the validity of the RPA
dispersion relation in the intermediate domain that can be answered
in principle at least by a direct numerical integration of the
coupled mode equations (II-10).  Not only do we test the assumptions
and approximations of the statistical analysis but, in Sec. VIIA, we
illustrate the meaning of the averaging procedure.  The fundamental
assumption that the RPA equations apply in the whole incoherent
domain is examined in Section (VII.B); finally the relation of the
statistical growth rates and thresholds to the growth rates and
thresholds of a particular model of induced laser beam incoherence
is examined in Section (VII.C).

\subsection{Ensemble Average Growth Rates and the
Distribution of Intensities} As has been noted at several points in
this article, the incoherent absolute growth rates for the average
intensity can be twice the rate expected on the basis of the growth
rate for the average amplitude if certain inequalities are
satisfied.  For the purely temporal problem (where no distinction
between absolute and convective is made), this factor of two arises
if the damping is weak, specifically if Max $( \nu_j) < \gamma_0^2 /
\Delta \omega_0$. However if this inequality is reversed, the
average amplitude and average intensity growth rates are equal (we
remind the reader that the definition of growth rates in Eq. III.9
removes the expected factor of two from the intensity growth rate).
In the space-time problem considered in this article, one again
notices a factor of two if $\mid V_1 \mid = \mid V_2 \mid$ in Eq.
(VI.8a) for purely temporal incoherence, $\Delta \omega_1 = \Delta
\omega_2 = \Delta \omega_0$.  If one realizes that these random
processes lead to a distribution of mode amplitudes f(a) after a
time T, then it is clear that the first moment $\langle a \rangle =
\int da f(a)$ and the second moment $\langle \mid a^2 \mid \rangle =
\int f(a)$ will be related only in special cases by $\langle \mid
a^2 \mid \rangle = \mid \langle a \rangle \mid^2$.  If the
distribution remains narrow then these two moments are simply
related.  However we will discover that, in some cases, that is just
when there is this factor of two at issue, the distribution is
broad. The second moment then depends on finding the distribution of
rare events where the wave amplitude grows to larger amplitude than
the mean amplitude. We anticipate the result that the width of the
distribution increases in time as does the mean so that the task of
computing the ensemble average $\langle \mid a^2 \mid \rangle$ is
increasingly more difficult as the sampling time is increased.

\subsubsection{Purely Temporal Evolution}
These remarks are now illustrated by considering the coupled mode
equations for the purely temporal problem which have the simple form
given by
$$
\left( {\partial \over \partial t} + \nu_1 \right) a_1 (t) =
\gamma_0 \tilde S (t) a_2^* (t) ,
$$
$$
\left( {\partial \over \partial t} + \nu_2 \right) a_2 (t) =
\gamma_0 \tilde S (t) a_1^* (t) . \eqno (VII.1)
$$
In previous work, an exact solution for the ensemble average mode
amplitude and intensity was found when $\tilde S (t)$ is a
Kubo-Anderson process (Section V.1). When $\Delta \omega_0 >
\gamma_0^2 \ \Delta \omega_0 > \nu_1, \nu_2$ the growth rates are
found to be $\gamma_{\langle a \rangle}$ and $\gamma_{\langle a^2
\rangle} = 2 \gamma_0^2 \ \Delta \omega_0$. Here, we are interested
in the fluctuations of the amplitude and intensity about the means
$\langle \mid a \mid \rangle$ and $\langle \mid a \mid ^2 \rangle$.
Some useful insight can be gained into this question of fluctuations
and its influence on the behaviors of $\mid a \mid$ and $\mid a
\mid^2$ by considering the simplest possible model, that of two
coupled undamped modes $a_1$ and $a_2$ described by Eqs. (VII.1)
where $\tilde S (t) = {\rm exp} (i \phi (t))$ and $\phi (t)$ is a
random phase with bandwidth $\Delta \omega_0$.  In this simple case
we will be able to calculate the evolution of the distribution
function f(a) of amplitudes.

Because $q = \mid a_1 \mid^2 - \mid a_2 \mid^2$ is a constant of the
motion, we can conveniently write:
$$
a_1 = q^{1/2} {\rm cosh} \theta e^{i \phi_1}
$$
$$
a_2 = q^{1/2} {\rm sinh} \theta e^{i \phi_2} \eqno (VII.2)
$$
from which the equations of motion (VII.1) imply
$$
{\partial\theta \over \partial t} = \gamma_0 {\rm cos} \psi \eqno
(VII.3)
$$
$$
{\partial\psi \over \partial t}  = -2 \gamma_0\, {\rm coth} 2
\theta\, {\rm sin}
 \psi + {\partial\phi \over \partial t}  \eqno (VII.4)
$$
where
$$
\psi \equiv \phi_2 - \phi_1 + \phi \eqno (VII.5)
$$

If we define a distribution function $f (\theta, \psi, t)$, it
evolves according to the Fokker-Planck equation:
$$
{\partial f \over \partial t} + {\partial \over \partial \theta}
\gamma_0 {\rm cos} \psi f - {\partial \over \partial \psi} 2
\gamma_0 {\rm coth}2 \theta\, {\rm sin} \psi f = \Delta \omega_0
{\partial^2 f \over \partial \psi^2} \eqno (VII.6)
$$

In the large bandwidth limit $\Delta \omega_0 >> \gamma_0$,
$\partial /
\partial t$., we can use a Chapman-Enskog like procedure, expanding
$f = f_0 + f_1 + . . .$ and to zero order obtain $\partial^2 f_0/
\partial \psi^2 = 0$, so that $f_0 = f_0 (\theta, t)$ independent of
$\psi$.  To first order we obtain
$$
f_1 = - {\gamma_0 \over \Delta \omega_0} {\rm cos} \psi \left(
{\partial f_0 \over \partial \theta} - 2 {\rm coth} 2 \theta f_0
\right) \eqno (VII.7)
$$
and to second order
$$
{\partial f_0 \over \partial t} - {\gamma_0^2 \over \Delta \omega_0}
{\rm cos}^2 \psi {\partial \over \partial \theta} \left( {\partial
\over
\partial
\theta} - {\rm coth} 2 \theta \right) f_0 - {\partial \over \partial
\psi} 2 \gamma_0 {\rm coth} 2 \theta\, {\rm sin} \psi f_1 = \Delta
\omega_0 {\partial^2 f_2 \over \partial \psi^2} \eqno (VII.8)
$$
Averaging over $\psi$, we obtain the consistency equation,
$$
{\partial f_0 \over \partial t} = {1 \over 2} {\gamma_0^2 \over
\Delta \omega_0} {\partial \over \partial \theta} \left( {\partial
\over \partial \theta} - {\rm coth} 2 \theta \right) f_0 \eqno
(VII.9)
$$

We are interested in the evolution of f from an initial condition
$f(t=0) = \delta (\theta - \theta_0)$.  If we assume $\theta >> 1$,
since we expect $\theta$ to increase we can approximate ${\rm coth}
2 \theta = 1$ in the above equation.  It is then readily solved
obtaining
$$
f = \left( 2 \pi \gamma_0^2 t / \Delta \omega_0 \right) {\rm exp}
\left( - \left( \theta - \theta_0 - \gamma_0^2 t/\Delta \omega_0
\right)^2 / \left( 2 \gamma_0^2 t / \Delta \omega_0 \right) \right)
; \eqno (VII.10)
$$
that is, $\theta$ is Gaussian distributed with mean $\theta_0 +
\gamma_0^2 t / \Delta \omega_0$ and variance $\gamma_0^2 t / \Delta
\omega_0$. This distribution is shown in Fig. 7.1 for $\Delta
\omega_0 = 10 \gamma_0$ and $\gamma_0t = 50$ together with some
numerically generated distributions. Clearly then  $\langle a^{-1}
{\partial a \over \partial t} \rangle = {\partial \over \partial t}
\langle \theta \rangle = \gamma_0^2 / \Delta \omega_0$ and $\langle
a \rangle \simeq q^{1/2} {\rm exp} \left( \theta_0 + \gamma_0^2 t /
\Delta \omega_0 \right)$. On the other hand, if we calculate
$\langle \mid a \mid^n \rangle \simeq q^{n/2} \langle {\rm exp}
\left( n \theta \right) \rangle$, we obtain
$$
\langle \mid a \mid ^n \rangle \sim \mid a \mid_0^n {\rm exp} \left(
n \left( n+2 \right) \gamma_0^2 t / 2 \Delta \omega_0 \right) .
\eqno (VII.11)
$$
For the special case n = 2 we observe that the growth rate for
$\langle \mid a \mid ^2 \rangle$ is $4 \gamma_0^2 / \Delta
\omega_0$, containing the factor of two noted earlier.  In
particular we see explicitly that the mean growth rate is definition
dependent because the distribution of growth rates is so broad.  To
compare with these analytic results, Eqs. (VII.1) were numerically
integrated for a large number of different independent realizations
of a Kubo-Anderson process such that $\gamma_{\langle a \rangle} T
>> 1$ and $\Delta \omega_0 = 10 \gamma_0$. In Fig. 6, the
distribution (broken line) of the intensity $\mid a \mid^2$ is shown
for $T \gamma_0 = 50$ and $2 \times 10^3$  realizations.  The
"measured" value of $\gamma_{\langle a \rangle}$ is 0.095 and
$\gamma_{\langle a^2 \rangle}$ is 0.16, and obviously the
distribution at $\mid a \mid^2 = {\rm exp} \left( 4 \gamma_0^2 T /
\Delta \omega_0 \right) = {\rm exp} \left( 20 \right) = 10^{8.7}$ is
not resolved. Moreover, the few events with $\mid a \mid^2 >
10^{7.2}$ that have grown at the rate greater than 0.165 $\gamma_0$
have determined $\gamma_{\langle a^2 \rangle}$.  In general, the
accurate numerical computation of random multiplicative processes in
which rare events determine the quantity of interest is
nontrivial[77].  On the other hand, the distribution of events that
have grown at the rate $\gamma_{\langle a \rangle}$ is well
resolved. When there is sufficient damping, i.e. Max $\left( \nu_1,
\nu_2 \right) \geq \gamma_0^2 / \Delta \omega_0$ the exact
theoretical results show the ensemble average growth rate
$\gamma_{\langle a^2 \rangle} \simeq \gamma_{\langle a \rangle}$.
The simulation results for $\nu_2 = \gamma_0$ also shown in Fig. 6
(dashed-dotted line), demonstrate that the damping has dramatically
narrowed the distribution of $\mid a \mid^2$ so that $\gamma_
{\langle a^2 \rangle} = 0.092$ and $f(a^2)$ is zero for $\mid a^2
\mid
> 10^5$.

Note that this behavior is not specific to the Kubo-Anderson
process. Numerical simulations with a randomly diffusing phase or a
K-A process with equally spaced elapsed times obtained similar
distributions of $\mid a^2 \mid$.

\subsubsection{Space-time Evolution of Absolute
Instabilities} Here, the incoherent absolute growth rate
$\gamma_{\langle \mid a^2 \mid \rangle}$ for temporal incoherence is
twice the "nominal" incoherent growth rate, $\gamma_{inc} \equiv
\gamma_0^2 / \Delta \omega$, if $\mid V_1 \mid = \mid V_2 \mid$.
Given the results just presented for the purely temporal problem,
this factor of two suggests that we might expect a broad
distribution to result from considering different realizations of a
temporal random process with bandwidth $\Delta \omega_0$. In this
case, we remind the reader that no absolute instability is allowed
by the average amplitude equations since the average amplitude
equations decouple in the RPA or Bourret approximation. It is also
interesting to observe that, unlike the purely temporal case,
damping on either mode does not affect the factor that multiplies
$\gamma_{inc}$ to obtain $\gamma_{\langle a^2 \rangle}$. This
factor, $q = 1 + 2 \sqrt{V_1 V_2} / \left( \mid V_1 \mid + \mid V_2
\mid \right)$, depends on the ratio of the group velocities and
approaches two when the group velocity ratio approaches one.  Thus,
in analogy with the purely temporal problem, we expect the
distribution of intensities of growing modes to broaden as the group
velocity ratio approaches one.

Before we proceed to test this hypothesis, we must be certain that
both the incoherent absolute growth rate $\gamma_{\langle a^2
\rangle}$ (VI.8a) is less than the coherent rate (VI.IIa) and that
the plasma length exceed sufficiently the threshold lengths (VI.31)
and (VI.29a). In addition the threshold set by the physical damping
and bandwidth must be exceeded.  These conditions are met by
choosing the parameters:  $\nu_j = 0$, $j = 1,2$, $\Delta \omega_0 =
10 \gamma_0$, and $L = 10 L_g$.  When the group velocity ratios $V_1
/ V_2 = -16$ and -100, the predicted incoherent absolute growth
rates $\gamma_{\langle \mid a^2 \mid \rangle}$ are 0.147 and 0.12 $
\gamma_0 $, respectively.  The predicted coherent absolute rates are
0.47 and 0.2 $\gamma_0$ respectively.

The distribution of $\mid a^2 \mid$ for $\Delta \omega_0 = 10
\gamma_0$ is shown in Fig. 7.2 for each velocity ratio.  Each event
in the distribution represents the ratio $a_1^2 (T) = \mid a_1
\left( X_0, T \right) \mid^2 / \mid a_1 \left( X_0, t_0 \right)
\mid^2$ where $\gamma_0 \left( T-t_0 \right) = 45$ and $X_0$ is the
initial position of a source at t = 0 in $a_2$.  The time $t_0$ is
chosen nonzero to avoid the influence of transients and was $20
\gamma_0^{-1}$ in all these cases.  This "initial" time is longer
than the time for the convective pulse to move across the system,
$t_c = 2L/ \left( \mid V_1 \mid + \mid V_2 \mid \right)$ but not
always longer than the time it takes the slow wave to transit the
length L.  However since the growth is measured at the source point,
the fact that in the entire plasma, the waves are not growing at the
asymptotic rate should not matter.

In these simulations, the numerically calculated ensemble average
growth rate $\gamma_{\langle a^2 \rangle}^{sim}$ was smaller than
the RPA prediction in all cases as expected.  When the velocity
ratio $V_1 / V_2 = -16$ and -100 the computed $\gamma_{\langle a^2
\rangle}^{sim}$ was 60\% and 75\% respectively of the RPA
prediction.  However, we were most interested n the width of the
distribution.  When $\mid V_1 / V_2 \mid$ was 16 and 100, the full
width at half maximum of the distribution was 7\% and 2\% of the
peak respectively, and thus displayed the narrowing expected. When
$V_1 / V_2 = -1$, the computed distribution was not significantly
broader than the distribution for $V_1 / V_2 = -16$.  Thus the
difference in growth rates between $V_1 / V_2 = -1$ and -16 may
arise from differences in the unresolved tail of the distribution.

We conclude from this study that the RPA growth rates for the
ensemble average intensity are correct but in a given simulation, or
perhaps a particular experiment, the measured growth rate may well
be a factor of two smaller.  However when the group velocity is
large, as is typically the case of physical interest, the RPA
predictions are more representative of the typical behavior.

\subsection{The RPA Conjecture}
The question here is whether the RPA equations are applicable in the
whole domain defined by the inequalities $\Delta \omega_1 > {\rm
Max} \left( \gamma_0, \nu_1 \right)$ and $\Delta \omega_2 > {\rm
Max} \left( \gamma_0, \nu_2 \right)$ or whether there is an
intermediate domain when one of the cross inequalities $\Delta
\omega_1 > \nu_2$ or $\Delta \omega_2 > \nu_1$ is not satisfied.  To
examined this case we look at the spatial Kubo-Anderson (K-A)
process analyzed previously$^{27}$ with zero temporal incoherence
and spatial correlation function with the property
$$
\langle S \left( x \right) S^* \left( x + \Delta x \right) \rangle_x
= {\rm exp} \left( - \Delta K \Delta x \right) . \eqno (VII.12)
$$
If we assume $\mid V_1 \mid >> \mid V_2 \mid$ and $V_1V_2 < 0$, then
the intermediate domain is described by the chain of inequalities,
$$
\Delta K V_1 > \nu_1 > \mid \Delta K V_2 \mid > \nu_2, \gamma_0 .
\eqno (VII.13)
$$
A physical example that might produce this set of inequalities is
two plasmon decay in the presence of stationary ion acoustic
turbulence where the shorter wavelength Langmuir wave is more
strongly damped and has a faster group velocity.  The choices in our
numerical solution were:

$$
 { \Delta K \mid V_1 V_2 \mid^{1/2} / \gamma_0 = 4 ,\hspace{0.2cm}
\nu_1 = 2 \gamma_0 , \hspace{0.2cm} \nu_2 = 0 , \hspace{0.2cm} V_1 /
V_2 = -16 , } \eqno(VII.14)
$$
for which the RPA equation absolute instability growth rate
$\gamma_{\langle a^2 \rangle} = 0.12 \gamma_0$ and the absolute
instability coherent growth rate (VI.11), $\gamma_{\rm coh}^{abs} =
0.35 \gamma_0$. The simulation region $L = 8 L_g, \left( L_g = \mid
V_1 V_2 \mid ^{1/2} / \gamma_0 \right)$ was larger than the
threshold length given by the maximum of Eq. VI.32 and VI.29.The
simulation was run until $\gamma_0 t = 50$ for 1000 independent
realizations of this spatial K-A process.  An initial value was
given to the undamped wave at t = 0 but the intensity at $\gamma_0T
= 50$ was measured relative to the intensity at $\gamma_0t_0 = 20$
because by that time the modes were observed to grow at this time
asymptotic rate.  The ensemble averaged rate $\gamma_{\langle a^2
\rangle}$ was found to be independent of whether its calculation was
based on the intensity at the source point $\mid a_1 \left( X_0, T
\right) \mid^2$ or the total mode "energy" $\int dx \mid a_1 \left(
x, T \right) \mid^2$.  The results support our conjecture that the
RPA equations apply in the whole incoherent domain, because the
measured intensity growth rate, $\gamma_{\langle a^2 \rangle}^{sim}
= 0.11 \gamma_0$ agreed with the RPA rate $\gamma_{\langle a^2
\rangle} = 0.12 \gamma_0$.  However, the distribution of intensity
at $\gamma_0T = 50$ (relative to $\gamma _0 t_0 = 20$) displayed in
Fig 8 showed that, in fully a third of the cases, no growth occurred
and that the remaining distribution consisted of a slowly decreasing
tail.  The maximum intensity in the distribution was achieved by a
mode that grew at close to the coherent rate.  In addition, the
faster growth rates were achieved in cases in which there was an
abnormally long distance between phase changes; in the maximum
growth case, this distance was 2.18 $L_g$.  As we remarked at the
outset of this section and in the introduction (Sec. I), the growth
of parametric instabilities with this model of spatial incoherence
was studied by Williams et al. using more sophisticated statistical
methods (quite different in nature to those used in this article)
with the result that the fastest growing mode grows at the rate
$\gamma_f$, given by Eq. I-9. This rate $\gamma_f$ depends on the
size of the system, which is understood on the basis that as the
system gets larger there is an increasing possibility that there
will be a large region with no phase change.  In our simulation this
size dependent factor ${\rm ln} \Delta KL = 3.47$ turns out to be
approximately equal to the numerical factor of 4 in the RPA growth
rate (Eq. VI.8a); all other factors are identical and $\Delta K
\rightarrow \Delta k_0$. Unfortunately it is not feasible to
simulate a system that doubles the size of this logarithmic factor
with a constant value of $\Delta K$.  Thus these simulations do not
distinguish between the RPA theory and the results of Williams et
al.  However, the simulations do support the conjecture that the
incoherent results apply in whole incoherent domain.

\subsection{Modeling of Induced Spatial Incoherence} One motivation for this
work is the induced spatial incoherence technique (ISI) for creating
smoothed laser beam intensity distributions in the focal plane.  In
this technique, statistically independent beamlets overlap and
thereby produce both intensity and phase variations. Only pumps with
phase variation alone have been considered up to this point in our
numerical examples.  Here we wish to show that the same statistical
methods are valid for intensity and phase varying pumps if the
conditions for the incoherent results to apply are met.

A model that exhibits both intensity and phase variation in one
dimension is given by,
$$
S \left( x,t \right) = {1 \over \sqrt N} \sum^N_{j=1} {\rm exp}
 \left\{ i\, \delta K_j x + i \phi_j \left( t \right) \right\} , \eqno (VII.5)
$$
when $\delta K_j$ is uniformly distributed on $(0, K_m)$ and $\phi_j
(t)$ is a random variable of time.  Each $\phi_i$ varies
independently with the same mean coherence time, $\left( \Delta
\omega_0 \right)^{-1}$. The spatial coherence length in this model,
$x_c = \pi / K_m$, is related to the focal length of the focussing
optic.  A shorter focal length results in a shorter coherence
length.

At any given point and instant of time, the intensity $\mid S \left(
x,t \right) \mid^2$, being the absolute square of the sum of N
complex numbers of length N$^{-1/2}$, is Gaussian distributed with
unit mean and variance.  Because the sum varies with position over a
length scale $x_c$, the instantaneous spatial intensity pattern is
quite spikey with peak to average variations of more than two quite
common.  However because the sum at a given spatial position also
varies in time with coherence time $\left( \Delta \omega_0
\right)^{-1}$, the time averaged intensity is a much smoother
function of space. Clearly if $\phi_j \left( t \right)$ is constant
in time, then the spikey intensity pattern of Eq. VII.15 is
stationary which can lead in the coherent domain to growth enhanced
relative to a uniform intensity. Even in the incoherent domain
without temporal bandwidth, we saw in the previous section that
there is a finite probability, especially in large systems, for the
phase to remain constant for a large distance.  With phase and
intensity variations, large fluctuations in growth rates are
expected in the absence of temporal bandwidth.  In the complementary
limit where $\Delta \omega_1 \geq \Delta \omega_2 > {\rm Max}
\left(\gamma_0, \nu_1, \nu_2 \right) $, the results of our
statistical analysis should apply.  For simplicity, we choose to
compute the convective growth for both temporal and spatial
incoherence in the incoherent limit and for large group velocity
ratios, $\mid V_1 / V_2 \mid >> 1$.  With a single term in the sum
in Eq. VII.15 and $\Delta \omega_1 = \Delta \omega_2 = \Delta
\omega_0$, the convective growth rate was in agreement with Eq.
(VI.7).  With no temporal bandwidth, $\Delta \omega_0 = 0$, the pump
intensity has striking amplitude variation with peak to average
variations of more than three to one.  Nonetheless if $\mid \Delta K
V_2 \mid > \overline \gamma_0$, the convective growth rate, is
reduced according to Eq. (VI.7) even if $\Delta \omega_0 = 0$.  As
the spatial incoherence weakens and the boundary $\Delta K V_2 =
\overline \gamma_0$ is approached for $\Delta \omega_0 = 0$ the
measured growth rate in the simulations diverges from Eq. (VI.7)
until, for $\overline \gamma_0 > \mid \Delta K V_2 \mid$, growth
rates in excess of $\overline \gamma_0$ can occur with large
fluctuations from case to case. Laser bandwidth, even if $\Delta
\omega_0 < \overline \gamma_0$, reduces both the fluctuations and
the growth rate to less than $\overline \gamma_0$.  Thus, some
bandwidth is needed in this model to provide a smooth transition
from the coherent to incoherent domain.  Other models such as the
one used in Section VII.B have the property that $\mid S \left( x, t
\right) \mid = 1$ so that the statement that pump incoherence can
only reduce the growth rate is indeed true.  Simulations in that
case do show a smooth transition from the coherent to incoherent
growth rate

\section{Conclusions and Applications}
A comprehensive treatment of the effects of pump wave temporal and
spatial incoherence has been presented for a homogeneous plasma.
Eschewing specific models of incoherence that allow exact solutions,
we have derived equations for ensemble averaged mode intensity and
amplitudes in the incoherent limit.  At each step in the process of
deriving these approximate statistical equations, the conditions
that must be satisfied are clearly stated.  Thus, the meaning of
incoherence in the context of parametric instability theory is
carefully defined. The set of inequalities that comprise this
definition form a major contribution of this work.  Of course, the
primary contribution of this work is the set of thresholds, growth
rates, and amplification rates for both coherent and incoherent,
convectively and absolutely unstable parametric interactions.

However, the majority of our readers are interested in the more
practical results concerning thresholds, growth rates, and
amplification coefficients as they pertain to laser plasma
interactions.  In a subsequent work,$^6$ we will apply the results
presented here to particular instabilities, and include both
backscatter and sidescatter geometries.  Here, we briefly consider
the examples of stimulated Raman and Brillouin backscatter for
parameters of interest to laser fusion.  We choose the electron
temperature, $T_e = 3 {\rm keV}$, the charge state Z = 40, the
atomic number A = 80, the laser wavelength, $\lambda_0 = 0.35 \mu
m$, the laser intensity I = 10$^{15}$ W/cm$^2$ and the electron
density, $N_e = 0.2 N_c$ for SRS and 0.25 N$_c$ for SBS. Here N$_c$
is the critical electron density defined by $4 \pi N_ce^2 / m_e =
\omega_0^2$.  For almost all calculations it is the ratio Z/A and
not Z or A by themselves that is important.

The nominal SBS growth rate is $\gamma_0^{srs} = 2.5 \times 10^{-3}
\omega_0$ whereas the SRS coherent convective threshold, Eq. (I-1)
is $\gamma_0^{conv} = 3.2 \times 10^{-4} \omega_0$.  Landau damping
was neglected because for $T_e = 3 {\rm keV}$, an electron density
can always be chosen to make it negligible.  Thus SRS is an order of
magnitude above threshold.  The laser bandwidth sufficient to
achieve convective stability can be found by using Eq. (VI.5).  We
find $\Delta \omega_c = 4.5 \times 10^{-2} \omega_0$.  On the other
hand, the coherent threshold for absolute instability given by Eq.
(VI.9a), $\gamma_{0 abs} = 1.5 10^{-3} \omega_0$, is exceeded by
less than a factor of two. However, as we have emphasized, in Sec.
VI, relatively large laser bandwidth is required to reduce the
absolute growth rate.  Satisfying the validity conditions for using
the incoherent formula, Eq. (VI.10), imposes the condition, $\Delta
\omega_0 / \omega_0 = 10^{-2}$, which is, coincidentally, the same
condition numerically that is required for absolute stability as
determined by Eq. (VI.9a).

The SBS thresholds are dependent on the ion acoustic damping rate
and thus the ion temperature, $T_i$.  However, even if $T_i = T_e$
the ion Landau damping is negligible for $Z > 10$.  Thus, the ion
acoustic damping rate is determined by the electrons and is given by
$\gamma_{ac} / \omega_0 = 2.3 \times 10^{-5}$.  Hence, the
convective coherent threshold $\gamma_{0 conv} / \omega_0 = 1.5
\times 10^{-4}$ is easily exceeded by the nominal SBS growth rate
$\gamma_0 = 7 \times 10^{-4} \omega_0$. In addition, the SBS
absolute instability coherent threshold $\gamma_{0 abs} = 3.4 \times
10^{-4} \omega_0$ is exceeded.  Here, the bandwidth required to
achieve convective stability is $\Delta \omega_c / \omega_0 = 2.2
\times 10^{-2}$. Moreover, the bandwidth required for absolute
stability is by accidental choice of parameters, the same as for
convective stability because although $\nu_1 << \nu_2$, nonetheless
$\left( V_1 / V_2 \right) \nu_2 > \nu_1$ and the incoherent absolute
threshold Eq. (VI.9a) reduces to the incoherent convective threshold
Eq. (VI.6).  At this point, we should warn the reader that the
results reported here assumed the validity of the coupled mode
equations which requires the mode frequencies $w_j > \Delta
\omega_0$ for j = 0,1,2.  However, we have in the case of SBS a
critical bandwidth for stability that exceeds the ion acoustic
frequency $\omega_a = 2 \times 10^{-3} \omega_0$.  At the very
least, one might expect additional stability when $\Delta \omega_0 >
\omega_a$.  For a partial answer to the effects that $\Delta
\omega_0 > \omega_a$ has, we considered SBS again for the purely
temporal problem without using the assumption that $\omega_a >>
\Delta \omega_0$.  We start with the equations,
$$
\left( {\partial \over \partial t} + \nu_1 - i \delta \omega \right)
a_1 = - i \gamma_0 a_2^* , \eqno (VIII.1a)
$$
$$
\left( {\partial^2 \over \partial t^2} + \omega_a^2 \right) a_2 = -
2 \gamma_0 \omega_a a_1^* , \eqno (VIII.1b)
$$
where $\delta \omega \equiv \omega_0 - \omega_1$, is the difference
between the pump frequency and the scattered light frequency.  Here
a$_{1,2}$ are the mode amplitudes for the scattered light and ion
acoustic waves respectively.  Two different dispersion relations are
obtained, as expected, by obtaining equations for the evolution of
average amplitude $\langle a_j \rangle$.  The ion acoustic mode
$\langle a_2 \rangle$ dispersion relation is not affected by
magnitude of $\Delta \omega_0 / \omega_a$ so that the growth rate,
$$
\gamma_{\langle a_2 \rangle} = - \nu_2 + {\gamma_0^2 \over \left(
\nu_1 + \Delta \omega_0 \right) } , \eqno (VIII.2)
$$
is a maximum when $\delta \omega = \omega_a$.  On the other hand,
the dispersion relation for $\langle a_1 \rangle$ is modified so
that the growth rate when $\Delta \omega_0 >> \omega_a, \nu_2$,
$$
\gamma_{\langle a_1 \rangle} = - \nu_1 + {3 \sqrt 3 \over 4}
{\omega_a \gamma_0^2 \over \Delta \omega_0^2} , \eqno (VIII.3)
$$
shows a stronger reduction with bandwidth than $\gamma_{\langle a_2
\rangle}$. Here the maximum growth rate for $\langle a_1 \rangle$
occurs for a bandwidth dependent real frequency $\omega_0 - \omega_1
= \Delta \omega_0 / \sqrt 3$. Thus, the threshold for mode $\langle
a_1 \rangle$ is increased when $\Delta \omega_0 > \omega_a$ by the
ratio $\Delta \omega_0 / \omega_a$. The actual threshold for the
coupled set VIII.1 is,
$$
\gamma_{0 conv}^2 = {\rm Min} \left( \nu_2 \left( \nu_1 + \Delta
\omega_0 \right) , \nu_1 \left( \nu_2 + \Delta \omega_0 \right) {\rm
Max} \left( 1 , {\Delta \omega_0 + \nu_2 \over \omega_a} \right)
\right) . \eqno (VIII.4)
$$
This equation (VIII.4) is a generalization of Eq. (VI.6).  In our
example, the threshold is not increased when $\Delta \omega_0 >
\omega_a$.

In our example, chosen for its relevance to laser fusion, the
critical bandwidth to reach convective or absolute stability has
been found to be 2-5\% of the laser frequency.  Such a bandwidth is
beyond the currently achievable with any ICF laser system.  However,
novel techniques are being pursued that are capable of achieving
these bandwidths[6].  In addition, there is the question of whether
the average laser intensity or the larger hotspot intensity should
be used in evaluating the thresholds.  The use of the larger
intensity would make the bandwidth requirements infeasible. However,
if bandwidth is combined with a beam smoothing scheme, use of the
mean intensity may be correct.  Recent calculations[66,67] have
shown that, for at least one such scheme, ISI, the use of the
average intensity is justified.

In addition to temporal bandwidth, there is also the effect of
spatial incoherence produced by beam smoothing techniques that lead
to a spread of wavevectors, $\Delta k_0$.  In the examples of
backscatter considered previously in this section for temporal
bandwidth, spatial incoherence will have little effect because
$\Delta k_{0**} / \Delta k_{0*} \sim f^{\#-1}$ where the f number of
the focussing optic, f\#, is typically large in ICF applications.
Furthermore, the wavenumber spread will not increase the convective
threshold because the effective bandwidth, $\mid \Delta k_0 \cdot
V_{g{\rm min}} \mid$, is in general small compared to $\nu_1,
\nu_2$, and $\gamma_0$. However absolute instabilities with
significant wavevectors perpendicular to the laser propagation axis
such as two plasmon decay may be influenced significantly by such a
spread since there $\Delta k_*V_{g*{\rm max}}$ will be the effective
bandwidth.

The work reported here is only a step in the application of a
statistical theory to particular instabilities.  Further progress
may require modifications or extensions to our work in order to
remove some of the limitations of our approximations.  We look
forward to new developments in the theory and even more to the
stimulus of experimental data.

\begin{acknowledgments}
The authors express their appreciation to their many colleagues who
have given freely of their ideas and criticisms, specifically B. B.
Afeyan, J. R.  Albritton, B. I. Cohen, W. L. Kruer, G. Laval and R.
Pellat.  D. Pesme acknowledges the support of the Centre National de
la Recherche Scientifique and the Plasma Physics Research Institute
of Lawrence Livermore National Laboratory and the University of
California at Davis through DOE Contract W-7405-ENG-48.  R. L.
Berger was supported by DOE contract DE-AC03-87DP10560 and is
grateful to Lawrence Livermore National Laboratory for its
hospitality during the course of this collaboration.  E. A. Williams
was supported by DOE contract W-7405-ENG-48.  Annick Bortuzzo-Lesne
was supported by a grant from Fondation Singer Polignac.  A.
Bourdier also acknowledges the interest of W. L. Kruer and the
hospitality of Lawrence Livermore National Laboratory where this
work began.  Two of the authors (R. L. Berger and E. A. Williams)
wish to thank M. Haines and C. Moser who organized the 1988 and 1989
CECAM Workshops on laser-plasma interactions which helped further
the advance of this work. This work was the subject of an internal
Lawrence Livermore National Laboratory report with report number
UCRL-JC-105479 November 19, 1990.
\end{acknowledgments}

\vfil \eject

\vskip 0.8 truein {\bf References} \vskip 0.2 truein

\begin{itemize}
\item[1.] K. Moncur, {\it Applied Optics}, {\bf 16}, 1449 (1977).

\item[2.] R. H. Lehmberg and S. P. Obenschain, {\it Opt.~Commun.} {\bf 46},
27 (1983).

\item[3.] Y. Kato, K. Mima, N. Miyanaaga, S. Arinaga, Y. Kitagawo, M. Nakatsuka,
and C. Yamanaka, {\it Phys.~Rev. Lett.}, {\bf 53}, 1057 (1984); Y.
Kato and K. Mima, {\it Applied Physics} {\bf 329}, 186 (1982).

\item[4.] S. Skupsky, R. W. Short, T. Kessler, R. S. Craxton, S. Letzring, and T.
M. Soures, {\it J.~Appl.~Phys.} {\bf 66}, 3456 (1989).

\item[5.] D. Veron, H. Ayral, C. Gouedard, D. Husson, J. Lauriou, O. Martin, B.
Meyer, M. Rostaing, and C. Sauteret, {\it Optics Communications},
{\bf 65} (1988).

\item[6.] D. M. Pennington, M. A. Henesian, R. B. Wilcox,
T. L. Wieland, D. Eimerl And H.T. Powell, ``A Novel Bandwidth Source
for Laser Experiments'', Technical Digest Series, Vol 18, CLEO '94
Anaheim, Ca.

\item[7.] D. Pesme, ``Effects of Temporal and Induced Spatial
Incoherence of Parametric Instabilities in Laser Plasma
Interactions'', 1987 Annual Technical Report CNRS-LULI, Ecole
Polytechnique. (available from NTIS, Springfield VA 22161:
 document PB 92-100312.)

\item[8.] K. Nishikawa, in ``Advances in Plasma Physics, Vol 6'', edited
by A. Simon and W. B. Thompson (Wiley N. Y. 1976)

\item[9.] G. M. Zaslavskii, V. S. Zakharov, {\it Sov.~Phys.~Tech.~
 Phys.},

{\bf 12}, 7 (1967).

\item[10.] G. E. Vekshtein, G. M. Zablavskii, {\it
Sov. Phys. Doklady}, {\bf 12}, 34 (1967).

\item[11.] E. J. Valeo and C. R. Oberman, {\it Phys.~Rev.~Lett.} {\bf 30},
1035 (1973).

\item[12.] S. Tamor, {\it Phys. Fluids} {\bf 16}, 1169 (1973).

\item[13.] J. J. Thomson, W. L. Kruer, S. E. Bodner, and J. S. DeGroot,
{\it Phys.~Fluids}, {\bf 17}, 849 (1974).

\item[14.] J. J. Thomson and J. I. Karush, {\it Phys. Fluids} {\bf 17}, 1608 (1974).

\item[15.] J. J. Thomson, {\it Nuclear Fusion}, {\bf 15}, 237 (1975).

\item[16.] G. Laval, R. Pellat, D. Pesme, A. Ramani, M. N. Rosenbluth, and
E. A. Williams, {\it Phys. Fluids}, {\bf 20}, 2049 (1977).

\item[17.] J. J. Thomson, {\it Phys. Fluids}, {\bf 21}, 2082 (1978).

\item[18.] W. L. Kruer, K. G. Estabrook, and K. H. Sinz.,
{\it Nucl. Fusion}, {\bf 13}, 952 (1973).

\item[19.] W. Kruer, E. Valeo, K. Estabrook, J. Thomson, B. Langdon, and B.
Lasinski, {\it Plasma Physics and Controlled Nuclear Fusion
Research}, {\bf Vol. II}, 525 (1974).

\item[20.] K. Estabrook, J. Harte, E. M. Campbell, F. Ze, D. W. Phillion,
M. D. Rosen, and J. T. Larsen, {\it Phys. Rev. Lett.} {\bf 46}, 724
(1981).

\item[21.] D. W. Forslund, J. M. Kindel and E. M. Lindman, {\it Phys
Fluids}, {\bf 18}, 1017 (1975).

\item[22.] K. Estabrook, W. L. Kruer, and B. F. Lasinski, {\it
Phys. Rev. Lett.}, {\bf 45}, 1399 (1980).

\item[23.] K. Estabrook and W. L. Kruer, {\it Phys. Fluids}, {\bf 26}, 1892

\item[24.] G. Bonnaud and C. Reisse, {\it Nuclear Fusion}, {\bf 26},
633 (1986).

\item[25.] C. Yamanaka, T. Yamanaka, T. Sasaki, and J. Mizui, {\it Phys. Rev.
Lett.}, {\bf 32}, 1038 (1974).

\item[26.] R. R. Johnson, P. M. Campbell, L. V. Powers, and D. C. Slater, in
proceedings of the Topical Meeting of Inertial Confinement Fusion,
7-9 February, 1978, San Diego (unpublished); D. C. Slater and D. J.
Tanner, Proceedings of the Ninth Annual Conference on the Anomalous
Absorption of Electromagnetic Waves.

\item[27.] C. E. Clayton, C. Joshi, A. Yasuda and F.
 F. Chen, {\it
Phys. Fluids}, {\bf 24}, 2312 (1981).

\item[28.] A. Mase, N. C. Luhmann, J. Holt, H. Huey, M. Rhodes,
W. F. DiVergilio, J. J. Thomson, and C. F. Randall, {\it Plasma
Physics and Controlled Nuclear Fusion Research} (IAEA, Vienna,
1981).

\item[29.] R. Giles, R. Fedosejevs, and A. A. Offenberger, {\it Physical Review
A}, 2{\bf 6}, 1113 (1982).

\item[30.] D. G. Colombant, W. M. Manheimer, and J. H. Gardner, Phys. Fluids 26,
3148 (1983).

\item[31.] V. V. Tamoikin and S. M. Fainstein, {\it Sov. Phys. JETP},
{\bf 35}, 115 (1972).

\item[32.] S. A. Akhmanov, Yu. E. Dyakov and L. I. Pavlov, {\it
Sov. Phys. JETP}, {\bf 39}, 249 (1974).

\item[33.] P. Kaw, R. White, D. Pesme, M. Rosenbluth, G. Laval, R. Varma, and
R. Huff, {\it Comments Plasma Phys. Controlled Fusion 2}, {\bf 11}
(1974).

\item[34.] D. R. Nicholson and A. N. Kaufman, {\it Phys. Rev. Lett.}, {\bf 33},
1207 (1974).

\item[35.] M. Y. Yu, P. K. Shukla and K. H. Spatschek, {\it Phys. Rev
A}, {\bf 12}, 656 (1975)

\item[36.] K. H. Spatschek, P. K. Shukla and M. Y. Yu, {\it
Phys. Lett. A}, {\bf 51}, 183 (1975).

\item[37.] D. R. Nicholson, {\it Phys. Fluids}, {\bf 19}, 889 (1976).

\item[38.] G. Laval, R. Pellat and D. Pesme, {\it Phys. Rev. Lett.},
{\bf 36}, 192 (1976).

\item[39.] E. A. Williams, J. R. Albritton, and M. N. Rosenbuth,
{\it Phys. Fluids}  {\bf 22}, 139 (1979).

\item[40.] E. Z. Gusakov and A. D. Piliya, {\it Sov.~J.~Plasma
Phys.}, {\bf 6}, 277 (1980).

\item[41.] E. Z. Gusakov and A. D. Piliya, {\it Sov.~J.~Plasma
Phys.}, {\bf 7}, 733 (1981).

\item[42.] E. Z. Gusakov and A. D. Piliya, {\it Sov.~J.~Plasma
Phys.}, {\bf 8}, 324 (1982).

\item[43.] W. Rozmus, A. A. Offenberger and R. Fedosejevs, {\it
Phys. Fluids}, {\bf 26}, 1071 (1983).

\item[44.] A. Bourdier and E. A. Williams, "Effects of Induced
Spatial Incoherence on Raman Scattering," {\it Laser Program Annual
Report 1986}, Lawrence Livermore National Laboratory, UCRL-50021-86
(1986).

\item[45.] E. A. Williams, R. L. Berger, and A. Bourdier, {
 \it Laser Program Annual Report 1987}, Lawrence Livermore National
Laboratory, UCRL-50021-87 (1988), p. 2-48.

\item[46.] A. Bortuzzo-Lesne, G. Laval, D. Pesme, and M. Casanova,
"Coefficient de retrodiffusion Brillouin stimulee lorsque l'onde
laser est incoherente," Annual Internal Report GILM, Ecole
Polytechnique (1984).

\item[47.] L. Lu, {\it Phys. Fluids}, {\bf 31}, 3362 (1988).

\item[48.] L. Lu, {\it Phys. Fluids}, {\bf B1}, 1605 (1989).

\item[49.] P. N. Guzdar, {\it Phys. Fluids}, {\bf B3}, 2882 (1991).

\item[50.] M. N. Rosenbluth, {Phys. Rev. Lett.} {\bf 29}, 565 (1972).

\item[51.] S. P. Obenschain, J. Grun, M. J. Herbst, K. J. Kearney, C. K. Manka,
E. A. McLean, A. N. Mostovych, J. A. Stamper, R. R. Whitlock, S. E.
Bodner, J. H. Gardner, and R.Lehmberg, {\it Phys. Rev. Lett.} {\bf
56}, 2807 (1986).

\item[52.] A. N. Mostovych, S. P. Obenschain, J. H. Gardner, J. Grun,
K. J. Kearney, C. K. Manka,, E. A. McLean, and C. J. Pawley, {\it
Phys. Rev. Lett.}, {\bf 59}, 1193 (1987).

\item[53.] S. P. Obenschain, C. J. Pawley, A. N. Mostovych, J. A. Stamper,
J. H. Gardner, A. J. Schmitt, and S. E. Bodner, {\it Phys. Rev.
Lett.} {\bf 62}, 768 (1989).

\item[54.] O. Willi, D. Bassett, A. Giulettii and S. J. Kartunnen,
{\it Opt. Comm.}, {\bf 70}, 487 (1989).

\item[55.] S. Coe, T. Afshar-rad, M. Desselberger, F. Khattak, O. Willi,
A. Giulietti, Z. Q. Lin, W. Yu, and C. Danson, {\it Europhys.
Lett.}, {\bf 10}, 31 (1989).

\item[56.] S. E. Coe, T. Afshar-rad and O. Willi, {\it
Europhys. Lett.}, {\bf 13}, 251 (1990).

\item[57.] O. Willi, T. Afshar-rad, S. E. Coe and A. Guilettii, {\it
Phys. Fluids}, {\bf B2}, 1318 (1990).

\item[58.] T. Afshar-rad, L. A. Gizzi, M. Desselberger, F. Khattak
and O. Willi, {\it Phys. Rev. Lett.}, {\bf 68}, 942 (1992).

\item[59.] T. Afshar-rad, S. E. Coe, O. Willi and M. Desselberger,
{\it Phys. Fluids}, {\bf B4}, 1301 (1992).

\item[60.] W. Seka, R. E. Bahr, R. W. Short, A. Simon, R. S. Craxton,
D. S. Montgomery and A. E. Rubenchik {\it Phys. Fluids}, {\bf B4},
2232 (1992).

\item[61.] C. Labaune, S. Baton, T. Jalinaud, H. A. Baldis and
D. Pesme {\it Phys. Fluids}, {\bf B4}, 2224 (1992).

\item[62.] J. D. Moody, H. A. Baldis, D. Montgomery, K. G. Estabrook,
S. Dixit and C. Labaune, {\it J. of Fusion Energy}, {\bf 12}, 323
(1993).

\item[63.] T. Jalinaud, S. D. Baton, C. Labaune and H. A. Baldis,
``Effets de lames de phases aleatoire sur les diffusions Brillouin
et Raman stimul\`ees en plasma preforme'', LULI 1992 Annual Report,
p56.

\item[64.] C. Labaune, S. D. Baton, T. Jalinaud, E. Schifano,
N. Renard, D. Pesme, H. A. Baldis, J. D. Moody and K. Estabrook,
``Effet du lissage optique par lame de phase aleatiore sur les
instabilities parametriques'', LULI 1992 Annual Report, p23.

\item[65.] J. D. Moody, H. A. Baldis, D. S. Montgomery, K. Estabrook,
R. L. Berger, E. A. Williams, W. L. Kruer and S. Dixit (submitted to
Phys. Plasmas (1995)).

\item[66.] R. L. Berger, {\it Phys. Rev. Lett.}, {\bf 65}, 1207
(1990).

\item[67.] P. N. Guzdar, {\it Phys. Fluids}, {\bf B3}, 776 (1991).

\item[68.] H. A. Rose, D. F. Dubois and D. Russell, {\it Sov. J. of
Plasma Physics}, {\bf 16}, 537 (1990).

\item[69.] A. Brissaud and U. Frisch, {\it J. Math Phys.}, {\bf 15}, 524
(1974); P. Kubo, ibid., {\bf 4}, 174 (1963).

\item[70.] R. Bourret, {\it Nuovo Cimento}, {\bf 26}, 1 (1962).

\item[71.] R. Z. Sagdeev and A. A. Galeev, {\it Nonlinear Plasma Theory}
(Benjamin, New York, 1969).

\item[72.] B. B. Kadomtsev, {\it Plasma Turbulence}
(Academic Press, New York, 1965).

\item[73.] V. N. Tsytovitch, {\it Nonlinear Effects in Plasma}
(Plenum, New York, 1970).

\item[74.]  R. C. Davidson, {\it Methods in Nonlinear Plasma Theory}
(Academic Press, New York, 1972).

\item[75.] A. Bers and R. J. Briggs, {\it Quarterly Progress Report No. 71},
Research Laboratory of Electronics, MIT, p. 122,(1963) unpublished).

\item[76.] V. P. Silin, {\it Zh. Eksp. Teor. Fiz.}, {\bf 48}, 1679 (1965)
[{\it Sov. Phys. JETP}, {\bf 21}, 1127 (1965)]; V. E. Zakharov, {\it
op. cit.} {\bf 62}, 1745 (1972)
 [{\it op. cit.}
{\bf 35}, 908 (1972)].

\item[77.] S. Redner, {\it Am. J. Phys.} {\bf 58}, 267 (1990).

\end{itemize}

\vfil \eject

\newpage
\begin{figure}
\includegraphics[width=15cm]{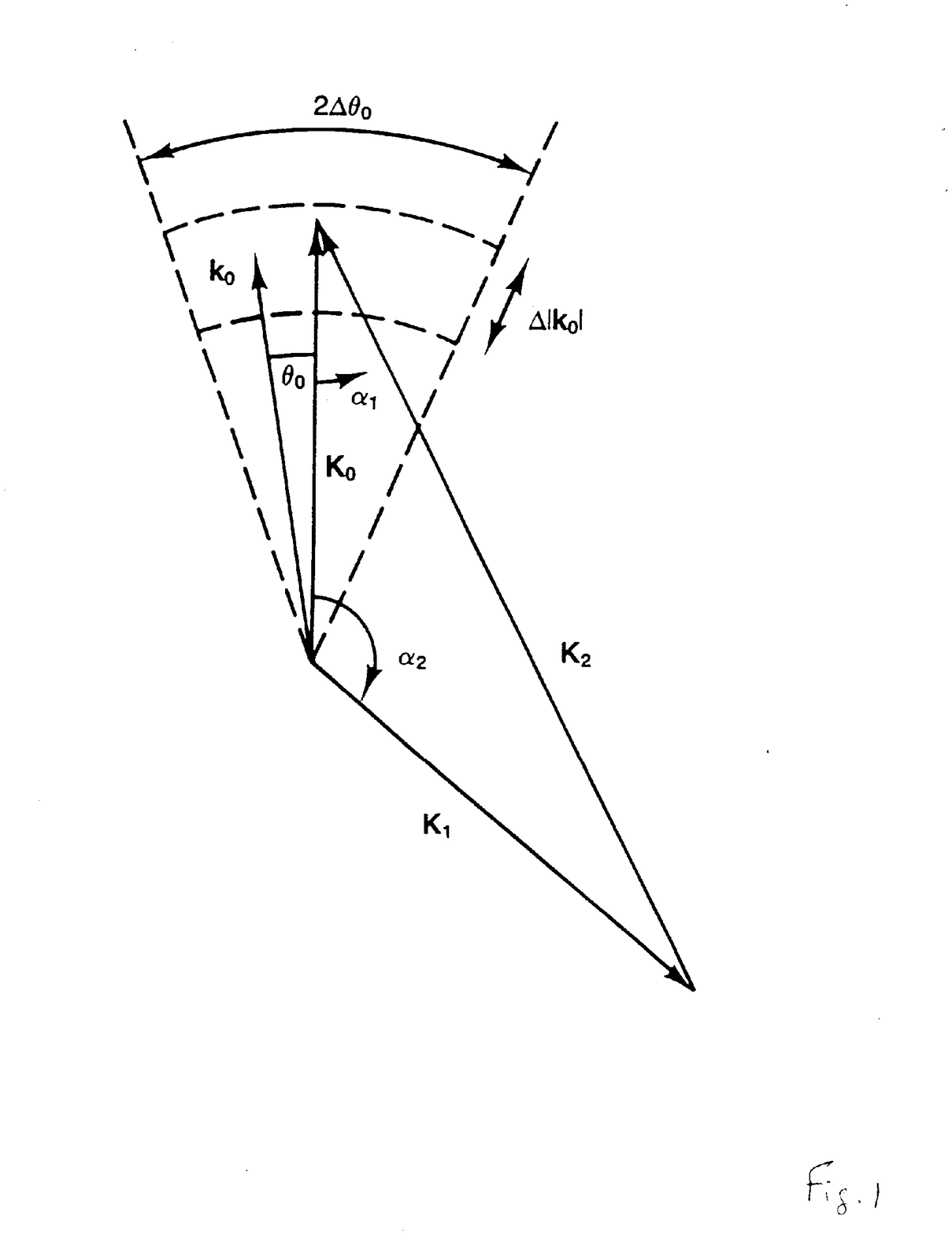}
\caption{Geometry of the scattering:  $K_0, K_1, K_2$ are the mean
wavenumbers of the pump wave and of the decay waves; the pump wave
vector $K_0$ varies in magnitude an amount $\Delta \mid k_0 \mid$
and lies in a cone of angle $\Delta \theta_0$ about the direction of
the mean, $K_0$.  Also shown are the angles $\alpha_2$ and
$\alpha_1$ which the mean wavevectors $K_1$ and $K_2$ make with the
mean pump wavevector $K_0$} \label{fig1}
\end{figure}
\vfill \eject

\newpage
\begin{figure}
\includegraphics[width=15cm]{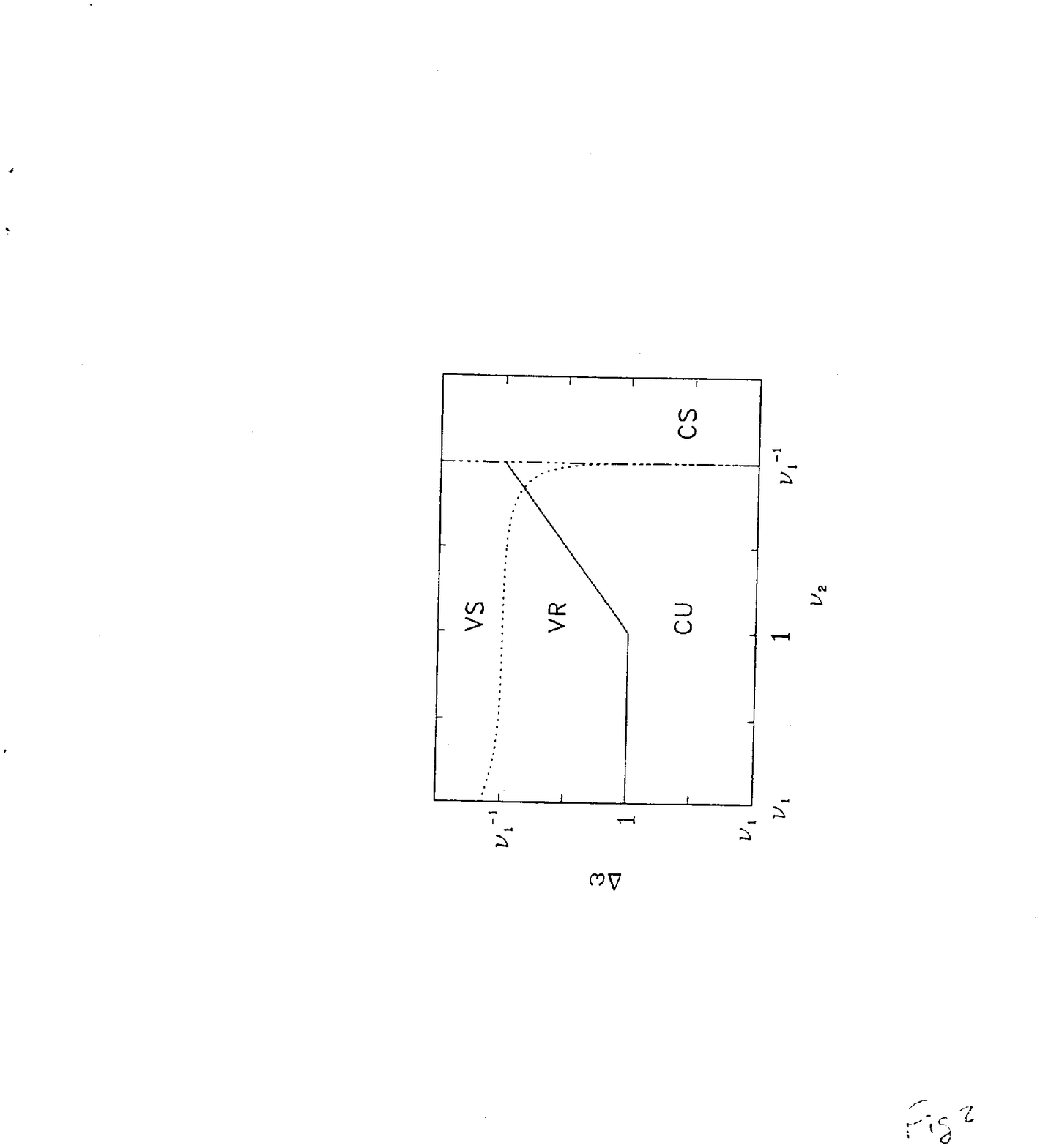}
\caption{Schematic stability diagram for the convective growth rate.
The solid line corresponds to the boundary of the domain VR where
the pump wave incoherence reduces the convective growth rate, and
the dotted line is the boundary of the domain VS where it completely
stabilizes the parametric coupling. The rightmost verticle line
divides the coherently stable (CS) from the coherently unstable (CU)
domains.  This diagram assumes $\nu_2 > \nu_1$ corresponding to the
ordering assumed in the main text.} \label{fig2}
\end{figure}
\vfill \eject

\newpage
\begin{figure}
\includegraphics[width=15cm]{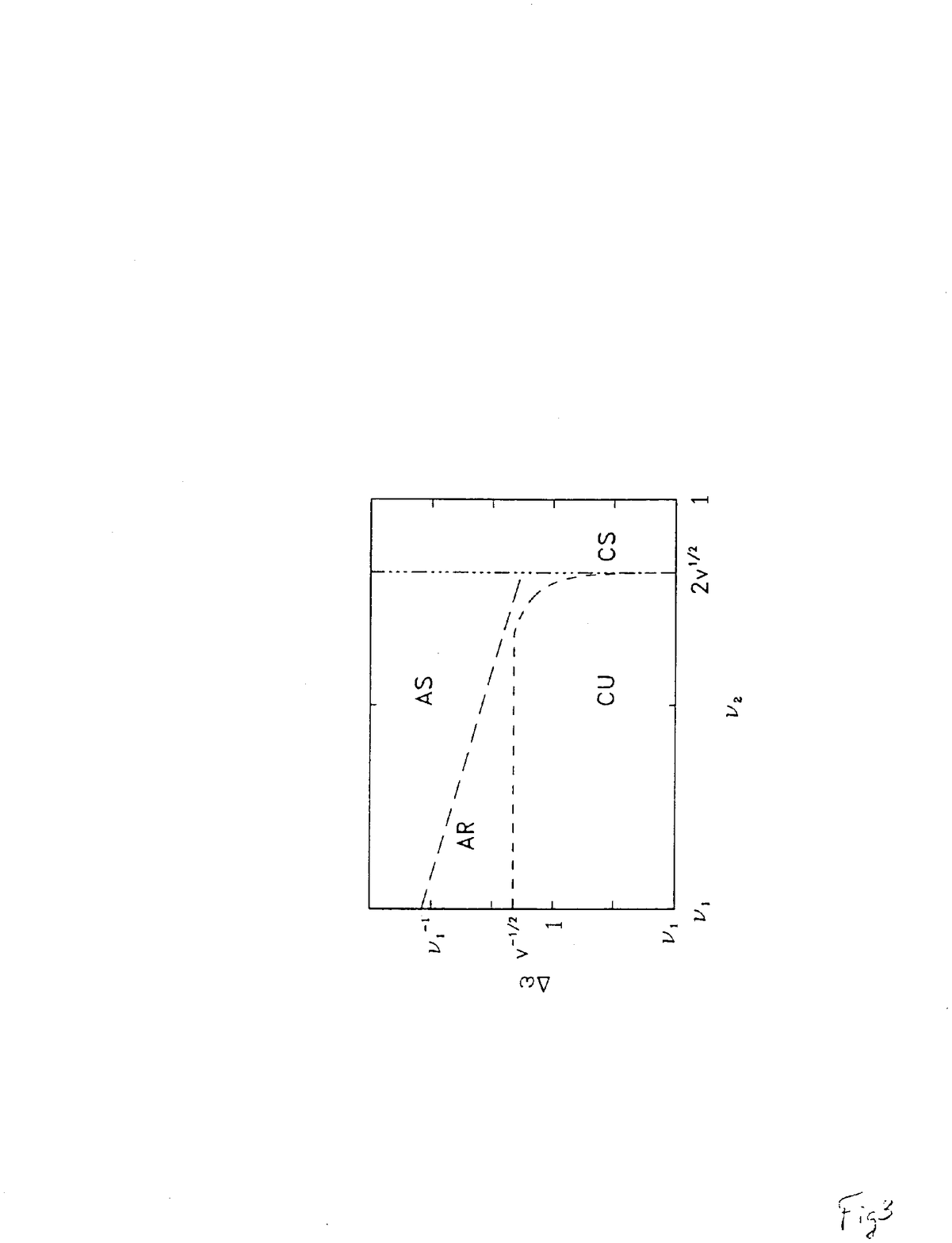}
\caption{Schematic stability diagram for the absolute growth rate.
The short-dashed line, between $\mid v \mid^{-1/2}$ on the vertical
axis and $2 \mid v \mid^{1/2}$ on the horizontal axis, represents
the boundary of the domain AR where the pump wave incoherence
reduces the absolute growth rate.  Here $v = \mid v_2 \mid$.  The
long-dashed line is the boundary of the domain AS where the pump
wave incoherence completely stabilizes the absolute growth. To the
right of the rightmost vertical line is the domain of coherent
stability.  This diagram assumes $\overline \nu_2 > \overline
\nu_1$.} \label{fig3}
\end{figure}
\vfill \eject

\newpage
\begin{figure}
\includegraphics[width=15cm]{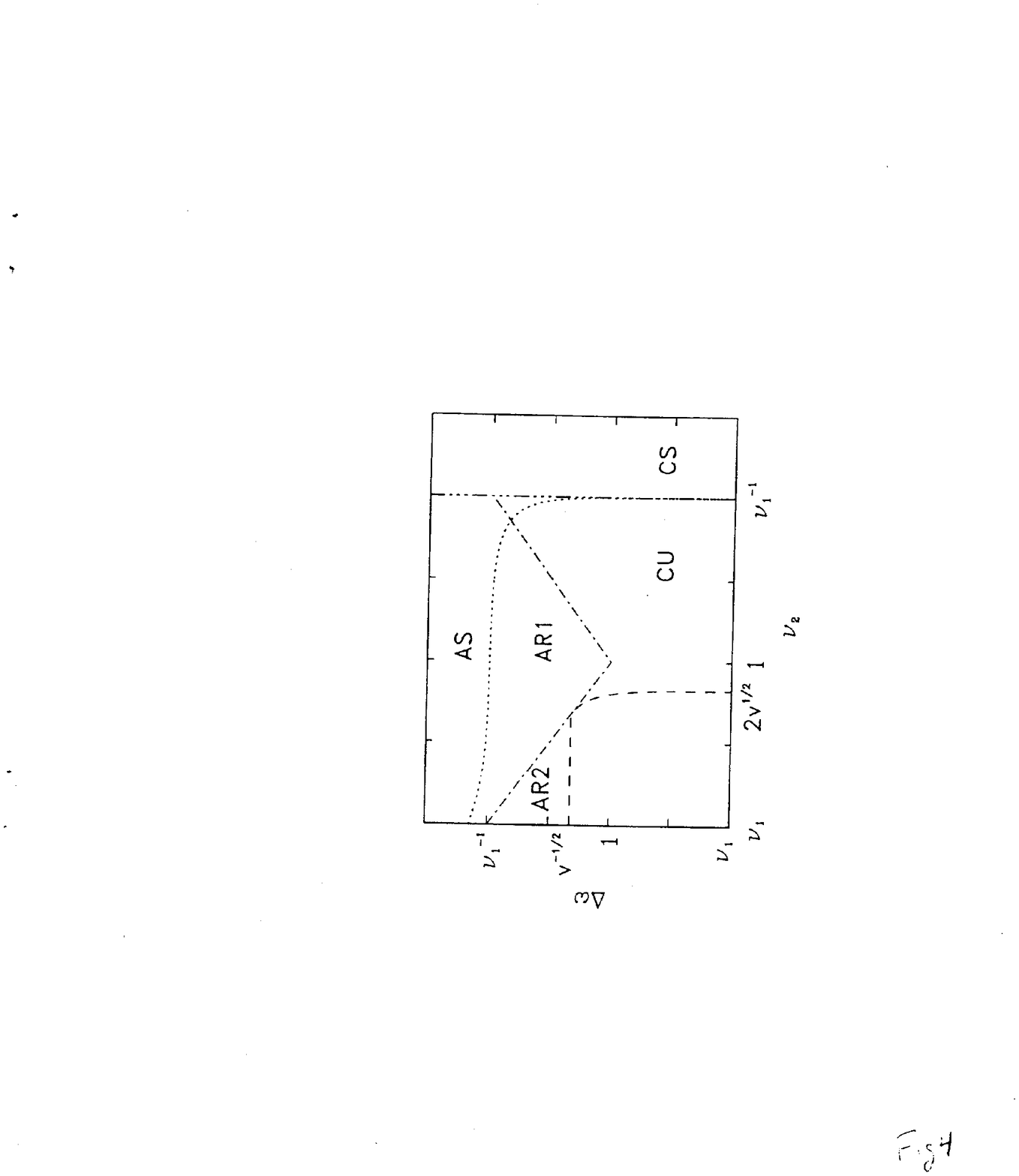}
\caption{Schematic stability diagram for the spatial amplification
growth rate.  The dotted-dashed lines represent the boundaries of
the domains AR1 and AR2 where the pump wave incoherence reduces the
spatial amplification rate; the dotted line of crosses is the
boundary of the domain AS (identical to the domain VS of Fig. 2)
where it completely stabilizes spatial amplification. The
short-dashed line (see Fig. 3) represents the threshold for the
existence of absolute instabilities in the case $V_1 V_2 < 0$.  The
rightmost vertical line separates the coherently stable (CS) and
unstable (CU) domains.  This diagram assumes $\overline \nu_2 >
\overline \nu_1$.} \label{fig4}
\end{figure}
\vfill \eject

\newpage
\begin{figure}
\includegraphics[width=15cm]{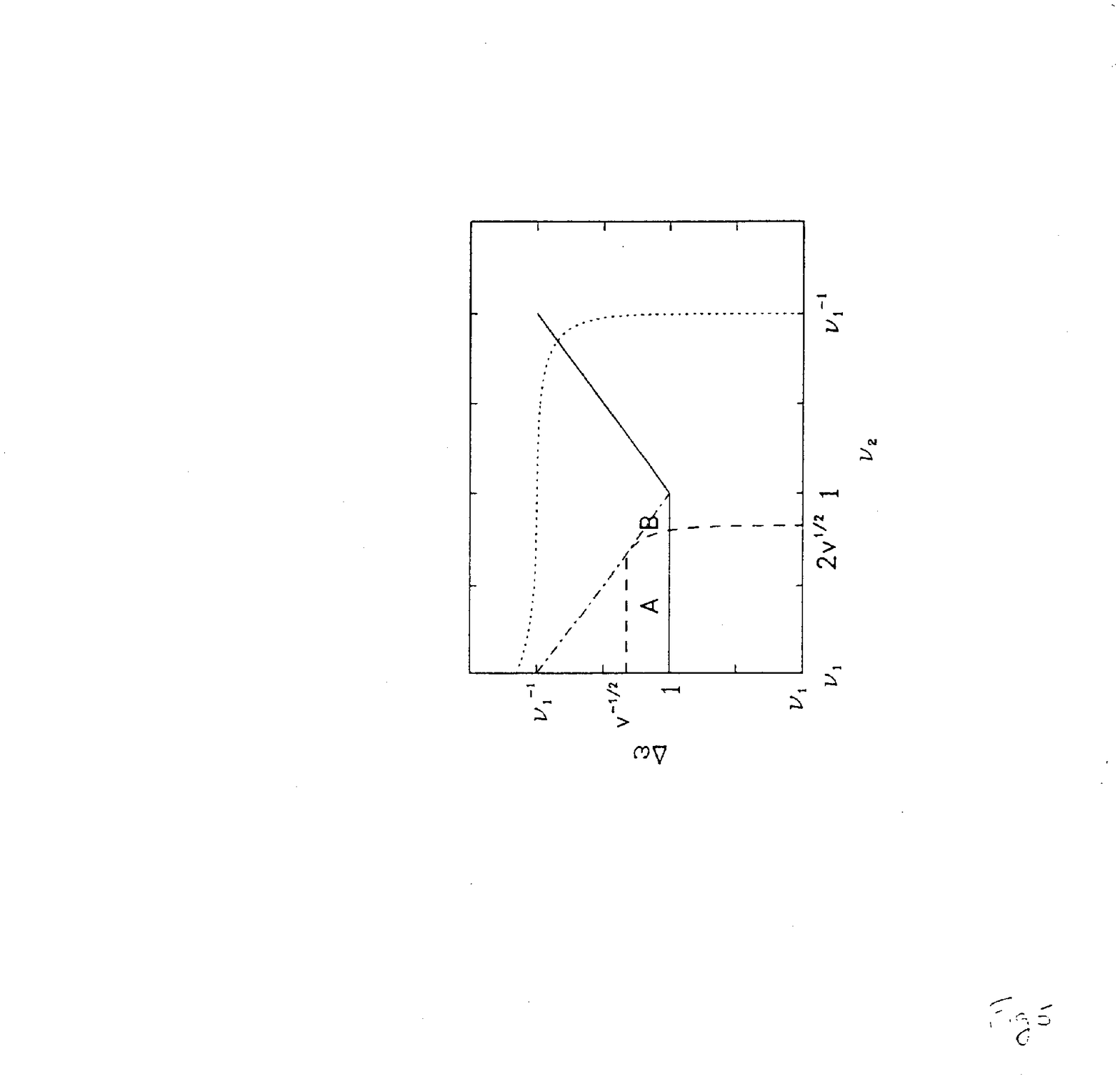}
\caption{Comparison of the short time and long time behavior.  The
domains A and B correspond to regimes for which the pump wave
incoherence reduces the initial parametric growth (convective
growth) whereas the long time behavior remains unaffected.  The
meaning of the solid and dotted, the short-dashed, and the
dotted-dashed lines are given in Figs. 2, 3, and 4 respectively. The
diagram assumes $\overline \nu_2 > \overline \nu_1$.} \label{fig5}
\end{figure}
\vfill \eject

\newpage
\begin{figure}
\includegraphics[width=15cm]{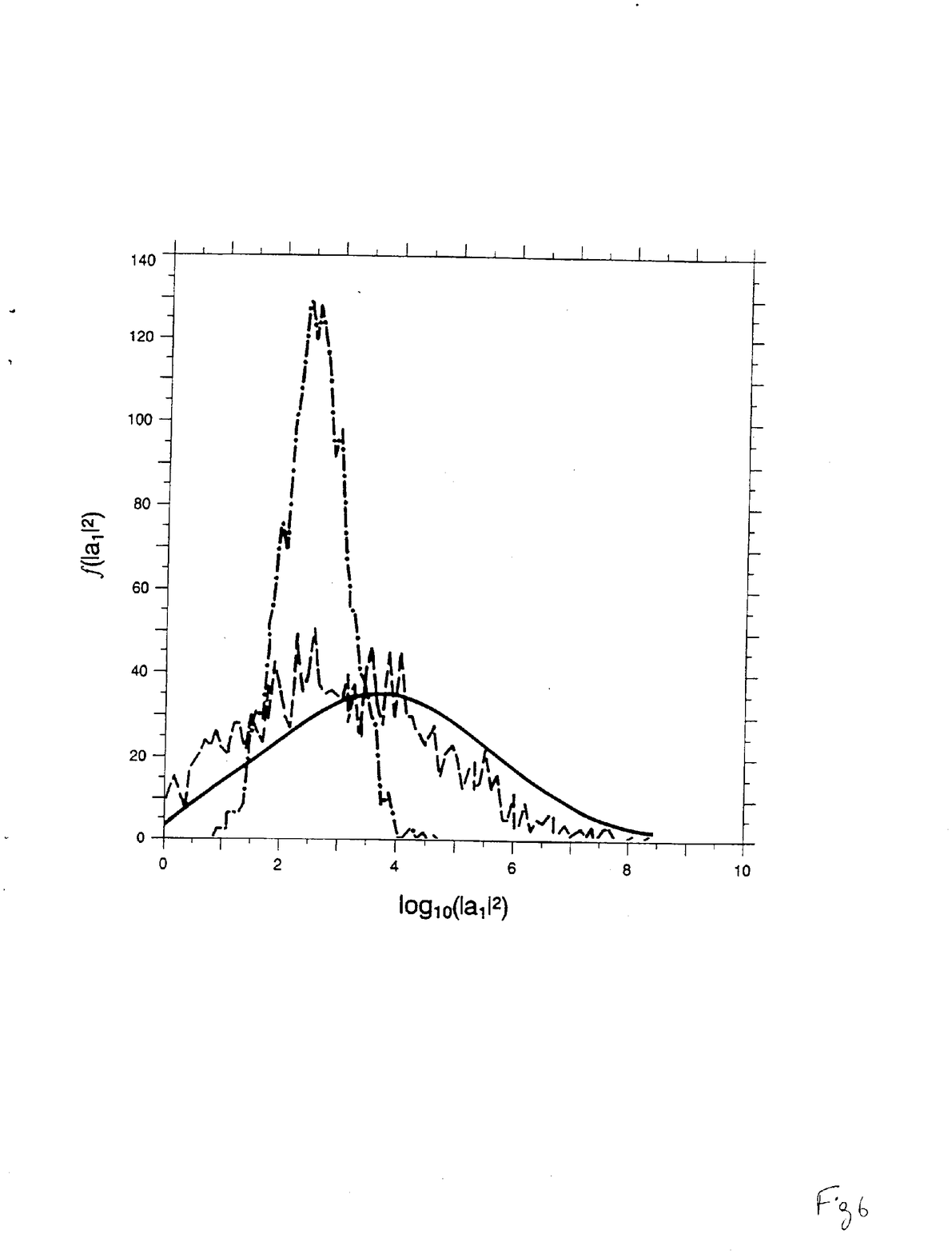}
\caption{Distribution of the intensity $\mid a_1 \mid^2$ for the
purely temporal problem.  The solid curve is the analytic solution,
and the broken curve is the numerically generated solution for
undamped waves.  The dashed-dotted curve is the numerical solution
when $\nu_2 = \gamma_0$.  The bandwidth is $\Delta \omega_0 = 10
\gamma_0$ and the abscissa is the base 10 logarithm of the intensity
$\mid a_1 \mid^2$ after an interval $\gamma_0 \Delta t = 50$.}
\label{fig6}
\end{figure}
\vfill \eject

\newpage
\begin{figure}
\includegraphics[width=15cm]{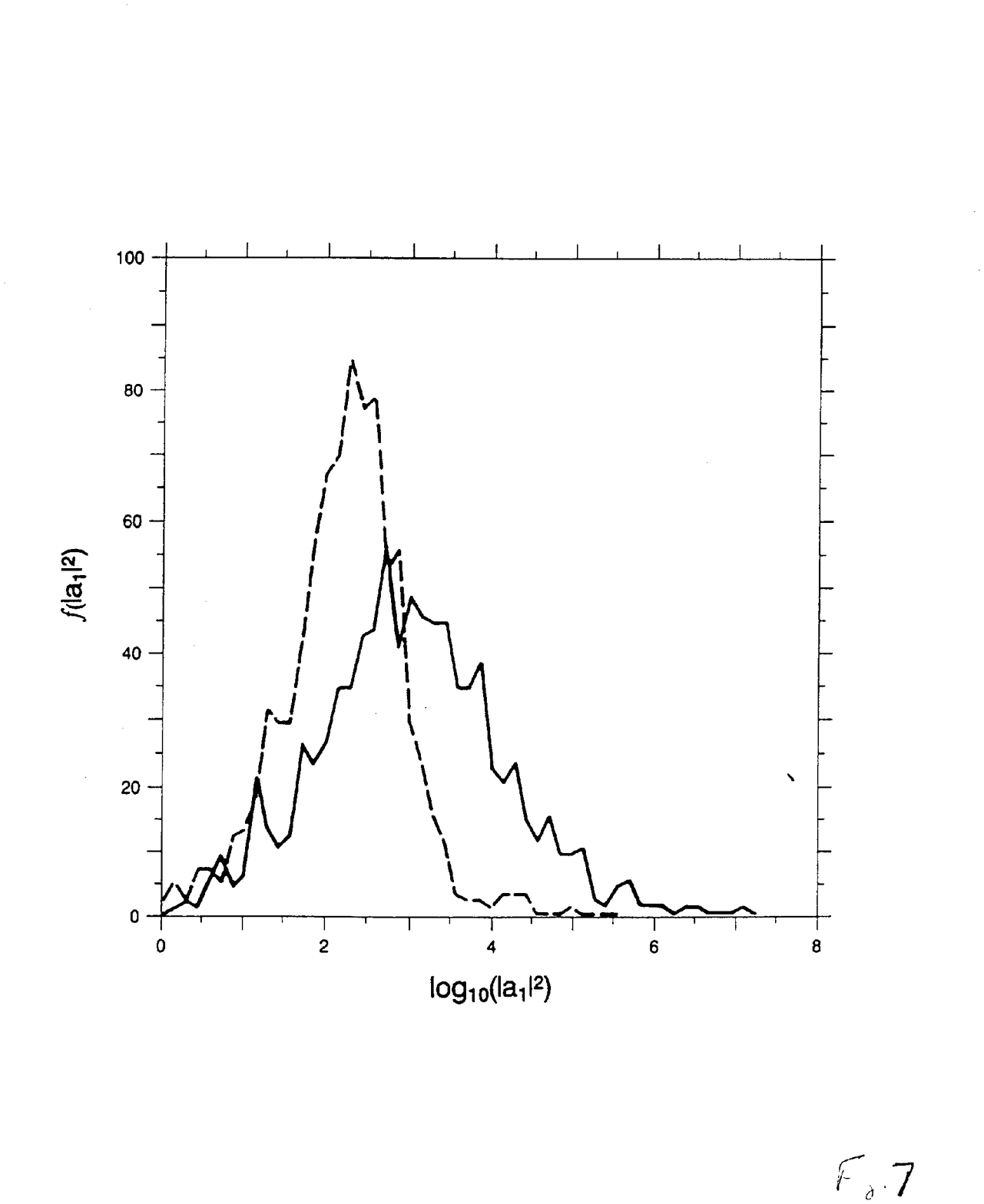}
\caption{Numerically generated distribution of the intensity $\mid
a_1 \mid^2$ for the space-time problem with purely temporal
bandwidth $\Delta \omega_0 = 10 \gamma_0$ after an interval
$\gamma_0 \Delta t = 45$. The solid curve has $\mid V_1 / V_2 \mid =
16$ and the dashed curve has $\mid V_1 / V_2 \mid = 100$.  The
abscissa is the base 10 logarithm of the intensity $\mid a_1
\mid^2$.} \label{fig7}
\end{figure}
\vfill \eject

\newpage
\begin{figure}
\includegraphics[width=15cm]{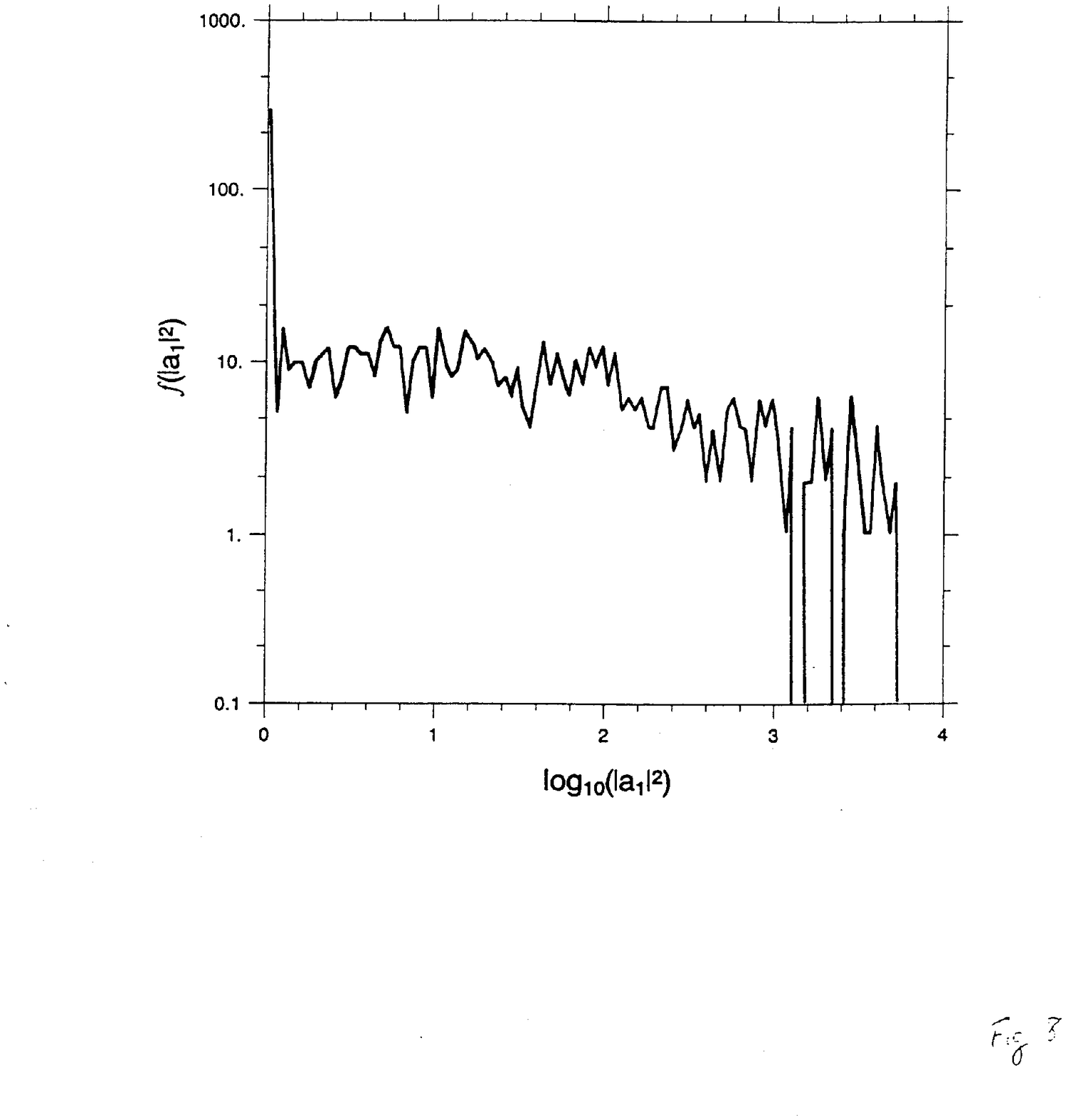}
\caption{Numerically generated distribution of the intensity $\mid
a_1 \mid^2$ for a spatially incoherent pump with $\Delta k_0 \mid
\sqrt { \mid V_1 V_2 \mid} / \gamma_0 = 4, \nu_1 = 2 \gamma_0$,
$\nu_2 = 0$, and $V_1 / V_2 = - 16$.  The abscissa is the base 10
logarithm of the intensity $\mid a_1 \mid^2$ after an interval
$\gamma_0 \Delta t = 30$.} \label{fig8}
\end{figure}
\vfill \eject

\tableofcontents

\end{document}